\documentclass[aps,prc,reprint,showpacs,superscriptaddress,nobibnotes]{revtex4-1}  % for a preview of print version 
\usepackage{amssymb}   % for math
\usepackage{graphicx}  % needed for figures
\usepackage{bm}        % for math
\usepackage{dcolumn}   % align numbers by decimal point in tables 
\graphicspath{{./}}
\setkeys{Gin}{clip=true,keepaspectratio=true} % defaults for graphicx
\begin{document}

   \title{Measurements of $d_{2}^{n}$ and $A_{1}^{n}$: Probing the neutron spin structure}

   %% From template for Version 4.1r of REVTeX, August 2010

% Group addresses by affiliation; use superscriptaddress for long
% author lists, or if there are many overlapping affiliations.
% For Phys. Rev. appearance, change preprint to twocolumn.
% Choose pra, prb, prc, prd, pre, prl, prstab, prstper, or rmp for journal
%  Add 'draft' option to mark overfull boxes with black boxes
%  Add 'showpacs' option to make PACS codes appear
%  Add 'showkeys' option to make keywords appear
%\preprint{}

%\title{Title of Paper}


%\thanks{}

\author{D.~Flay}
\email{flay@umass.edu}
\affiliation{Temple University, Philadelphia, PA 19122}
\affiliation{University of Massachusetts, Amherst, MA 01003}

\author{M.~Posik}
\affiliation{Temple University, Philadelphia, PA 19122}

\author{D.~S.~Parno}
\affiliation{Carnegie Mellon University, Pittsburgh, PA 15213}
\affiliation{Center for Experimental Nuclear Physics and Astrophysics, University of Washington, Seattle, WA 98195}

\author{K.~Allada}
\affiliation{University of Kentucky, Lexington, KY 40506}

\author{W.~R.~Armstrong}
\affiliation{Temple University, Philadelphia, PA 19122}
\affiliation{Argonne National Lab, Argonne, IL 60439}

\author{T.~Averett}
\affiliation{College of William and Mary, Williamsburg, VA 23187}

\author{F.~Benmokhtar}
% \affiliation{Carnegie Mellon University, Pittsburgh, PA 15213}
\affiliation{Duquesne University, Pittsburgh, PA 15282}

\author{W.~Bertozzi}
\affiliation{Massachusetts Institute of Technology, Cambridge, MA 02139}

\author{A.~Camsonne}
\affiliation{Thomas Jefferson National Accelerator Facility, Newport News, VA 23606}

\author{M.~Canan}
\affiliation{Old Dominion University, Norfolk, VA 23529}

\author{G.~D.~Cates}
\affiliation{University of Virginia, Charlottesville, VA 22904}

\author{C.~Chen}
\affiliation{Hampton University, Hampton, VA 23187}

\author{J.-P.~Chen}
\affiliation{Thomas Jefferson National Accelerator Facility, Newport News, VA 23606}

\author{S.~Choi}
\affiliation{Seoul National University, Seoul, South Korea}

\author{E.~Chudakov}
\affiliation{Thomas Jefferson National Accelerator Facility, Newport News, VA 23606}

\author{F.~Cusanno} \thanks{Deceased}
\affiliation{INFN, Sezione di Roma, I-00161 Rome, Italy}
\affiliation{Istituto Superiore di Sanit\`a, I-00161 Rome, Italy}

\author{M.~M.~Dalton}
\affiliation{University of Virginia, Charlottesville, VA 22904}

\author{W.~Deconinck}
\affiliation{Massachusetts Institute of Technology, Cambridge, MA 02139}

\author{C.~W.~de~Jager}
\affiliation{Thomas Jefferson National Accelerator Facility, Newport News, VA 23606}
\affiliation{University of Virginia, Charlottesville, VA 22904}

\author{X.~Deng}
\affiliation{University of Virginia, Charlottesville, VA 22904}

\author{A.~Deur}
\affiliation{Thomas Jefferson National Accelerator Facility, Newport News, VA 23606}

\author{C.~Dutta}
\affiliation{University of Kentucky, Lexington, KY 40506}

\author{L.~El~Fassi}
\affiliation{Rutgers, The State University of New Jersey, Piscataway, NJ 08855}
\affiliation{Mississippi State University, MS 39762}

\author{G.~B.~Franklin}
\affiliation{Carnegie Mellon University, Pittsburgh, PA 15213}

\author{M.~Friend}
\affiliation{Carnegie Mellon University, Pittsburgh, PA 15213}

\author{H.~Gao}
\affiliation{Duke University, Durham, NC 27708}

\author{F.~Garibaldi}
\affiliation{INFN, Sezione di Roma, I-00161 Rome, Italy}

\author{S.~Gilad}
\affiliation{Massachusetts Institute of Technology, Cambridge, MA 02139}

\author{R.~Gilman}
\affiliation{Thomas Jefferson National Accelerator Facility, Newport News, VA 23606}
\affiliation{Rutgers, The State University of New Jersey, Piscataway, NJ 08855}

\author{O.~Glamazdin}
\affiliation{Kharkov Institute of Physics and Technology, Kharkov 61108, Ukraine}

\author{S.~Golge}
\affiliation{Old Dominion University, Norfolk, VA 23529}

\author{J.~Gomez}
\affiliation{Thomas Jefferson National Accelerator Facility, Newport News, VA 23606}

\author{L.~Guo}
\affiliation{Los Alamos National Laboratory, Los Alamos, NM 87545}

\author{O.~Hansen}
\affiliation{Thomas Jefferson National Accelerator Facility, Newport News, VA 23606}

\author{D.~W.~Higinbotham}
\affiliation{Thomas Jefferson National Accelerator Facility, Newport News, VA 23606}

\author{T.~Holmstrom}
\affiliation{Longwood University, Farmville, VA 23909}

\author{J.~Huang}
\affiliation{Massachusetts Institute of Technology, Cambridge, MA 02139}

\author{C.~Hyde}
\affiliation{Old Dominion University, Norfolk, VA 23529}
\affiliation{Universit\'e Blaise Pascal/IN2P3, F-63177 Aubi\`ere, France}

\author{H.~F.~Ibrahim}
\affiliation{Cairo University, Giza 12613, Egypt}

\author{X.~Jiang}
\affiliation{Rutgers, The State University of New Jersey, Piscataway, NJ 08855}
\affiliation{Los Alamos National Laboratory, Los Alamos, NM 87545}

\author{ G.~Jin}
\affiliation{University of Virginia, Charlottesville, VA 22904}

\author{J.~Katich}
\affiliation{College of William and Mary, Williamsburg, VA 23187}

\author{A.~Kelleher}
\affiliation{College of William and Mary, Williamsburg, VA 23187}

\author{A.~Kolarkar}
\affiliation{University of Kentucky, Lexington, KY 40506}

\author{W.~Korsch}
\affiliation{University of Kentucky, Lexington, KY 40506}

\author{G.~Kumbartzki}
\affiliation{Rutgers, The State University of New Jersey, Piscataway, NJ 08855}

\author{J.~J.~LeRose}
\affiliation{Thomas Jefferson National Accelerator Facility, Newport News, VA 23606}

\author{R.~Lindgren}
\affiliation{University of Virginia, Charlottesville, VA 22904}

\author{N.~Liyanage}
\affiliation{University of Virginia, Charlottesville, VA 22904}

\author{E.~Long}
\affiliation{Kent State University, Kent, OH 44242}

\author{A.~Lukhanin}
\affiliation{Temple University, Philadelphia, PA 19122}

\author{V.~Mamyan}
\affiliation{Carnegie Mellon University, Pittsburgh, PA 15213}

\author{D.~McNulty}
\affiliation{University of Massachusetts, Amherst, MA 01003}

\author{Z.-E.~Meziani}
\email{meziani@temple.edu}
\affiliation{Temple University, Philadelphia, PA 19122}

\author{R.~Michaels}
\affiliation{Thomas Jefferson National Accelerator Facility, Newport News, VA 23606}

\author{M.~Mihovilovi\v{c}}
\affiliation{Jo\v{z}ef Stefan Institute, Ljubljana, Slovenia}

\author{B.~Moffit}
\affiliation{Massachusetts Institute of Technology, Cambridge, MA 02139}
\affiliation{Thomas Jefferson National Accelerator Facility, Newport News, VA 23606}

\author{N.~Muangma}
\affiliation{Massachusetts Institute of Technology, Cambridge, MA 02139}

\author{S.~Nanda}
\affiliation{Thomas Jefferson National Accelerator Facility, Newport News, VA 23606}

\author{A.~Narayan}
\affiliation{Mississippi State University, MS 39762}

\author{V.~Nelyubin}
\affiliation{University of Virginia, Charlottesville, VA 22904}

\author{B.~Norum}
\affiliation{University of Virginia, Charlottesville, VA 22904}

\author{Nuruzzaman}
\affiliation{Mississippi State University, MS 39762}

\author{Y.~Oh}
% \affiliation{Seoul National University, Taegu 702-701, Republic of Korea}
\affiliation{Seoul National University, Seoul 151-742, South Korea} 

\author{J.~C.~Peng}
\affiliation{University of Illinois at Urbana-Champaign, Urbana, IL 61801}

\author{X.~Qian} 
\affiliation{Duke University, Durham, NC 27708}
\affiliation{Kellogg Radiation Laboratory, California Institute of Technology, Pasadena, CA 91125}

\author{Y.~Qiang}
\affiliation{Duke University, Durham, NC 27708}
\affiliation{Thomas Jefferson National Accelerator Facility, Newport News, VA 23606}

\author{A.~Rakhman}
\affiliation{Syracuse University, Syracuse, NY 13244}

\author{R.~D.~Ransome}
\affiliation{Rutgers, The State University of New Jersey, Piscataway, NJ 08855}

\author{S.~Riordan}
\affiliation{University of Virginia, Charlottesville, VA 22904}

\author{A.~Saha} \thanks{Deceased}
\affiliation{Thomas Jefferson National Accelerator Facility, Newport News, VA 23606}

\author{B.~Sawatzky}
\affiliation{Temple University, Philadelphia, PA 19122}
\affiliation{Thomas Jefferson National Accelerator Facility, Newport News, VA 23606}

\author{M.~H.~Shabestari}
\affiliation{University of Virginia, Charlottesville, VA 22904}

\author{A.~Shahinyan}
\affiliation{Yerevan Physics Institute, Yerevan 375036, Armenia}

\author{S.~\v{S}irca}
\affiliation{University of Ljubljana, SI-1000 Ljubljana, Slovenia}

\author{P.~Solvignon}
\affiliation{Argonne National Lab, Argonne, IL 60439}
\affiliation{Thomas Jefferson National Accelerator Facility, Newport News, VA 23606}

\author{R.~Subedi}
\affiliation{University of Virginia, Charlottesville, VA 22904}

\author{V.~Sulkosky}
\affiliation{Massachusetts Institute of Technology, Cambridge, MA 02139}
\affiliation{Thomas Jefferson National Accelerator Facility, Newport News, VA 23606}

\author{W.~A.~Tobias}
\affiliation{University of Virginia, Charlottesville, VA 22904}

\author{W.~Troth}
\affiliation{Longwood University, Farmville, VA 23909}

\author{D.~Wang}  %Diancheng
\affiliation{University of Virginia, Charlottesville, VA 22904}

\author{Y.~Wang}
\affiliation{University of Illinois at Urbana-Champaign, Urbana, IL 61801}

\author{B.~Wojtsekhowski}
\affiliation{Thomas Jefferson National Accelerator Facility, Newport News, VA 23606}

\author{X.~Yan}
\affiliation{University of Science and Technology of China, Hefei 230026, People's Republic of China}

\author{H.~Yao}
\affiliation{Temple University, Philadelphia, PA 19122}
\affiliation{College of William and Mary, Williamsburg, VA 23187}

\author{Y.~Ye}
\affiliation{University of Science and Technology of China, Hefei 230026, People's Republic of China}

\author{Z.~Ye}
\affiliation{Hampton University, Hampton, VA 23187}

\author{L.~Yuan}
\affiliation{Hampton University, Hampton, VA 23187}

\author{X.~Zhan}
\affiliation{Massachusetts Institute of Technology, Cambridge, MA 02139}

\author{Y.~Zhang} %Yi
\affiliation{Lanzhou University, Lanzhou 730000, Gansu, People's Republic of China}

\author{Y.-W.~Zhang} %Yawei
\affiliation{Lanzhou University, Lanzhou 730000, Gansu, People's Republic of China}
\affiliation{Rutgers, The State University of New Jersey, Piscataway, NJ 08855}

\author{B.~Zhao}
\affiliation{College of William and Mary, Williamsburg, VA 23187}

\author{X.~Zheng}
\affiliation{University of Virginia, Charlottesville, VA 22904}

\collaboration{Jefferson Lab Hall A Collaboration}
\noaffiliation

   \begin{abstract}

We report on the results of the E06-014 experiment performed at Jefferson Lab 
in Hall A, where a precision measurement of the twist-3 matrix element $d_2$ 
of the neutron ($d_{2}^{n}$) was conducted.  The quantity $d_{2}^{n}$ represents 
the average color Lorentz force a struck quark experiences in a deep inelastic electron 
scattering event off a neutron due to its interaction with the hadronizing remnants.
This color force was determined from a linear combination of the third moments 
of the $^3$He spin structure functions, $g_1$ and $g_2$, after nuclear corrections
had been applied to these moments.  
The structure functions were obtained from a measurement of the unpolarized 
cross section and of double-spin asymmetries in the scattering of a longitudinally 
polarized electron beam from a transversely and a longitudinally polarized 
$^{3}$He target.
The measurement kinematics included two average $Q^{2}$ bins of 
$3.2$\,GeV$^{2}$ and $4.3$\,GeV$^{2}$, and Bjorken-$x$  $0.25 \leq x \leq 0.90$ 
covering the deep inelastic and resonance regions.
We have found that $d_2^n$ is small and negative for $\left<Q^{2}\right> = 
3.2$\,GeV$^{2}$, and even smaller for $\left<Q^{2}\right> = 4.3$\,GeV$^{2}$, 
consistent with the results of a lattice QCD calculation.
The twist-4 matrix element $f_{2}^{n}$ was extracted by combining our measured 
$d_{2}^{n}$ with the world data on the first moment in $x$ of $g_{1}^{n}$, $\Gamma_{1}^{n}$.
We found $f_{2}^{n}$ to be roughly an order of magnitude larger than $d_{2}^{n}$.  
Utilizing the extracted $d_{2}^{n}$ and $f_{2}^{n}$ data, we separated the Lorentz 
color force into its electric and magnetic components, $F_{E}^{y,n}$ and $F_{B}^{y,n}$, 
and found them to be equal and opposite in magnitude, in agreement with the predictions 
from an instanton model but not with those from QCD sum rules.  Furthermore, using the 
measured double-spin asymmetries, we have extracted the virtual photon-nucleon asymmetry 
on the neutron $A_{1}^{n}$, the structure function ratio $g_{1}^{n}/F_{1}^{n}$, 
and the quark ratios $(\Delta u + \Delta \bar{u})/(u + \bar{u})$ and 
$(\Delta d + \Delta \bar{d})/(d + \bar{d})$.  These results were found to be 
consistent with DIS world data and with the prediction of the constituent quark 
model but at odds with the perturbative quantum chromodynamics predictions at large $x$.   
       
\end{abstract}


   \maketitle

   \tableofcontents

   %===============================================================================
\section{Introduction} \label{sec:intro} 
\subsection{Overview of nucleon structure}
%===============================================================================

Experiments utilizing the scattering of leptons from nucleons have been instrumental 
in uncovering the complex structure of subatomic matter over the past half century.  
In the mid-1950s, elastic scattering of electrons from hydrogen revealed that the 
proton is not a point-like particle but has internal structure~\cite{Mcallister:1956ng}; 
in the 1970s, deep-inelastic scattering (DIS) of electrons from hydrogen showed 
that point-like particles, labeled ``partons,'' are the underlying constituents 
of the proton~\cite{Bloom:1969kc,*Breidenbach:1969kd}.  These partons were later 
identified as quarks and gluons in the modern theory of strong interactions, 
quantum chromodynamics (QCD)~\cite{Gross:1973id,*Gross:1973ju,*Gross:1974cs}. 

Since the late 1970s, scattering of polarized lepton beams from polarized nucleons 
and polarized light nuclear targets (deuterium and $^3$He) has given us the opportunity 
to probe the spin structure of the nucleon encoded in the $g_{1}$ and $g_{2}$ 
spin-structure functions.  In particular, worldwide DIS studies focusing on $g_{1}$ 
as a function of both Bjorken-$x$ and $Q^{2}$ allowed the determination of the 
fraction of the proton spin that is carried by the quarks~\cite{Kuhn:2008sy,Aidala:2012mv} 
and by the gluons~\cite{Adare:2014hsq,Adamczyk:2014ozi}.  Here, $x$ is interpreted 
as the fractional momentum of the parent nucleon carried by the struck quark in 
the infinite momentum frame, and $Q^{2} \equiv -q^{2}$ is the four-momentum 
transferred to the target squared. 

Early theoretical work~\cite{Shuryak:1981pi,*Jaffe:1989xx,*Jaffe:1990qh} has shown 
that the $g_{1}(x,Q^2)$ and $g_{2}(x,Q^2)$ spin-structure functions contain information 
on quark-gluon correlations.  These dynamical effects are accessible through the 
$Q^2$-variations of these functions beyond those of the calculable perturbative 
QCD (pQCD) radiative corrections~\cite{Dokshitzer:1977sg,*Gribov:1972ri,*Altarelli:1977zs}. 
In fact, they appear in an expansion of both the measured $g_1$ spin-structure function 
and its moments in $x$ in powers of $1/Q^{2}$, but only at higher order.  In contrast, 
in the measured $g_{2}$ spin-structure function, quark-gluon interactions are accessible 
at leading order in a similar expansion and thus suffer no $1/Q^{2}$ suppression.  
This makes measurements of $g_{2}$ particularly sensitive and important for studying 
multi-parton correlations in the nucleon.  

Studies of the moments in $x$ of spin-structure functions have resulted in fundamental 
tests of QCD like that of the Bjorken sum rule~\cite{Bjorken:1966jh,*Bjorken:1969mm}; here, 
not only do they offer an opportunity to test our understanding of pQCD beyond the 
simple partonic picture, but they also allow for measured observables to be tested 
against {\it ab~initio} calculations of lattice QCD.  While there is a wealth of data 
available for $g_1$, fewer data exist for $g_2$---especially in the valence region. 
This region provides the dominant contribution to higher moments. 
These moments are of interest because the contribution arising from the lower-$x$ region 
of integration, where the structure functions are unknown, is small. Thus these higher 
moments offer robust experimental results relevant for a comparison with 
lattice QCD, for example.  Finally, it is worth noting that high-precision data of 
the $g_1$ nucleon structure function in the valence region of deep inelastic 
scattering---namely $x \ge 0.6$---are still sparse, and every new data set with good 
precision offers a real possibility to test nucleon models in a domain sensitive to 
those models' parameters.  

%===============================================================================
\subsection{The $g_{2}$ structure function and quark-gluon correlations} \label{sec:g2-intro} 
%===============================================================================

While the polarized structure function $g_{2}$ has no clear interpretation in the 
quark-parton model~\cite{Manohar:1992tz}, it is known to contain quark-gluon 
correlations, and can be decomposed as:

\begin{equation} \label{eqn:g2_full} 
   g_{2}\left(x,Q^{2}\right) = g_{2}^{\text{WW}}\left(x,Q^{2}\right) + \bar{g}_{2}\left(x,Q^{2}\right), 
\end{equation} 

\noindent where $\bar{g}_{2}$ is the component of $g_{2}$ that contains the 
quark-gluon correlations~\cite{Jaffe:1989xx}, given by~\cite{Cortes:1991ja}:

\begin{eqnarray}
   \bar{g}_{2}\left(x,Q^{2}\right) &=& \int_{x}^{1} \frac{\partial}{\partial y} 
                                       \left [ \frac{m_q}{M}h_T\left(x,Q^{2}\right) + \xi \left(y,Q^{2}\right) \right ] 
                                       \frac{dy}{y}.  
\end{eqnarray}

\noindent Here, $h_T$ denotes the transversity distribution in the 
nucleon~\cite{Filippone:2001ux}, $\xi$ the quark-gluon correlation function, 
$m_q$ the quark mass of flavor $q$ and $M$ the nucleon mass.  The quantity 
$g_{2}^{\text{WW}}$ in Eq.~\ref{eqn:g2_full} is the Wandzura-Wilczek 
term, which is fully determined from the knowledge of 
the $g_1$ structure function~\cite{Wandzura:1977qf}: 

\begin{eqnarray}
   g_{2}^{\text{WW}}\left(x,Q^{2}\right) &=& - g_{1}\left(x,Q^{2}\right) 
                                          + \int_{x}^{1} \frac{g_{1}\left(y,Q^{2}\right)}{y} dy.  
\end{eqnarray}
 
Under the operator product expansion (OPE)~\cite{Wilson:1969zs}, one can access the 
effects of quark-gluon correlations via the third moment of a linear combination
of $g_{1}$ and $g_{2}$: 

\begin{eqnarray} \label{eqn:d2-sf-intro} 
   d_{2}\left(Q^{2}\right)  &=& 3 \int_{0}^{1} x^{2} \bar{g}_{2}\left(x,Q^{2}\right) dx\nonumber \\        
   &=& \int_{0}^{1} x^{2} \left[ 2g_{1}\left(x,Q^{2}\right) + 3g_{2}\left(x,Q^{2}\right) \right] dx.                    
\end{eqnarray} 

\noindent Because of the $x^{2}$-weighting, $d_{2}$ is particularly sensitive to the 
large-$x$ behavior of $\bar{g}_{2}$.  The quantity $d_{2}$ is related to a specific 
twist-3 ($\tau = 3$) matrix element consisting of local operators of quark and gluon 
fields~\cite{Ehrnsperger:1993hh,Filippone:2001ux,Burkardt:2008ps}:  
% perhaps a more original or earlier citation is called for here like Schafer, Manckiewicz and Ji etc..

\begin{eqnarray} \label{eqn:d2_matrix_elem} 
   2 M P^{+}P^{+} S^{x} d_{2} &=& g \langle P,S \vert \bar{\psi}(0) \gamma^{+} G^{+y}(0) \psi(0) \vert P,S \rangle,  
\end{eqnarray}

\noindent where $P$ denotes the nucleon momentum, $S$ its spin, $\psi$ the quark field, and 
$g$ the QCD coupling constant.  The $+$ superscript indicates the equation is expressed in 
light-cone coordinates.  In analogy to the electromagnetic Lorentz force $F^{y}$ 
that acts on a charged particle, the gluon field 
$G^{+y} = \left( B^{x} - E^{y} \right)/\sqrt{2} = F^{y}$, where $B^{x}$ 
and $E^{y}$ are the transverse components of the color magnetic and color electric field, 
respectively; the $z$ direction is defined by the three-momentum transfer of the virtual photon~\cite{Burkardt:2008ps}.   

There are two interpretations of $d_{2}$ in the literature.  
The first connects $d_{2}$ with color electromagnetic fields induced in a transversely 
polarized nucleon probed by a virtual photon.  These induced color fields (appearing 
in Eq.~\ref{eqn:d2_matrix_elem}) are represented as color polarizabilities 
$\chi$~\cite{Filippone:2001ux}:  

\begin{eqnarray}
   \chi_{E}\vec{S} &=& \frac{1}{2M^{2}} \langle P,S \vert \psi^{\dagger} g \vec{a} \times \vec{E} \psi \vert P,S \rangle  \label{eqn:chi-e}\\ 
   \chi_{B}\vec{S} &=& \frac{1}{2M^{2}} \langle P,S \vert \psi^{\dagger} g \vec{B} \psi \vert P,S \rangle, \label{eqn:chi-b}  
\end{eqnarray}

\noindent where $\vec{a}$ denotes the velocity of the struck quark.  Then, $d_{2}$ can 
be expressed as: 

\begin{equation}
   d_{2} = \frac{1}{4}\left( \chi_{E} + 2\chi_{B} \right).   
\end{equation} 

A second, more recent interpretation shows that the matrix element connected to 
$d_{2}$ represents an average color Lorentz force $F^{y}$ acting on the struck quark due to 
the remnant di-quark system at the instant it is struck by the virtual 
photon (cf. Eq.~\ref{eqn:d2_matrix_elem}):

\begin{eqnarray} \label{eqn:d2_force}  
   F^{y}(0) &\equiv& \langle P,S \vert \bar{\psi}(0) \gamma^{+} G^{+y}(0) \psi(0) \vert P,S \rangle \\ 
            &=& - M^{2} d_{2},  
\end{eqnarray}

\noindent where the last equality is true only in the rest frame of the nucleon~\cite{Burkardt:2008ps}. 

Combining measurements of $d_{2}$ with the twist-4 matrix element $f_{2}$
allows the extraction of the color electric and magnetic forces 
$F_{E}^{y}$ and $F_{B}^{y}$~\cite{Burkardt:2008ps}: 
 
\begin{eqnarray}
   d_{2} &=& - \frac{1}{M^{2}}\left(  F_{E}^{y} + F_{B}^{y} \right) \\ 
   f_{2} &=& - \frac{2}{M^{2}}\left( 2F_{E}^{y} - F_{B}^{y} \right).  
\end{eqnarray}

\noindent The quantity $f_{2}$ is sensitive to quark-gluon correlations, since 
it is expressed as a matrix element similar to $d_{2}$, containing a mixed quark-gluon field 
operator~\cite{Shuryak:1981kj,Shuryak:1981pi,Edelmann:2000,Osipenko:2004xg}.  
The $f_{2}$ matrix element cannot be measured directly, but can be extracted from $g_{1}$ 
data by utilizing a twist expansion of $\Gamma_{1}$, the first moment of $g_{1}$:   

\begin{eqnarray} \label{eqn:gamma1} 
   \Gamma_{1} &\equiv& \int_{0}^{1} g_{1} dx \nonumber \\ 
              &=&      \mu_{2} + \frac{M^{2}}{9Q^{2}}\left( a_{2} + 4d_{2} + 4f_{2} \right) 
                     + \frac{\mu_{6}}{Q^{4}}\nonumber \\
                     &+& \mathcal{O}\left( \frac{1}{Q^{6}} \right) + \ldots.  
\end{eqnarray} 

\noindent For simplicity the $Q^{2}$-dependence of the structure functions, 
matrix elements and $\mu_n$ terms has been omitted in Eq.~\ref{eqn:gamma1}.  
The quantity $a_{2} = \int x^{2} g_{1} dx$ is the third moment of $g_{1}$, a 
twist-2 matrix element that has connections to target mass corrections.  
The term $\mu_{6}$ is a higher-twist ($\tau > 4$) term.  The quantity $\mu_{2}$ 
is the twist-2 contribution, given as:

\begin{equation} \label{eqn:mu_2}
   \mu_{2}\left( Q^{2} \right) = C_{ns}\left(Q^{2}\right) \left( - \frac{1}{12}g_{A} + \frac{1}{36}a_{8} \right) 
                               + C_{s}\left(Q^{2}\right)\frac{1}{9}\Delta \Sigma, 
\end{equation}

\noindent where $C_{ns}$ and $C_{s}$ denote the non-singlet and singlet Wilson 
coefficients~\cite{Larin:1997qq}, $g_{A}$ the flavor-triplet axial charge, 
$a_{8}$ the octet axial charge and $\Delta\Sigma \equiv \Delta\Sigma \left( Q^{2} = \infty \right)$, 
the renormalization group invariant definition of the singlet axial current. 
This definition of $\Delta\Sigma$ is used to factorize all of the $Q^{2}$ dependence 
into the Wilson coefficients, as was done in Refs.~\cite{Osipenko:2004xg,Meziani:2004ne}.  
The $f_{2}$ matrix element can be extracted from Eq.~\ref{eqn:gamma1} by first 
subtracting $\mu_{2}$ from $\Gamma_{1}$ and then fitting the result as a
function of $1/Q^{2}$.

In practice, in order to access the spin-structure functions $g_1$ and $g_2$, we measure experimental asymmetries: 

\begin{eqnarray}
   A_{\parallel} &\equiv& \frac{\sigma^{\downarrow\Uparrow} - \sigma^{\uparrow\Uparrow}}{
                                \sigma^{\downarrow\Uparrow} + \sigma^{\uparrow\Uparrow}}   \\
                 &=& \frac{4\alpha^{2}}{M Q^{2}} \frac{(1-y)(2-y)}{2y^{2}\sigma_{0}}  \times \nonumber \\  
                 & & \frac{(1-y)\sin \theta}{1 + (1-y)\left[ \cos \theta + \tan \left(\theta/2\right)\sin \theta \right] } \times \nonumber \\
                 & & \left\{ y \frac{1 + (1-y)\cos\theta}{(1-y)\sin\theta}g_{1} - 2\tan\left( \theta/2 \right)g_{2} \right\} \nonumber  
                     \label{eqn:apara_intro}\\ 
                 &{\rm and}& \nonumber   \\ 
    A_{\perp}    &\equiv& \frac{\sigma^{\downarrow\Rightarrow} - \sigma^{\uparrow\Rightarrow}}{
                                \sigma^{\downarrow\Rightarrow} + \sigma^{\uparrow\Rightarrow}}  \\  
                 &=& \frac{4\alpha^{2}}{M Q^{2}} \frac{(1-y)(2-y)}{2y^{2}\sigma_{0}} \times \nonumber \\ 
                 & & \frac{(1-y)\sin \theta}{1 + (1-y)\left[ \cos \theta + \tan \left(\theta/2\right)\sin \theta \right] } \times \nonumber \\
                 & & \left\{ y g_{1} + 2 g_{2} \right\}. \nonumber \label{eqn:aperp_intro} 
\end{eqnarray}

\noindent The quantity $\sigma^{s S}$ denotes the polarized cross section for electron spin $s$ and target spin $S$.  
The $\uparrow$ ($\downarrow$) indicates the electron spin parallel (antiparallel) to its momentum, 
and $\Uparrow$ ($\Downarrow$) indicates the target spin parallel (antiparallel) to the electron beam 
momentum.  The $\Leftarrow$ ($\Rightarrow$) indicates the target spin perpendicular to the beam 
momentum, pointing away from (towards) the side of the beamline on which the scattered 
electron is detected.  The quantity $y = (E - E')/E$ is the fractional energy transferred to the target, 
with $E$ being the electron beam energy and $E'$ the scattered electron energy, with $E$ and $E'$ measured
in the laboratory frame. The quantity $\alpha$ denotes the electromagnetic coupling constant and $\theta$ 
the electron scattering angle.  The quantity $\sigma_{0}$ is the unpolarized electron 
scattering cross section.  The dependence of $g_{1}$, $g_{2}$ and $\sigma_{0}$ on $x$ and $Q^{2}$ has 
been suppressed for simplicity.  

The two spin structure functions $g_{1}$ and $g_{2}$ can be expressed in terms of the experimental 
observables $A_{\parallel}$, $A_{\perp}$ and $\sigma_{0}$ by combining and inverting 
Eqs.~\ref{eqn:apara_intro} and~\ref{eqn:aperp_intro}.  Then the expression for $d_{2}$ in 
Eq.~\ref{eqn:d2-sf-intro} can be re-written in terms of those experimental observables: 

\begin{eqnarray} \label{eqn:d2_exp} 
  d_{2} &=& \int_0^1 dx \frac{MQ^2}{4{\alpha}^2} \frac{x^2 y^2}{\left( 1 - y\right) \left( 2 - y\right)}
                        \sigma_0 \times  \\
        & & \Bigg[ \left( 3 \frac{1 + \left( 1 - y \right)\cos \theta}{\left( 1 - y \right)\sin \theta} 
                 + \frac{4}{y} \tan \left( \theta/2 \right)\right)A_{\perp}  \nonumber \\  
        & & + \left( \frac{4}{y} - 3 \right)A_{\parallel} \Bigg].  \nonumber  
\end{eqnarray}

\begin{figure}[hbt]
   \centering 
   \includegraphics[width=0.5\textwidth]{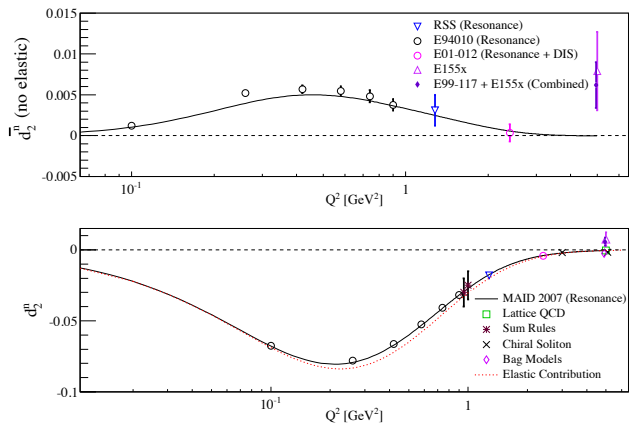}
   \caption{(Color online) The world $d_{2}^{n}$ data as a function of $Q^{2}$.  
            Upper panel: Data and models without the elastic contribution.  Bottom panel:
            Data and models with the elastic contribution included.  The experimental
            resonance data from JLab E94-010~\protect\cite{Amarian:2003jy} and RSS~\protect\cite{Slifer:2008xu},
            along with resonance plus DIS data from E01-012~\protect\cite{Solvignon:2013yun}
            are shown at $Q^{2} \lesssim 3$\,GeV$^{2}$, while at larger $Q^{2}$ DIS data from 
            SLAC E155x~\protect\cite{Anthony:2002hy} and combined data from JLab E99-117 and SLAC E155x~\protect\cite{Zheng:2004ce}
            are shown.  The solid curve is from a MAID~\protect\cite{Drechsel:2007if} calculation, which is 
            dominated by the resonance contribution.  Model calculations for $Q^{2} \approx 1$\,GeV$^{2}$ 
            from a QCD sum rule approach from Ref.~\protect\cite{Stein:1994zk} (offset lower in $Q^{2}$) 
            and Ref.~\protect\cite{Balitsky:1989jb} are shown.
            A chiral soliton model~\protect\cite{Weigel:1996jh,*Weigel:2000gx} and a bag model~\protect\cite{Song:1996ea} 
            are also given.  Additionally, a lattice QCD~\protect\cite{Gockeler:2005vw} 
            calculation is shown.  
            The model calculations that include the elastic contribution are shown in the 
            lower panel only.  We added the elastic contribution to the MAID model in 
            the lower panel.  The elastic contribution to $d_{2}^{n}$ is given in the  
            lower panel by the dashed curve, evaluated using the Cornwall-Norton (CN) 
            moments (see Appendix~\ref{sec:ope_app}) where the 
            Riordan~\protect\cite{Riordan:2010id} and Kelly~\protect\cite{Kelly:2004hm} parameterizations 
            are used for $G_{E}^{n}$ and $G_{M}^{n}$, respectively.   
           }
   \label{fig:d2n_current}
\end{figure}

The prior world data for $d_{2}^{n}$ as a function of $Q^{2}$~\footnote{In this paper, natural units are used.} 
are presented in Fig.~\ref{fig:d2n_current}.  The top panel shows measured data and model calculations 
without the elastic contribution, while the bottom panel shows the same data and models with the elastic 
contribution included.  Resonance measurements from JLab E94-010~\cite{Amarian:2003jy} and RSS~\cite{Slifer:2008xu},
along with resonance plus DIS data from E01-012~\cite{Solvignon:2013yun}, are shown at 
$Q^{2} \lesssim 3$\,GeV$^{2}$.  At large $Q^{2}$ towards 5\,GeV$^{2}$ are DIS measurements from 
SLAC E155x~\cite{Anthony:2002hy} and the combined data from JLab E99-117 and SLAC E155x~\cite{Zheng:2004ce}.
In the latter data set, $d_{2}^{n}$ was evaluated by combining the $g_{2}^{n}$ data from JLab E99-117 
with the $g_{2}$ data of SLAC E155x, and $\bar{g}_{2}$ was assumed to be $Q^{2}$-independent and
to follow $\bar{g}_{2} \propto (1-x)^{m}$ with $m = 2$ or 3 for $x \gtrsim 0.78$~\cite{Anthony:2002hy}, 
for which there were no data from either experiment~\cite{Zheng:2004ce}.  

The solid curve in Fig.~\ref{fig:d2n_current} is from a MAID~\cite{Drechsel:2007if} calculation, 
which uses phenomenelogical fits to electro- and photoproduction data for the nucleon, extending 
from the single-pion production threshold to the resonance/DIS boundary at $W = 2$\,GeV.  The major 
resonances are modeled using Breit-Wigner functions to construct the production channels.  The bottom panel 
displays the results of additional model calculations from a QCD sum rule approach~\cite{Stein:1994zk,Balitsky:1989jb}, 
which in general uses dispersion relations, combined with the OPE, to interpolate 
between the perturbative and non-perturbative regimes of QCD.  The two calculations presented at 
$Q^{2} \approx 1$\,GeV$^{2}$ use a three-quark field with~\cite{Stein:1994zk} 
(offset lower in $Q^{2}$ in Fig.~\ref{fig:d2n_current}) and without~\cite{Balitsky:1989jb} a gluon 
field. A chiral soliton model~\cite{Weigel:1996jh,*Weigel:2000gx} is shown, where the nucleon is 
described as a non-linear dynamical system consisting of ``mesonic lumps''~\cite{Weigel:1996jh} 
governed by a $U(1)\times SU(2)_{L}\times SU(2)_{R}$ chiral symmetry.  Another model displayed is 
a bag model, in which the quarks are confined to a nucleon ``bag.'' Here, the confinement mechanism of QCD 
is simulated using quark-gluon and gluon-gluon interactions~\cite{Song:1996ea}.  The model also 
includes generalized spin-dependent effects via an explicit symmetry-breaking parameter~\cite{Song:1994ip}.  
A lattice QCD calculation~\cite{Gockeler:2005vw} is also presented, which solves the dynamical 
QCD equations non-perturbatively on a discretized lattice.  The model calculations that include 
the elastic contribution are shown in the lower panel only.  We added the elastic contribution 
to the MAID model in the lower panel.  Our measurement focused on the moderately large-$Q^{2}$ 
region of $3 < Q^{2} < 5$\,GeV$^{2}$, where the elastic contribution is seen to be small 
(lower panel of Fig.~\ref{fig:d2n_current}) and where a theoretical interpretation in terms of 
twist-3 contributions is cleaner.
  
While bag~\cite{Song:1996ea,Stratmann:1993aw,Ji:1993sv} and soliton~\cite{Weigel:1996jh,*Weigel:2000gx} 
model calculations of $d_{2}$ for the neutron yield numerical values consistent with 
those of lattice QCD~\cite{Gockeler:2005vw}, prior experimental data differ 
by roughly two standard deviations in the large $Q^{2}$-range.  This is illustrated by
the data for $Q^{2} \approx 5$\,GeV$^{2}$ in Fig.~\ref{fig:d2n_current}.  This situation 
called for a dedicated experiment for the neutron, JLab E06-014.  For the proton 
$d_{2}$, the measurements and models are in better 
agreement~\cite{Anthony:2002hy,Stein:1994zk,Balitsky:1989jb,Weigel:1996jh,*Weigel:2000gx,Song:1996ea,Ji:1997gs,Gockeler:2005vw}.  
These data sets will be further extended by a recent measurement~\cite{SANE} whose precision
results are expected in the near future.  Under the assumption of isospin symmetry, 
combining the neutron and proton data would then allow a flavor decomposition to 
determine the average color force felt by the up and down quarks in the proton. 
Measurements of $d_{2}$ access similar forces as those that cause quark confinement. 
Consequently, such measurements are important for understanding the dynamics of the 
constituents of the nucleon.     

Our measurements of the unpolarized cross section $\sigma_{0}$ and the double-spin
asymmetries $A_{\parallel}$ and $A_{\perp}$ allow the extraction of $d_{2}$, and 
in turn, $f_{2}$.  Combining our results for these higher-twist matrix elements, 
we obtain the color electric and magnetic forces $F_{E}^{y}$ and $F_{B}^{y}$.  
Utilizing our data on $g_{1}$, we also evaluate the twist-2 
matrix element $a_{2}$ and test it against lattice QCD calculations. 

%===============================================================================
\subsection{$A_{1}$ and flavor decomposition} \label{sec:a1-intro} 
%===============================================================================

The measurement of the double-spin asymmetries $A_{\parallel}$ and $A_{\perp}$ 
required for the extraction of $d_{2}$ also gives access to the virtual
photon-nucleon asymmetry $A_{1}$ and the polarized to unpolarized structure-function 
ratio $g_{1}/F_{1}$: 

\begin{eqnarray}  
   A_{1}           &=& \frac{1}{D(1 + \eta\xi)}A_{\parallel} - \frac{\eta}{d(1 + \eta\xi)}A_{\perp}  \label{eqn:a1_exp} \\
   \frac{g_1}{F_1} &=& \frac{1}{d'} \left( A_{\parallel} + \tan\frac{\theta}{2}A_{\perp} \right), \label{eqn:g1f1_exp}   
\end{eqnarray}

\noindent where $F_{1}(x,Q^{2})$ denotes the unpolarized structure function  
and $D$ the virtual photon depolarization factor.  
This quantity, along with $\eta$, $d$, $\xi$ and $d'$ are defined as:

 \begin{eqnarray}
    D    &=& \frac{E - \epsilon E'}{E(1 + \epsilon R)}    \\ 
    \eta &=& \frac{\epsilon\sqrt{Q^{2}}}{E - \epsilon E'} \\ 
    d    &=& D \sqrt{ \frac{2\epsilon}{1 + \epsilon} }    \\  
    \xi  &=& \eta \frac{1 + \epsilon}{2\epsilon}          \\  
    d'   &=& \frac{(1 - \epsilon)(2 - y)}{y(1 + \epsilon R)}, 
 \end{eqnarray} 

\noindent where $R \equiv \sigma_{L}/\sigma_{T}$, the ratio of longitudinally 
to transversely polarized photoabsorption cross sections~\footnote{We use the 
parameterization for this ratio in our analysis from Ref.~\cite{Abe:1998ym}.}.   
and $\epsilon$ denotes the ratio of the longitudinal to transverse polarization 
of the virtual photon:

\begin{equation}
   \epsilon = \left[ 1 + 2 \left(1 + \gamma^{2}\right)\tan^{2}\frac{\theta}{2} \right]^{-1},
\end{equation}

\noindent with $\gamma^{2} = \left( 2Mx\right)^{2}/Q^{2}$. 

\begin{figure}[hbt]
   \centering 
   \includegraphics[width=0.5\textwidth]{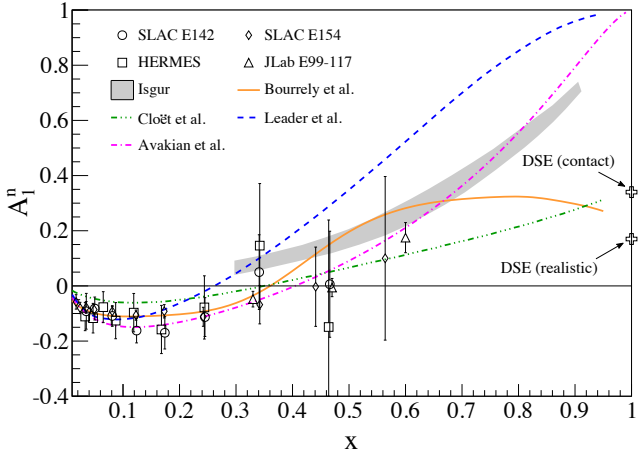}
   \caption{(Color online) World data for $A_{1}^{n}$ from SLAC E142~\protect\cite{Anthony:1996mw} and 
            E154~\protect\cite{Abe:1997qk}, HERMES~\protect\cite{Ackerstaff:1997ws}, and 
            JLab E99-117~\protect\cite{Zheng:2003un,Zheng:2004ce}, compared to 
            various models, including a pQCD-inspired global analysis (dashed)~\protect\cite{Leader:1997kw}, 
            a statistical quark model from Bourrely {\it et al.} (solid)~\protect\cite{Bourrely:2015kla}, 
            a pQCD parameterization including OAM from Avakian {\it et al.} (dash-dotted)~\protect\cite{Avakian:2007xa} 
            and a CQM model from Isgur (gray band)~\protect\cite{Isgur:1998yb}.  Also plotted is an NJL-type model 
            from Clo\"{e}t {\it et al.} (dash triple-dotted)~\protect\cite{Cloet:2005pp}.  Predictions from 
            Dyson-Schwinger Equation (DSE) treatments by Roberts {\it et al.}~\protect\cite{Roberts:2013mja} are shown at $x = 1$. 
            }
   \label{fig:a1n-world}
\end{figure}

The $A_{1}$ asymmetry is particularly sensitive to the way that the quark spins combine 
to give the nucleon spin.  Therefore, $A_{1}$ is a good discriminator for various
model calculations that aim to describe the spin structure of the nucleon.  
Figure~\ref{fig:a1n-world} shows the previous world data using $^{3}$He targets from 
SLAC E142~\cite{Anthony:1996mw} and E154~\cite{Abe:1997qk}, HERMES~\cite{Ackerstaff:1997ws}, and 
JLab E99-117~\cite{Zheng:2003un,Zheng:2004ce} compared to various models.  The SLAC E143~\cite{Abe:1998wq} 
data, which used NH$_{3}$ and ND$_{3}$ targets, has been omitted from the plot due to their 
large uncertainties.  It is seen that the relativistic constituent quark model (RCQM)~\cite{Isgur:1998yb} describes 
the trend of the data reasonably well. The pQCD parameterization with hadron helicity 
conservation (HHC)~\cite{Leader:1997kw} (dashed)---assuming quark orbital angular 
momentum to be zero---does not describe the data adequately. However, the pQCD parameterization 
allowing for quark orbital angular momentum to be non-zero~\cite{Avakian:2007xa} 
(dash-dotted) is in good agreement with the data, suggesting the importance 
of quark orbital angular momentum in the spin structure of the nucleon.  The statistical 
quark model (solid)~\cite{Bourrely:2015kla}, which interprets the 
constituent partons as fermions (quarks) and bosons (gluons), adequately describes the 
trend of the world data after fitting its parameters to a subset of the available data.  
A modified NJL model from Clo\"{e}t {\it et al.} (dash triple-dotted)~\cite{Cloet:2005pp} 
is shown to fit the data accurately in the large-$x$ region.  
This NJL-type model imposes constraints for confinement such that unphysical thresholds 
for nucleon decay into quarks are excluded.  Nucleon states are obtained by 
solving the Faddeev equation using a quark-diquark approximation, including scalar and 
axial-vector diquark states.  Relatively recent predictions come from 
Dyson-Schwinger Equation (DSE) treatments by Roberts {\it et al.}~\cite{Roberts:2013mja}, 
which reveal non-pointlike diquark correlations in the nucleon due to dynamical chiral symmetry 
breaking.  In these calculations Roberts {\it et al.} employ two different types of 
dressed-quark propagators for the Faddeev equation: one where the mass 
term is momentum-independent, and the other where the mass term carries a momentum 
dependence.  This yields two different sets of results, referred to as {\it contact} 
and {\it realistic}, respectively.  The predictions for the two approaches are 
shown at $x = 1$ (Fig.~\ref{fig:a1n-world}).  We note the contrast between the DSE 
predictions and those from pQCD and constituent quark models, where the latter 
two predict $A_{1}^{n} \rightarrow 1$ as $x \rightarrow 1$.  The measurement 
presented here provides more contiguous coverage over the region of $0.27 < x < 0.60$ 
compared to the JLab E99-117 measurement~\cite{Zheng:2003un,Zheng:2004ce}.  

\begin{figure}[!ht]
   \centering
   \includegraphics[width=0.5\textwidth]{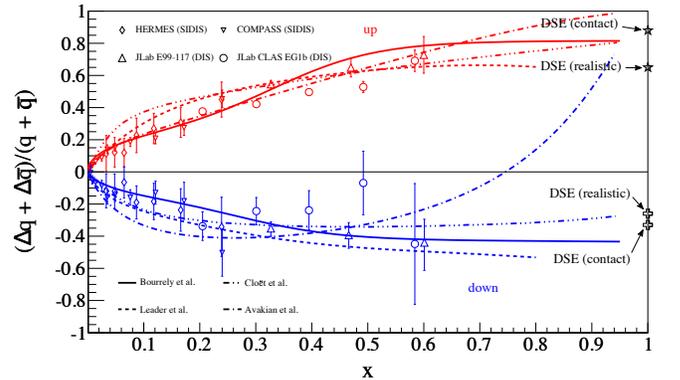}
   \caption{(Color online) The world data for the up and down quark polarized-to-unpolarized PDF
            ratios.  The data shown are from HERMES~\protect\cite{Airapetian:2004zf}, a semi-inclusive DIS measurement,
            and JLab E99-117~\protect\cite{Zheng:2004ce}, and CLAS EG1b~\protect\cite{Dharmawardane:2006zd}, both of which
            are DIS measurements.  Theoretical curves are from an NLO QCD analysis from Leader {\it et al.}~\protect\cite{Leader:2006xc} (dashed)
            and pQCD-inspired fit from Avakian {\it et al.}~\protect\cite{Avakian:2007xa} (dash-dotted).  The solid curve shows a statistical
            quark model from Bourrely {\it et al.}~\protect\cite{Bourrely:2015kla} and the dash triple-dotted curve shows
            a modified NJL model calculation from Clo\"{e}t {\it et al.}~\protect\cite{Cloet:2005pp}.  The open stars (crosses) at $x = 1$
            indicate the DSE calculations from Roberts {\it et al.}~\protect\cite{Roberts:2013mja} for up (down) quarks.  }
   \label{fig:flavor-sep-world-data}
\end{figure}

Even more than $A_{1}$, the polarized-to-unpolarized quark parton distribution
function (PDF) ratios for the up quark ($u$), given by $\left( \Delta u + \Delta \bar{u}\right)/(u + \bar{u})$
and the down quark ($d$), given by $\left(\Delta d + \Delta \bar{d}\right)/(d + \bar{d})$, allow 
a high level of discrimination between theoretical models that describe the quark-spin 
contribution to nucleon spin.  Such ratios may be extracted from measurements of $g_{1}/F_{1}$ 
at leading order in $Q^{2}$ according to:     

\begin{eqnarray} 
   \frac{\Delta u + \Delta \bar{u}}{u + \bar{u}} &=&   \frac{4}{15}\frac{g_1^{p}}{F_{1}^{p}}\left( 4 +   R^{du} \right)  \\ 
                                                 & & - \frac{1}{15}\frac{g_1^{n}}{F_{1}^{n}}\left( 1 + 4 R^{du} \right)  \nonumber \label{eqn:up}  \\ 
   \frac{\Delta d + \Delta \bar{d}}{d + \bar{d}} &=&   \frac{4}{15}\frac{g_1^{n}}{F_{1}^{n}}\left( 4 +   \frac{1}{R^{du}}\right)  \\  
                                                 & & - \frac{1}{15}\frac{g_1^{p}}{F_{1}^{p}}\left( 1 + 4 \frac{1}{R^{du}}\right) \nonumber \label{eqn:down},    
\end{eqnarray}

\noindent where $R^{du} \equiv (d + \bar{d})/(u + \bar{u})$.  Earlier 
experimental data for $(\Delta u + \Delta \bar{u})/(u + \bar{u})$ and 
$(\Delta d + \Delta \bar{d})/(d + \bar{d})$ are shown in Fig.~\ref{fig:flavor-sep-world-data}, 
where the data in the upper (lower) part of the figure represent the up (down) 
quark ratio.  The data shown are from HERMES~\cite{Airapetian:2004zf} and 
COMPASS~\cite{Alekseev:2010ub}, both semi-inclusive DIS measurements, and JLab 
experiments E99-117~\cite{Zheng:2004ce} and CLAS EG1b~\cite{Dharmawardane:2006zd}, 
both of which are inclusive DIS measurements.  The semi-inclusive DIS data from HERMES 
and COMPASS are constructed from their published polarized PDF data, where we used 
the same unpolarized PDF parameterizations as were applied in the original 
analyses: CTEQ5L~\cite{Lai:1999wy} for the HERMES data, and MRST2006~\cite{Martin:2006qz} 
for the COMPASS data.  The uncertainties are thus slightly larger than could be 
achieved from the raw data.  The dashed curve represents a next-to-leading order 
(NLO) QCD global analysis that includes target mass corrections and higher-twist 
effects~\cite{Leader:2006xc}, and the dashed-dotted curve represents a pQCD calculation 
that includes orbital angular momentum effects~\cite{Avakian:2007xa}.  The solid curve 
shows the statistical quark model~\cite{Bourrely:2015kla}, and the dash triple-dotted 
curve is a modified NJL model~\cite{Cloet:2005pp}.  At $x = 1$, DSE calculations~\cite{Roberts:2013mja} 
are indicated by open stars (crosses) for the up (down) quark ratios. Clearly, both 
pQCD models predict that $\Delta q/q \rightarrow 1$ at large $x$, which implies that the positive 
helicity state of the quark (quark spin aligned with the nucleon spin) must 
dominate as $x \rightarrow 1$.  The data for $(\Delta u + \Delta \bar{u})/(u + \bar{u})$ 
are consistent with this prediction; however, we note that the current 
$(\Delta d + \Delta \bar{d})/(d + \bar{d})$ data show no sign of turning 
positive as we approach the large $x$ region.  The Avakian {\it et al.} 
calculation fits the down quark data better, but still has a zero-crossing 
at $x \sim 0.75$.  The data in Fig.~\ref{fig:flavor-sep-world-data} 
imply that in general, the up quark spins tend to be parallel to  
the nucleon spin, whereas the down quark spins are antiparallel to the 
nucleon spin.  The trend of the down quark data, supported by the model of  
Avakian {\it et al.}, suggests that quark orbital angular momentum might 
play an important role in the spin of the nucleon.  The experiment presented here
aims to provide more complete kinematic coverage for the down quark, especially 
in the large-$x$ region approaching $x \sim 0.6$, where the predictions of the pQCD 
models start to contrast with those of the constituent quark models and the DSE calculations. 

%===============================================================================
\subsection{Outline of the paper} 
%===============================================================================

The body of this paper is structured as follows: in Section~\ref{sec:exp} 
we discuss the experimental setup and the performance of the polarized electron
beam and of the particle detectors for JLab E06-014; in Section~\ref{sec:target}, 
we discuss the polarized $^{3}$He target; in Section~\ref{sec:data-analysis}, the 
data analysis to obtain the cross sections and asymmetries is presented.  
The nuclear corrections required to extract the neutron results for $d_{2}$, $a_{2}$, 
$A_{1}$ and $g_{1}/F_{1}$ are also discussed.  In Section~\ref{sec:results} the 
results of the experiment are presented.  In particular, $^{3}$He results for the 
unpolarized cross section, double-spin asymmetries, $g_{1}$, $g_{2}$, $A_{1}$ and 
$g_{1}/F_{1}$ are given in Section~\ref{sec:res_he3}.  In Section~\ref{sec:res_n} 
the results for the neutron $d_{2}$ and $a_{2}$ are presented.  Following this, 
the analysis necessary to obtain the twist-4 matrix element $f_{2}$, leading to the 
extraction of the color forces $F_{E}^{y}$ and $F_{B}^{y}$ on the neutron, is discussed in 
Section~\ref{sec:res_cf}.  The quantities $A_{1}$ and $g_{1}/F_{1}$ on the neutron are 
presented in Sections~\ref{sec:a1n-results} and~\ref{sec:g1f1n-results}, respectively.  
The flavor separation analysis to obtain $(\Delta u + \Delta \bar{u})/(u + \bar{u})$ 
and $(\Delta d + \Delta \bar{d})/(d + \bar{d})$ is discussed and the results are 
presented in Section~\ref{sec:res_flavor}.  Concluding remarks are given in Section~\ref{sec:conclusion}.  
Appendix~\ref{sec:dis_app} gives an overview of the DIS kinematics, structure functions 
and cross sections, while Appendix~\ref{sec:ope_app} discusses the details of the 
operator product expansion.  Fits to unpolarized nitrogen cross sections and positron 
cross sections measured in this experiment, used in correcting the measured $e-^{3}$He cross section, 
are presented in Appendix~\ref{sec:fits-app}.  Also presented in that appendix are 
fits to world proton data on $g_{1}/F_{1}$ and $A_{1}$, needed for the nuclear corrections. 
Details for the world $\Gamma_{1}^{n}$ data and fitting the higher-twist component
of $\Gamma_{1}^{n}$ are given in Appendix~\ref{sec:gamma1-world-app}. 
The systematic uncertainties for all results presented in this paper are tabulated 
in Appendix~\ref{sec:syst-err-tables}.

   %===============================================================================
\section{The Experiment} \label{sec:exp} 
%===============================================================================

\begin{figure}[hbt]
   \centering 
   \includegraphics[width=0.5\textwidth]{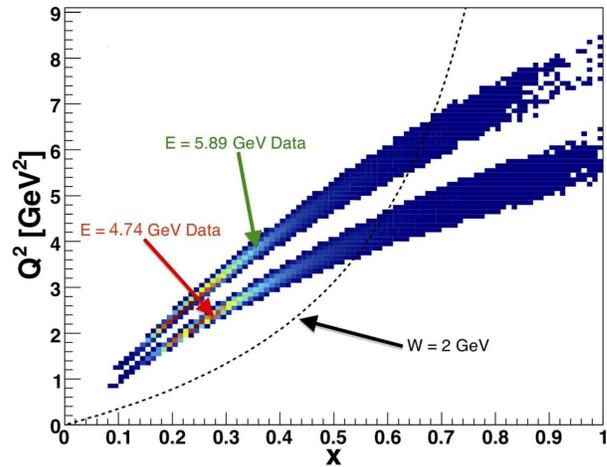}
   \caption{(Color online) The E06-014 kinematic coverage in $Q^2$ and $x$. The lower 
            band represents the $E = 4.74$\,GeV data set and the upper band the 
            $E = 5.89$\,GeV one. The black dashed line shows $W = 2$\,GeV. 
            The regions to the left and right of this line correspond to DIS and resonance 
            kinematics, respectively.
            }
   \label{fig:coverage}
\end{figure}

The E06-014 experiment ran in Hall A of Thomas Jefferson National Accelerator Facility 
(Jefferson Lab or JLab) for six weeks in five run periods from February to March of 2009,
consisting of a commissioning run using 1.2\,GeV electrons, a 5.89\,GeV run using polarized
electrons, a 4.74\,GeV run using unpolarized electrons, and finally runs using polarized 
electrons at energies of 5.89\,GeV and 4.74\,GeV.  The data at 4.74\,GeV and 5.89\,GeV 
were the production data sets, which covered the resonance and deep inelastic 
valence quark regions, in a kinematic region of $0.25 \leq x \leq 0.9$ and 
$2 \text{ GeV}^2 \leq Q^2 \leq 6$ $\text{GeV}^2$, shown in Fig.~\ref{fig:coverage}. 

Polarized electrons were scattered from a polarized $^3$He target, which acts as an 
effective polarized neutron target~\cite{Friar:1990vx}.  The scattered electrons were 
detected independently in the Left High-Resolution Spectrometer (LHRS) and in the BigBite 
Spectrometer, that were oriented at a scattering angle of $\theta = 45^{\circ}$ to the 
left and right of the beamline, respectively.  The unpolarized cross section $\sigma_0$ 
was extracted from the LHRS data and the double-spin asymmetries $A_{\parallel}$ and $A_{\perp}$ 
were obtained from the BigBite data.  The matrix element $d_{2}$ was computed using 
Eq.~\ref{eqn:d2_exp}, and the virtual photon asymmetry $A_{1}$ and structure function 
ratio $g_{1}/F_{1}$ were extracted according to Eqs.~\ref{eqn:a1_exp} and~\ref{eqn:g1f1_exp}, respectively.   

The measurement with the BigBite spectrometer consisted of twenty evenly 
spaced, continuous bins in $x$ with a bin width of 0.05 for each beam energy;
of these, seven were discarded because of insufficient statistics.  
The statistics in all bins for a given beam energy were recorded simultaneously. 
The LHRS data were acquired in nine unevenly spaced bins in the scattered 
electron momentum $p$ for the $E = 4.74$\,GeV run and eleven unevenly spaced bins 
for the $E = 5.89$\,GeV run, covering a range of $0.6 \leq p \leq 1.7$\,GeV as 
listed in Tables~\ref{tab:lhrs-bins-4} and~\ref{tab:lhrs-bins-5}.  The statistics in 
the LHRS were recorded sequentially.  For the $d_{2}$ extraction, the measured 
cross sections were interpolated and extrapolated to match the binning of the BigBite data.  

\begin{table}[h!]
\centering
\caption{Kinematic bins for the LHRS for the 4.74 GeV run.  The LHRS
         momentum setting is labeled as $p_{0}$.}
\label{tab:lhrs-bins-4}
\begin{ruledtabular}
\begin{tabular}{ccc}
 $p_0$ $\left( \text{GeV} \right)$ & $x$ & $Q^2$ $\left( \text{GeV}^2 \right)$ \\
\hline
0.60 & 0.215 & 1.66 \\ 
0.80 & 0.301 & 2.22 \\ 
1.12 & 0.458 & 3.10 \\ 
1.19 & 0.496 & 3.30 \\ 
1.26 & 0.536 & 3.49 \\ 
1.34 & 0.584 & 3.71 \\ 
1.42 & 0.634 & 3.93 \\ 
1.51 & 0.693 & 4.18 \\ 
1.60 & 0.755 & 4.43 \\ 
\end{tabular}
\end{ruledtabular}
\end{table}

\begin{table}[h!]
\centering
\caption{Kinematic bins for the LHRS for the 5.89 GeV run.  The LHRS
         momentum setting is labeled as $p_{0}$.}
\label{tab:lhrs-bins-5}
\begin{ruledtabular}
\begin{tabular}{ccc}
 $p_0$ $\left( \text{GeV} \right)$ & $x$ & $Q^2$ $\left( \text{GeV}^2 \right)$ \\
\hline
0.60 & 0.209 & 2.07 \\ 
0.70 & 0.248 & 2.42 \\ 
0.90 & 0.332 & 3.11 \\ 
1.13 & 0.437 & 3.90 \\ 
1.20 & 0.471 & 4.14 \\ 
1.27 & 0.506 & 4.38 \\ 
1.34 & 0.542 & 4.62 \\ 
1.42 & 0.584 & 4.90 \\ 
1.51 & 0.634 & 5.21 \\ 
1.60 & 0.686 & 5.52 \\ 
1.70 & 0.746 & 5.87 \\ 
\end{tabular}
\end{ruledtabular}
\end{table}

The experimental run plan optimized its statistics on the $d_{2}$ integral 
(Eq.~\ref{eqn:d2_exp}) in order to minimize the error on $d_{2}$, {\it not} on the 
structure functions $g_{1}$ and $g_{2}$.  After the extraction of $d_{2}^{^{3}\text{He}}$, 
nuclear corrections were applied (Sec.~\ref{sec:nuclear_cor}) to obtain $d_{2}^{n}$.

%===============================================================================
\subsection{The polarized electron beam}
%===============================================================================

The high-energy longitudinally polarized electron beam is provided by the 
Continuous Electron Beam Accelerator Facility (CEBAF) at JLab~\cite{Leemann:CEBAF01}.  
Polarized electrons are produced by shining circularly polarized laser light on 
a strained superlattice GaAs photo-cathode.  This produces electrons with a 
polarization of up to $\sim 85$\% at currents up to $\sim 200$\,$\mu$A.  High-energy 
electrons are achieved by two superconducting radio-frequency (RF) linear accelerators 
connected by two magnetic recirculating arcs. The beam may be circulated around the 
racetrack accelerator up to a maximum of five times to achieve an energy of 
$\sim 6$\,GeV~\cite{Leemann:CEBAF01}.  

%===============================================================================
\subsection{Beam helicity}
%===============================================================================

To control certain systematic errors associated with the electron beam polarization 
during the experiment, the helicity of the electrons was flipped every 33\,ms. 
This time frame was referred to as a helicity window, and successive windows were 
separated by master pulse signals.  Each window had a definite helicity state
in which the electron spin was either parallel ($+$) or anti-parallel ($-$) to the
beam direction.  Helicity windows were organized into quartets, taking the form 
$+--+$ or $-++-$.  The helicity state of the first window of the quartet was decided 
by a pseudo-random number generator, and in turn defined the helicity state for the 
remaining windows.  A signal indicating the helicity of each window was sent to the 
data acquisition (DAQ) systems. 

At the electron source an insertable half-wave plate (IHWP) can be placed in the 
path of the laser illuminating the strained GaAs source to reverse the helicity of the 
extracted polarized electrons relative to the helicity signal.  This was done for 
about half of the statistics to minimize possible systematic effects due to the 
helicity bit.  The asymmetry in the amount of charge delivered with the two helicity 
states was found to be negligible~\cite{MPosikThesis}; this was accomplished using 
a feedback loop and a specialized data acquisition system developed by a previous 
JLab experiment~\cite{Aniol:2004hp}.

To determine the actual sign of the electrons' helicity state for each window
type, a measurement of the quasi-elastic $^{3}$He asymmetry was made and compared
to a theoretical calculation~\cite{DParnoThesis}.  For more details, see 
Section~\ref{sec:asym_ana}.        

%===============================================================================
\subsection{Hall A overview}
%===============================================================================

\begin{figure}[hbt]
    \centering 
    \includegraphics[width=0.5\textwidth]{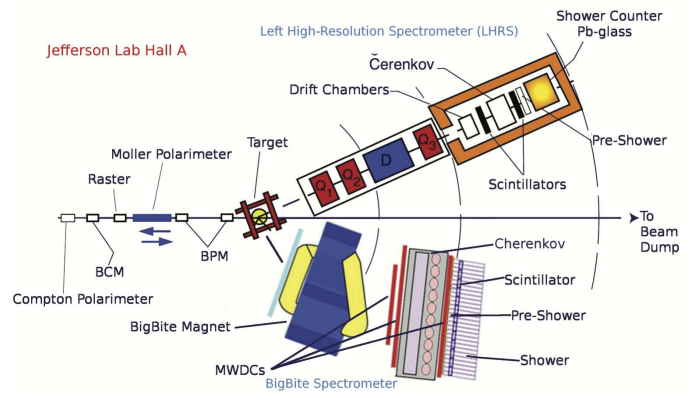}
    \caption{(Color online) Overhead view of the experimental setup for E06-014.  
             The longitudinally polarized electron beam enters from the left and 
             scatters from the longitudinally or transversely polarized $^{3}$He target, 
             which is discussed in Section~\ref{sec:target}.  
             The M\o ller and Compton polarimeters provide beam polarization 
             measurements, presented in Section~\ref{sec:beam_pol}.  The LHRS 
             and BigBite spectrometers are positioned at 45$^{\circ}$ with respect to 
             the beamline and detect scattered electrons.    
             }
    \label{fig:floor-plan}
 \end{figure}

The layout of the Hall A hardware for this experiment is shown in Fig.~\ref{fig:floor-plan}.  
Along the beamline are beam diagnostic tools, like the beam current monitors (BCMs), 
beam position monitors (BPMs), and the M\o ller and Compton polarimeters.
A polarized $^{3}$He target was utilized as an effective polarized neutron target. 
Scattered electrons were measured independently in the LHRS and the BigBite 
spectrometers, each equipped with a gas \v{C}erenkov detector and electromagnetic 
calorimeters for particle identification (PID) purposes.  In the LHRS quadrupole 
and dipole magnets are used to focus charged particles into the detector stack, 
while a single dipole magnet bends charged particles into the BigBite detector stack. 
In each spectrometer wire drift chambers are used to reconstruct particle tracks.  
Each of these elements will be described in the following sections.       
 
%===============================================================================
\subsection{The Hall A beamline}
%===============================================================================

The beamline in Hall A contains a number of important diagnostic components: 
BCMs, BPMs and the polarimetry apparatus.  We first discuss the BCMs and BPMs 
in Section~\ref{sec:bcm-bpm}, followed by the beam polarization measurements 
in Section~\ref{sec:beam_pol}.  The measurement of the beam energy is presented
in Section~\ref{sec:beam_energy}.  

%===============================================================================
\subsubsection{Beam charge and position monitoring} \label{sec:bcm-bpm}
%===============================================================================

The experiment ran at beam currents of $\sim$15\,$\mu$A.  Fluctuations about 
the required value and beam trips, due to difficulties in the accelerator 
or in the other two experimental halls, make it important to monitor the beam current.  
To this purpose two BCMs, which are resonant RF cavities, are utilized.  
These cavities, stainless-steel cylinders with a $Q$-factor of $\sim 3000$, 
were tuned to the fundamental beam frequency of 1.497\,GHz.  The two BCMs 
were located 25\,m upstream of the target, where one cavity was denoted as upstream 
and the other as downstream, based on their relative positions along the beamline.  
Each produced a voltage signal that was proportional to the measured current.  
Three copies of the signal were recorded, each amplified by a different gain 
factor (1, 3 or 10), resulting in six signals altogether (three for each cavity)~\cite{HallANIM}. 
Each copy of the signal was amplified by its assigned gain and then sent to a 
voltage-to-frequency converter.  These signals were calibrated using a Faraday cup~\cite{DParnoThesis}.  
Each signal was read out by scalers in the LHRS and BigBite spectrometers.

For accurate vertex reconstruction and proper momentum calculation for each
detected electron, the position of the electron beam in the plane transverse 
to the nominal beam direction at the target was needed.  The measurement of
the beam position was accomplished through the use of two BPMs.  They each consisted of 
four antenna arrays placed $\sim 7.5$\,m and $\sim 1.3$\,m upstream of the target.  
Pairs of wires were positioned at $\pm 45^{\circ}$ relative to the horizontal and 
vertical directions in the hall.  The signal induced in the wires by the beam was 
inversely proportional to the distance from the beam to the wires, and was recorded 
by analog-to-digital converters (ADCs).  The differences between the signals 
in pairs of wires in a given plane yields a positional resolution of 
100\,$\mu$m~\cite{KAlladaThesis}.  Combining the measurements of the two BPMs
yields the trajectory of the beam; extrapolating these data gives the position
at the target.  The BPMs were calibrated using wire scanners called harps.  
A single harp was located immediately downstream of each BPM.  Harp measurements 
allow the relative position measurements from the BPMs to be tied to the 
Hall A coordinate system.  They interfered with the beam, so dedicated runs called 
``bull's eye'' scans were needed.  A ``bull's eye'' scan consisted of five 
measurements with $(x,y)$ data points in the plane perpendicular to the beam momentum 
with the beam positioned at different locations.  Four of these points described 
the corners of a 4\,mm by 4\,mm square, and the fifth data point measured the 
square's center~\cite{DParnoThesis}.

In order to avoid damage to the glass target cell due to beam heating, the 
beam was rastered (scanned) at high speeds (17--24\,kHz) across a large rectangular 
cross section $\left( \approx 4 \times 6\text{ mm}^{2} \right)$ at the target.  
This rectangular distribution was achieved by two dipole magnets (one for vertical, 
one for horizontal) located 23\,m upstream of the target~\cite{HallANIM}.  

%===============================================================================
\subsubsection{Beam polarization measurement} \label{sec:beam_pol}
%===============================================================================

The polarization of the electron beam was measured using two different polarimeters, 
a M\o ller and a Compton polarimeter. M\o ller polarimetry utilizes scattering the 
polarized electron beam from polarized atomic electrons in a magnetized iron foil.  
The scattering rate is proportional to the beam and foil 
polarizations~\cite{HallANIM,Kresnin,Glamazdin:Moller1999}.  Such a measurement 
required the insertion of a magnetized foil into the beam path which inhibited normal 
data-taking.  A total of seven M\o ller measurements were made during the course 
of the experiment.  This method has sub-percent statistical accuracy, but a sizable 
systematic uncertainty mainly due to uncertainty in the target foil polarization.  
The total relative systematic uncertainty on the M\o ller measurement during this 
experiment was $\sim 2$\%.  

The Compton polarimeter utilized $\vec{e}$-$\vec{\gamma}$ scattering to determine 
the polarization of the electron beam as the interaction is sensitive 
to the relative polarizations of the electrons and photons~\cite{Lipps1,Lipps2}.  
The newly commissioned polarimeter consisted of a magnetic chicane which deflected 
the electron beam towards a photon source and deflected unscattered electrons 
back towards the original beam path.  At the center of the chicane was the 
photon source, a 700\,mW laser at a wavelength of 1064\,nm.  The laser output was 
400--500\,W with a resonant Fabry-P\'{e}rot cavity~\cite{Jorda:ComptonCavity98}. 
The laser polarization for the left- and right-circular polarization states was 
$99\% \pm 0.02\%$ during the experiment~\cite{DParnoThesis}.  There was also an 
electromagnetic calorimeter, a Gd$_{2}$SiO$_{5}$ (GSO) crystal doped with cerium, 
for detecting scattered photons~\cite{Friend:2012}.  The electron detector was 
not used in this experiment.  

The electron polarization was extracted from an asymmetry in the rate of scattering 
circularly polarized photons from the longitudinally polarized electrons, between 
two unique spin configurations: electron and photon spins parallel and antiparallel.  
The energy-weighted, integrated asymmetry was measured in a new integrating DAQ 
and then combined with the polarimeter's theoretically calculated analyzing power 
to determine the electron beam polarization~\cite{Friend:2012,Parno:2012xa}. 
Since Compton polarimetry is a non-invasive measurement, polarization measurements 
could be performed in parallel with data-taking.  
 
Combining the results of the M\o ller and Compton measurements for the three production 
run periods with polarized beam resulted in a beam polarization of 
$74\% \pm 1\%$ ($E = 5.89$\,GeV), $79\% \pm 1\%$ ($E = 5.89$\,GeV) and $63\% \pm 1\%$ 
($E = 4.74$\,GeV)~\cite{DParnoThesis}.  

%===============================================================================
\subsubsection{Beam energy measurement} \label{sec:beam_energy}
%===============================================================================

The beam energy was monitored throughout the experiment using the so-called Tiefenback 
method~\cite{Tiefenback}, which combined BPM measurements and the estimated 
integral of the magnetic field produced by the Hall A arc magnets.  This method 
was calibrated against an invasive ``Arc Energy'' measurement.  This measurement 
used the results of a detailed field mapping of all nine arc dipoles (including the 
reference one) after following a controlled excitation.  In the actual Arc Energy 
measurement, all nine dipoles were excited following the same curve and the field 
was measured in the ninth dipole.  The actual deflection of the beam was then measured 
and the beam energy was computed from the deviation from the nominal bend angle 
of 34.3$^{\circ}$.  The uncertainty on such a measurement was 
$\delta E/E \approx 2 \times 10^{-4}$~\cite{DMarchandThesis}.  Arc measurements were not 
performed during this experiment but were done for the immediately preceding experiment, 
E06-010~\cite{Qian:2011py}.  Their arc measurement was used as a reference for the 
Tiefenback measurements.  The arc measurement conducted during E06-010 for $\sim 6$\,GeV 
beam energies yielded a value of $5889.4 \pm 0.5_{\text{stat}} \pm 1_{\text{syst}}$\,MeV, 
while the Tiefenback measurement yielded $5891.3 \pm 2.5_{\text{syst}}$\,MeV~\cite{Qian:2011py}. 
In our data analysis we used the Tiefenback measurements without correcting  
for the difference relative to the arc measurement, which was $\ll 1\%$. 

%===============================================================================
\subsection{The spectrometers} \label{sec:spectrometers} 
\subsubsection{The Left High-Resolution Spectrometer}
%===============================================================================

The Hall A high-resolution spectrometers were designed for in-depth studies 
of the structure of nuclei and nucleons.  The LHRS has high resolution
in both the momentum and angle reconstruction of the scattered particles, 
in addition to the capability of running at high luminosity.  

At the entrance of the LHRS there are two superconducting quadrupoles, for focusing
the charged particles, followed by a superconducting dipole magnet that bends 
the charged particles upwards through a nominal $45^{\circ}$ bending angle. After 
this, the particles pass through a third quadrupole before entering the detector 
stack.  The LHRS has an angular acceptance of 6\,msr, for a horizontal (vertical) 
angular resolution of 0.5\,mrad (1\,mrad).  The momentum acceptance is 10\% with a 
momentum resolution of $10^{-4}$.  The designed maximum central momentum is 
$4$\,GeV~\cite{HallANIM}. 

\begin{figure}[hbt]
    \centering
    \includegraphics[width=0.45\textwidth]{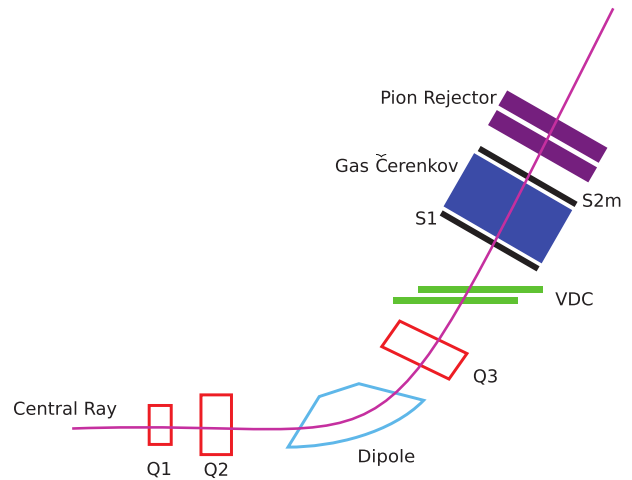}
    \caption{(Color online) The layout of the Left High-Resolution Spectrometer 
             in Hall A of Jefferson Lab during E06-014.  Drawing not to scale.   
             }
    \label{fig:lhrs-layout}
 \end{figure}

For E06-014 the LHRS detector stack was composed of a number of sub-packages, 
located in the shield hut at the end of the magnet configuration.  The detector 
sub-packages included vertical drift chambers (VDCs), which provided tracking 
information for scattered particles, and the S1 and S2m scintillating planes served 
as the main trigger.  Finally, the gas \v{C}erenkov and the pion rejector 
yielded particle identification (PID) capabilities.  The layout of the spectrometer 
is shown in Fig.~\ref{fig:lhrs-layout}. 

The VDCs allowed precise reconstruction of particle trajectories.  Each chamber 
had two wire planes containing 368 sense wires, spaced 4.24\,mm apart~\cite{HallANIM}; 
the wires were oriented orthogonally with respect to one another.  The two wire planes 
lay in the horizontal plane of the laboratory, thus oriented at 45$^{\circ}$ with respect 
to the central (scattered) particle trajectory.  Gold-plated Mylar high-voltage planes 
were placed above and below each wire plane at an operating voltage of -$4$\,kV, thus 
setting up an electric field between the high-voltage planes.  This defined a ``sense region'' 
for each wire plane.  The chambers were filled with a mixture of 62\% argon and 38\% 
ethane by weight.  Traversing particles ionized the gas mixture; the ionization 
electrons drifted along the field lines to the closest sense wires, triggering 
a ``hit'' signal in the wires.  A central track passing through at an angle of 45$^{\circ}$ 
fired five sense wires on average, resulting in a positional resolution of 
$\sim 100$\,$\mu$m and an angular resolution of $\sim 0.5$\,mrad~\cite{HallANIM}.  

The gas \v{C}erenkov had ten spherical mirrors, each with a focal length of 80\,cm, 
stacked in two columns of five.  Each mirror was viewed by a photomultiplier tube (PMT), 
placed 45\,cm from the mirror.  The chamber was filled with CO$_2$ gas at STP with 
an index of refraction of 1.00043~\cite{GrupenShwartz}.  This yielded a momentum 
threshold for triggering the gas \v{C}erenkov of $\sim 17$\,MeV for electrons and 
$\sim 4.8$\,GeV for pions.

Incident particles were also identified using their energy deposits in the lead
glass shower calorimeter, called a pion rejector.  It was composed of two layers
of thirty-four lead-glass blocks, the first 14.5\,cm $\times$ 14.5\,cm $\times$ 30\,cm,
the second 14.5\,cm $\times$ 14.5\,cm $\times$ 35\,cm made of the material SF-5,
which has a radiation length of 2.55\,cm~\cite{BachNeuroth}.  The blocks were
stacked so that the long dimensions of the blocks were transverse with respect to
the direction of the scattered particle from the target.  The gaps between the
blocks in the first layer were compensated for by a slight offset in the second 
layer of blocks.  

Since electrons and heavier particles like pions have different energy deposition
distributions in electromagnetic calorimeters, we can distinguish between the two 
particle distributions, where electrons tend to leave most (if not all) of their 
energy in the calorimeter, while pions act like minimum ionizing particles (MIPs), 
leaving only a small amount of energy in the calorimeter.  The energy loss of a 
MIP can be approximated by 1.5\,MeV per g/cm$^{\text{2}}$ traversed~\cite{Green}.
With the density of SF-5 being $\sim 4$\,g/cm$^{\text{3}}$~\cite{GrupenShwartz}, 
pions deposited $\sim 175$\,MeV in the calorimeter (both layers of the pion
rejector taken together).  As a result there are two distinct peaks in the energy 
distribution with good separation in the calorimeter: one due to pions and the 
other due to electrons.  This allows the selection of electrons in the analysis 
while rejecting pions.   

\begin{figure}[hbt]
   \centering
   \includegraphics[width=0.5\textwidth]{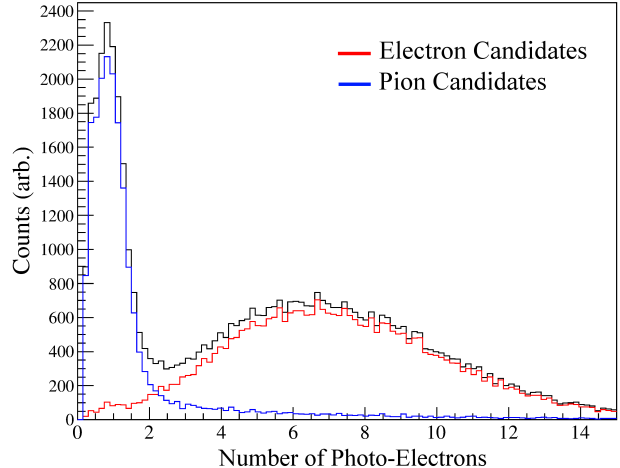}
   \caption{(Color online) A typical signal distribution in the LHRS gas \v{C}erenkov (black curve).
            Electron (pion) candidates were selected by placing cuts on the energy deposited 
            in the pion rejector, described by $E/p > 0.54$ ($E/p < 0.54$) and for the energy deposited
            in the first layer of the pion rejector to be greater than (less than) $200$\,MeV.  
            Electrons are indicated by the distribution centered at $\sim 6.5$ photo-electrons, 
            while pion candidates have their distribution peaked at $\sim 1$ photo-electron.  
            }
   \label{fig:lhrs-pid-gc}
\end{figure}

\begin{figure}[hbt]
   \centering 
   \includegraphics[width=0.5\textwidth]{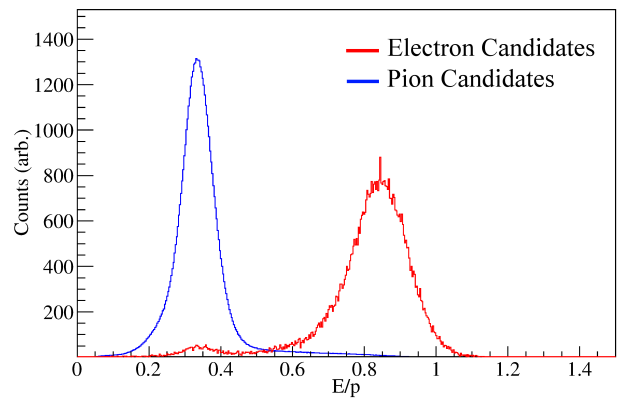}
   \caption{(Color online) A typical signal distribution in the LHRS pion rejector layers, 
            where the particle's total deposited energy divided by its reconstructed momentum is plotted.
            Electron (pion) candidates are shown by the distributions on the right (left), 
            as selected by placing a cut on the gas \v{C}erenkov signal to be greater than 
            (less than) two photo-electrons.
            }
   \label{fig:lhrs-pid-pr}
\end{figure}

Figure~\ref{fig:lhrs-pid-gc} shows a typical signal distribution in the gas \v{C}erenkov.
Electron (pion) candidates are indicated by the distributions centered at 
$\sim 6.5$ photo-electrons ($\lesssim 2$ photo-electron) in Fig.~\ref{fig:lhrs-pid-gc}, 
that are obtained by placing cuts on the pion rejector signals.  While scattered
electrons yielded an ADC signal corresponding to the main photo-electron peak 
in the gas \v{C}erenkov, pions may also influence the ADC spectrum. 
This occurs because pions could have ionized the atoms of the gaseous medium in 
the \v{C}erenkov, producing electrons with enough energy to trigger the detector.  
Such electrons are called $\delta$-rays, or {\it knock-on electrons}.  The distribution 
of these electrons has a peak at the one-photo-electron peak (leftmost peak in Fig.~\ref{fig:lhrs-pid-gc})
with a long tail underneath the multiple (main) photo-electron peak. 
These knock-on electrons can effectively be removed in the analysis because on average
they deposited less energy in the pion rejector.  To identify electrons the ratio
$E/p$ of the energy deposited in the pion rejector and the reconstructed momentum was 
required to be greater than 0.54, as illustrated in Fig.~\ref{fig:lhrs-pid-pr} 
(Section~\ref{sec:det-perf}).  Additionally, events that deposited less than 200\,MeV
in the first layer of the pion rejector were removed from the analysis, as they were 
likely to be pions or knock-on electrons.

There were two planes of plastic scintillating material, labeled S1 and S2m.
S1 was composed of six horizontal scintillating paddles with 
36\,cm $\times$ 29.3\,cm $\times$ 0.5\,cm active area.  Each paddle was viewed 
by a 5.1\,cm-diameter photomultiplier tube (PMT) on each end.  The paddles overlapped 
by 10\,mm, oriented at a small angle with the S1 plane. The S2m plane consisted of 
sixteen non-overlapping paddles with dimensions of 43.2\,cm $\times$ 14\,cm $\times$ 5.1\,cm.  
The timing resolution of the PMTs used for each plane was $\sim 50$\,ps~\cite{XQianThesis}.

When a paddle absorbed ionizing radiation, it emitted light which traveled down
the length of the paddle and is collected by the PMTs attached at each end.  
The timing information encoded in the PMTs' TDCs is utilized in the formation of 
the LHRS main trigger, discussed in Section~\ref{sec:daq}.

%===============================================================================
\subsubsection{The BigBite Spectrometer} \label{sec:bigbite} 
%===============================================================================

\begin{figure}[hbt]
   \centering 
   \includegraphics[width=0.5\textwidth]{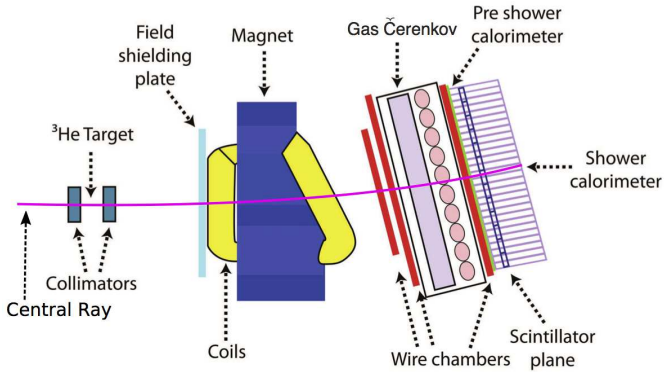}
   \caption{(Color online) The layout of the BigBite Spectrometer in Hall A of Jefferson Lab.  
            Drawing not to scale.  The central ray drawn here is a for a path similar to what 
            a 1.7\,GeV electron would take through the magnet.  Figure modified from~\protect\cite{XQianThesis}. 
            }
   \label{fig:bb-layout}
\end{figure}

The BigBite spectrometer is a large-acceptance spectrometer, able to detect 
particles over a wide range in scattering angle and momentum.  BigBite consists of 
one large dipole magnet, capable of producing a maximum magnetic field of $\sim 1.2$\,T. 
The magnet entrance was located 1.5\,m from the target center, resulting in an 
angular acceptance of about 64\,msr.  Charged particles with momenta of $\sim 0.5$\,GeV 
entering the magnet near its optical axis are then deflected roughly $25^{\circ}$ 
for a total trajectory of 64\,cm when the field is 0.92\,T~\cite{deLange:BBgeneralNIKHEF}.    
The momentum range covered by the spectrometer at full field had a lower bound of 
roughly 0.6\,GeV.  In its standard configuration, the magnet bent negatively charged 
particles upwards into the detector stack, while positively charged particles were 
deflected downwards.  The large acceptance of the spectrometer allowed the detection of 
both negatively and positively charged particles.  The detector stack for E06-014 
included multi-wire drift chambers for particle tracking, a newly installed gas \v{C}erenkov, 
a scintillator plane and an electromagnetic calorimeter, composed of a pre-shower 
and shower calorimeter.  The gas \v{C}erenkov, scintillator plane and the pre-shower 
and shower calorimeters were used for PID purposes.  The schematic layout of BigBite 
is shown in Fig.~\ref{fig:bb-layout}.

The Multi-Wire Drift Chambers (MWDCs) were utilized for particle tracking, in much 
the same way as described for the VDC planes in the LHRS.  There were three chambers, 
each filled with a 50--50 mixture of argon and ethane gas.  Each chamber had three 
pairs of wire planes, giving a total of eighteen planes in all.  Each of the eighteen 
planes was perpendicular to the detector's central ray (Fig.~\ref{fig:bb-layout}), 
bounded by cathode planes 6\,mm apart from one another.  Halfway between the cathode 
planes was a plane of wires, composed of alternating field and sense wires.  
The field wires and the cathode planes were held at the same constant high voltage, 
producing a nearly symmetric potential in the region close to the sense wires.  
Each pair of wire planes had a different orientation so as to optimize track 
reconstruction in three dimensions.  The two so-called X-planes (X, X') ran horizontally 
(in detector coordinates), while the U and V planes were oriented at $+30^{\circ}$ 
and $-30^{\circ}$ with respect to the X-planes, respectively.  The wires in each 
plane were 1\,cm apart and the primed planes (X', U', V') were offset from their 
unprimed counterparts by 0.5\,cm.  This allowed the tracking algorithm to determine 
if the track passed above or below a given wire in the X plane based upon which wire registered 
a hit in the X' plane, for example.  This alignment resulted in a positional resolution 
of less than 300\,$\mu$m~\cite{MPosikThesis}. 

\begin{figure}[hbt]
   \centering
   \includegraphics[width=0.48\textwidth]{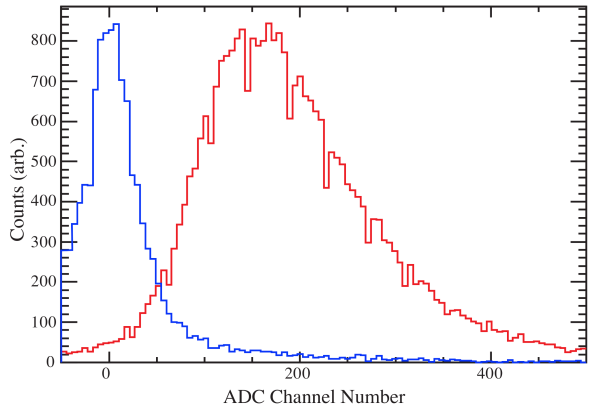}
   \caption{(Color online) Signal distributions in the BigBite gas \v{C}erenkov.
            Electron (pion) events are shown by the right (left) distributions.  
            }
   \label{fig:bigbite-pid-gc}
\end{figure}

The gas \v{C}erenkov, which was constructed by Temple University specifically for 
this experiment~\cite{BBGCNIM:2016}, included twenty spherical mirrors, each with a focal length 
of 58\,cm, stacked in two columns of ten.  The chamber was filled with the gas C$_{4}$F$_{8}$O, 
which has an average index of refraction of 1.00135~\cite{MPosikThesis}.  \v{C}erenkov 
light incident on each mirror was reflected onto a corresponding secondary flat mirror.  
This mirror then directed the \v{C}erenkov light onto the face of a corresponding PMT.  
To boost the amount of light collected, each PMT was fitted with a cone similar to 
a Winston cone~\cite{Winston}.  This extended the effective diameter of each 
PMT collection area from five inches to eight inches.  The PMTs were recessed 5\," 
within their shielding in order to reduce the effects of the BigBite magnetic field.  
The resulting gap between the PMT face and the edge of the shielding was filled with 
a cylindrical lining of Anomet UVS reflective material, so as to direct light 
incident upon this region onto the PMT face.  Figure~\ref{fig:bigbite-pid-gc} shows 
a typical signal distribution in the gas \v{C}erenkov.  Electron (pion) events are 
indicated by the right (left) distributions.  Electron events for a given PMT were 
identified by selecting those events that had a hit in their corresponding TDC 
with a projected track from the target that fell within the PMT's geometrical 
acceptance.  

The calorimeter was composed of two layers of lead-glass blocks.  The first layer 
was the pre-shower, composed of the material TF-5, which has a radiation length of 
2.74\,cm.  The pre-shower was located 85\,cm from the first drift chamber plane.  
It contained 54 blocks of dimensions 35\,cm $\times$ 8.5\,cm $\times$ 8.5\,cm.  
They were organized in two columns of 27 rows.  The long dimension of each block 
was oriented transverse with respect to scattered particles coming from the target.  
The shower layer was composed of the material TF-2, which has a radiation length 
of 3.67\,cm  and was located 15\,cm behind the pre-shower and 1\,m from the first 
drift chamber.  It had 189 blocks of the same dimensions as the blocks in the 
pre-shower, but they were organized in seven columns and 27 rows. The long dimension 
of the block was oriented along the scattered particle path, ensuring the capture of 
a large amount of the electromagnetic shower of the particle~\footnote{The lead-glass 
blocks used had a Moli\`{e}re radius that contained $\gtrsim 90$\% of the total 
deposited energy of an incident particle; therefore, the blocks were grouped into a 
clustering scheme to capture more of the deposited energy~\cite{MPosikThesis}. 
See section~\ref{sec:daq}.}.   

The plane located between the pre-shower and the shower consisted of a scintillator 
plane composed of 13 paddles of plastic scintillator, each of which had a PMT at 
each end with a timing resolution of 0.3\,ns.  Each paddle had the dimensions 
17\,cm $\times$ 64\,cm $\times$ 4\,cm.  The first dimension was transverse with 
respect to the scattered particles, while the short dimension was along the 
scattered particle path.  This resulted in an active area of 221\,cm $\times$ 64\,cm.  
This plane provided an additional source of pion rejection to complement the gas 
\v{C}erenkov and the shower calorimeter, as the charged pions left a significant 
signal in the low end of the ADC spectrum via knock-on electrons~\cite{MPosikThesis}.

%===============================================================================
\subsection{Data acquisition and data processing} \label{sec:daq}
%===============================================================================

In this experiment, the CEBAF Online Data Acquisition (CODA)~\cite{CODA} system 
was used to process the various trigger signals and data coming from the LHRS and 
BigBite spectrometers, beamline and target equipment.  The LHRS and the BigBite 
detector systems were run independently with a total of 5\,TB of data recorded.  

Eight triggers were configured for E06-014, summarized in Table~\ref{tab:triggers}.
The T8 trigger was used for troubleshooting purposes only.  It was a 1024\,Hz clock,
injected into the data stream to ensure that the electronics were working correctly.  
The T5 trigger was the coincidence (coin.) trigger between the LHRS and BigBite, 
used for optics calibration purposes.

\begin{table}[h!]
\centering
\caption{Triggers used during E06-014.}
\label{tab:triggers}
\begin{ruledtabular}
\small{\begin{tabular}{ccl}
% \hline
\multicolumn{1}{c}{Trigger}         &
\multicolumn{1}{c}{Spectrometer(s)} &
\multicolumn{1}{l}{Description}     \\
\hline
 T1 & BigBite       & Low shower threshold                       \\ 
 T2 & BigBite       & Coin. of T6 and T7                         \\ 
 T3 & LHRS          & Coin. of S1 and S2m                        \\ 
 T4 & LHRS          & Coin. of either S1 or S2m and \v{C}erenkov \\ 
 T5 & LHRS, BigBite & Coin. of T1 and T3                         \\ 
 T6 & BigBite       & High shower threshold                      \\ 
 T7 & BigBite       & Gas \v{C}erenkov                           \\ 
 T8 & LHRS, BigBite & 1024 Hz Clock                              \\ 
\end{tabular}}
\end{ruledtabular}
\end{table}

The generation of the main LHRS trigger (T3) required a hit in both scintillating 
planes S1 and S2m, where a hit in a single plane corresponded to a signal in the two 
PMTs affixed to a paddle (left and right sides) in a plane.  Thus, a T3 trigger 
corresponded to a pulse detected in four PMTs, two in the S1 plane and two in the S2m plane. 
The timing of this trigger was set by the leading edge of the TDC signal recorded 
in the PMT attached to the right side of the S2m scintillator paddles~\cite{KAlladaThesis}.  
The second LHRS trigger was the T4 trigger.  The only difference between the T3 and 
T4 triggers was that a T4 was generated when there was a coincidence between either S1 
or S2m and the gas \v{C}erenkov detector, {\it without} generating a T3 trigger.  
The T4 trigger was used to study the efficiency of the T3 trigger, as these events were 
potentially good events since they generated a signal in the gas \v{C}erenkov.  It was found 
that the efficiency of the T3 trigger was $99.95\%$ over the course of the experiment~\cite{DFlayThesis}.

\begin{figure}[hbt]
   \centering
   \includegraphics[width=0.5\textwidth]{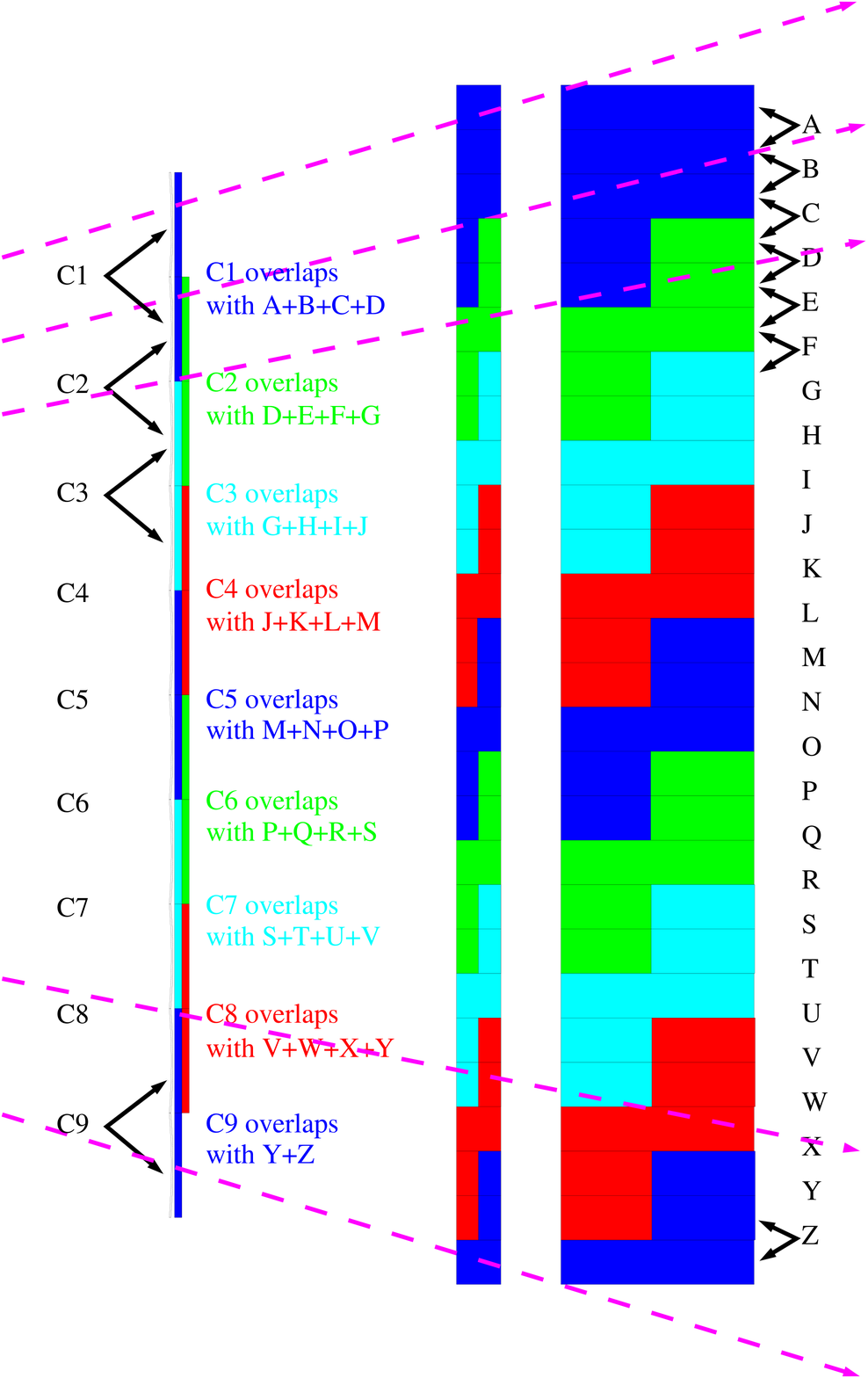}
   \caption{(Color online) The geometrical overlap for the main trigger for the 
            BigBite spectrometer.  The \v{C}erenkov mirrors are represented by the 
            leftmost column, where the cluster groupings are labeled by C$i$, with $i = 1,\ldots,9$.  
            The pre-shower blocks are shown as the middle column of colored blocks, 
            while the shower blocks are the long blocks in the rightmost column.    
            The calorimeter cluster groupings are labeled with the letters A--Z.
            The dashed lines indicate typical electron paths at the 
            extremes of the acceptance of the BigBite spectrometer.  
            }
   \label{fig:t2-trigger}
\end{figure}

The BigBite spectrometer had four dedicated triggers, T1, T2, T6, and T7.  The T1 
and T6 triggers involved taking the hardware (voltage) sum of the calorimeter blocks 
belonging to the cluster with the largest signal, where a cluster for the pre-shower 
and shower calorimeters was defined as two adjacent rows of calorimeter blocks.  There are 26 
clusters each for the pre-shower and shower calorimeters.  The sum of the pre-shower 
and shower signals was then formed and sent to a discriminator.  If this signal was greater 
than $\sim 300$--$400$\,MeV ($\sim 500$--$600$\,MeV), then the T1 (T6) trigger was formed.  
The T7 trigger was formed in a manner similar to the T1 and T6 triggers, but using the 
\v{C}erenkov detector instead of the pre-shower or shower calorimeter.  The \v{C}erenkov signals 
from two adjacent rows of mirrors (four mirrors in total) were summed, resulting in nine 
overlapping mirror clusters.  If this sum was larger than the set threshold value 
($> 1$--$1.5$ photo-electrons), then the T7 trigger was formed.  The main trigger for 
the BigBite spectrometer, T2, imposed a geometric constraint on the incident particle 
track by requiring a coincidence between the geometrically overlapping regions in 
the gas \v{C}erenkov and the calorimeter.  An example of an event that generated a T2 
trigger is illustrated in Fig.~\ref{fig:t2-trigger}: a particle that triggered  
cluster C1 in the gas \v{C}erenkov would also have to trigger at least one of the 
clusters A--D in the calorimeter.  Similar coincidences were imposed for the 
eight other groupings that could form a T2 trigger. 

The raw data were processed by the Hall A Analyzer~\cite{Analyzer}, which is 
based on ROOT~\cite{ROOT}.  Specific C++ classes have been written to interpret 
the data recorded by the various detectors and their sub-detectors.  For instance, 
there are classes that convert the ADC signals registered in a calorimeter block 
into the corresponding amount of energy deposited.  There are also classes that 
handle the computation of a particle's path (or track) through the LHRS (and BigBite) 
up to its focal plane and its reconstructed vertex position back at the target. 
The optics for BigBite required special attention, as discussed 
in~\cite{XQianThesis,MPosikThesis,DParnoThesis}. 

%===============================================================================
\subsection{Particle identification} \label{sec:det-perf} 
%===============================================================================

The LHRS and the BigBite spectrometers each utilized a gas \v{C}erenkov detector
and a double-layered lead-glass shower calorimeter for PID purposes.  In this 
experiment, PID corresponds to distinguishing electrons from pions, that constituted 
the primary background.  

The PID performance of each detector was characterized by the efficiencies of the 
conditions (or cuts) placed on the corresponding observable.  Before PID cut efficiencies
were evaluated, the sample distribution of events to be studied was selected using
data quality criteria (such as removing beam trips) and conditions to remove events 
that may have originated in the target's glass endcaps~\cite{DFlayThesis,MPosikThesis}.  
The electron cut efficiency $\varepsilon_e$ is defined as the ratio of the number of events 
that pass a given cut to the size of the electron event sample defined by another detector.  
For the gas \v{C}erenkov, the electron sample was chosen by using the calorimeter, 
and vice-versa.  To characterize how well a given detector can reject pions, the 
rejection factor $f_{\pi,\text{rej}}$ is evaluated.  It is defined as the ratio of 
the size of the selected pion sample to the number of events misidentified as 
electrons for a given cut.  The PID cuts were chosen such that the pion rejection 
was maximized while the highest electron efficiency was maintained. 

In the momentum acceptance range of the experiment, $0.6$\,GeV $\leq p \leq 1.7$\,GeV,
the electron cut efficiency for the LHRS gas \v{C}erenkov was found to be
$\varepsilon_e^{\text{cer}} \approx$ 96\% for a cut of greater than two photoelectrons 
in the ADC.  For the LHRS pion rejector, $\varepsilon_e^{\text{pr}} \approx$ 99\% 
for $E/p > 0.54$.  These efficiencies are critical for the LHRS data since they 
contribute directly in the determination of the unpolarized cross section (Section~\ref{sec:xs-ana}).
The pion rejection factor was found to be $\sim 660$ for both the gas \v{C}erenkov 
and pion rejector, resulting in a combined rejection of 
$f_{\pi,\text{rej}} \approx 4 \times 10^{5}$~\cite{DFlayThesis}.  As a result, 
the pion contamination in the final electron sample was negligible.    

PID studies were also conducted for the data recorded by BigBite. Here, the pion 
rejection factor was determined to be better than $2 \times 10^{4}$ when combining 
the pion rejection capabilities of the gas \v{C}erenkov~\footnote{The details of the performance 
of the gas \v{C}erenkov will be discussed in an upcoming publication~\cite{BBGCNIM:2016}.}, 
pre-shower and shower calorimeters, and the scintillator plane~\cite{MPosikThesis}.  
Unlike the cross section analysis using the LHRS data, the electron cut efficiencies 
do not play a role in the asymmetry extraction that is performed using the BigBite 
data; the efficiencies cancel in the asymmetry defintion (Section~\ref{sec:asym_ana}).

   %===============================================================================
\section{Polarized $^{3}$He target}  \label{sec:target} 
%===============================================================================

\begin{figure}[hbt]
   \centering 
   \includegraphics[width=0.5\textwidth]{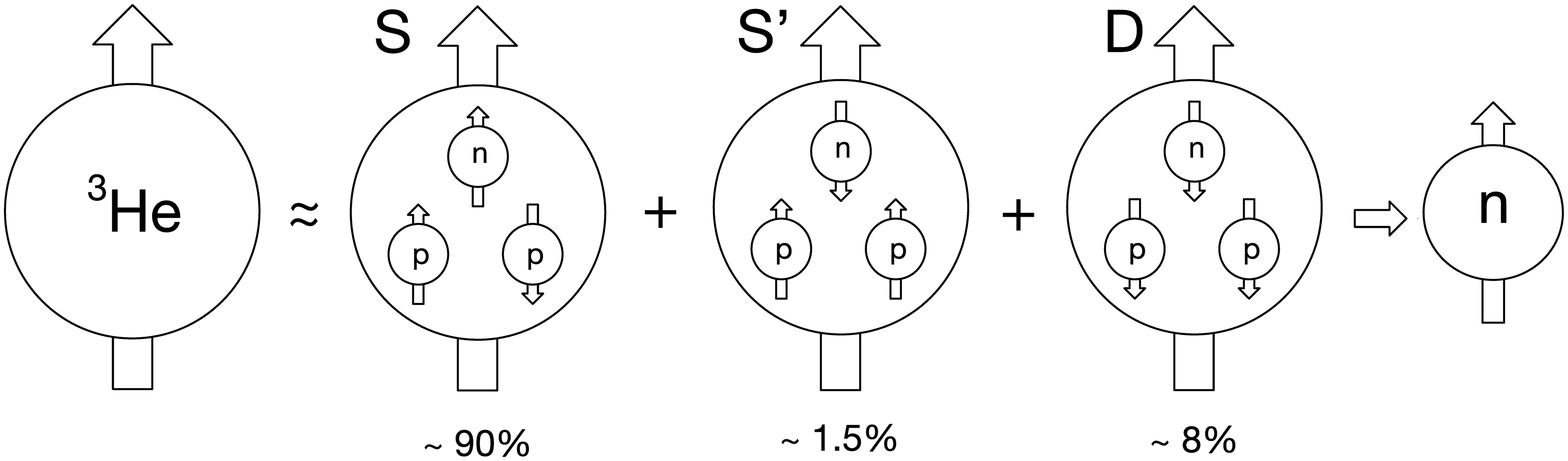}
   \caption{$^{3}$He ground states.  The dominant state is the $S$
            state, where $\sim 90$\% of the polarization is carried
            by the neutron.  In this state, the protons pair to $s = 0$.   
            }
   \label{fig:pol-he3}
\end{figure}

Since the lifetime of the neutron is less than 15 minutes~\cite{Wietfeldt:2014rea} 
outside the nucleus, a free-neutron target is not practical.  $^{3}$He, a spin-1/2 
nucleus consisting of two protons and a neutron, is a candidate for a polarized 
neutron target.  Deuterium, a spin-1 nucleus consisting of a proton and a neutron, 
is another option.  Both nucleons in deuterium have their spins aligned with the 
nuclear spin.  However, large corrections due to the proton result in large 
uncertainties when using a deuterium target.  When $^{3}$He is polarized, there 
are three principal states in play: $\sim 90$\% of the time the nucleus is in the 
symmetric $S$ state; $\sim 1.5$\% of the time the nucleus is in the $S'$ state, 
and $\sim 8$\% of the time the nucleus is in the $D$ state, see Fig.~\ref{fig:pol-he3}.  
In the $S$ state, the spins of the protons are antiparallel to one another, resulting 
in the neutron carrying the majority of the $^{3}$He polarization~\cite{Friar:1990vx}.  
As a result, a polarized $^{3}$He target can be used as an effective polarized neutron target.

In this experiment, polarized $^{3}$He $\left(^{3}\vec{\text{He}}\right)$ was used 
to study the electromagnetic structure and the spin structure of the neutron.  
Two major methods exist to polarize $^{3}$He nuclei. The first one uses the 
metastable-exchange optical pumping technique~\cite{Colegrove}, while the second 
method utilizes both spin-exchange~\cite{Happer:1972zz,*Appelt} and optical pumping~\cite{Chupp:1987zz}, 
dubbed {\it hybrid spin-exchange optical pumping}.  

$^{39}$K atoms were optically pumped using 795-nm circularly polarized 
laser light, inducing the D1 transition in $^{85}$Rb: $5^{2}S_{1/2}$ ($m = -1/2$) 
$\rightarrow 5^{2}P_{1/2}$ ($m = +1/2$), in accordance with the selection rule of 
$\Delta L = +1$.  The excited $^{85}$Rb electrons decay from the $p$ orbital to 
the $s$ orbital with equal probabilities for the $m = \pm 1/2$ sub-states, but the 
excitation only occurs for the $m = -1/2$ initial state of the $s$ orbital; this 
results in the selective population of the $m = 1/2$ state of the $s$ orbital.  Second, the 
polarization of the $^{85}$Rb atoms was transferred to the $^{39}$K atoms via 
spin-exchange binary collisions~\cite{Happer:1972zz}.  In the third and final step, 
the polarization of the $^{85}$Rb and $^{39}$K atomic electrons was transferred to 
the $^{3}$He nuclei via the hyperfine interaction, where the nuclear spin of $^{3}$He 
takes part in the process~\cite{Walker:1997zzc,*Newbury}.  The use of $^{39}$K greatly 
decreases the spin-relaxation rate for collisions involving $^{3}$He, resulting in an 
increase in the spin-exchange efficiency of the polarization process~\cite{Walker2}.  

As the atomic electrons decayed to the ground state, photons were emitted. 
These photons were typically unpolarized, and therefore reduced the efficiency 
of the pumping process.  To minimize this effect for the alkali atoms, a small 
amount of N$_{2}$ buffer gas was added to the cell.  The excitation energy of the 
alkali atoms was passed to the rotational and vibrational modes of the buffer gas 
via collisions, reducing the emission of photons~\cite{Appelt}. 

%===============================================================================
\subsection{Setup}
%===============================================================================

\begin{figure}[hbt]
    \centering
    \includegraphics[width=0.5\textwidth]{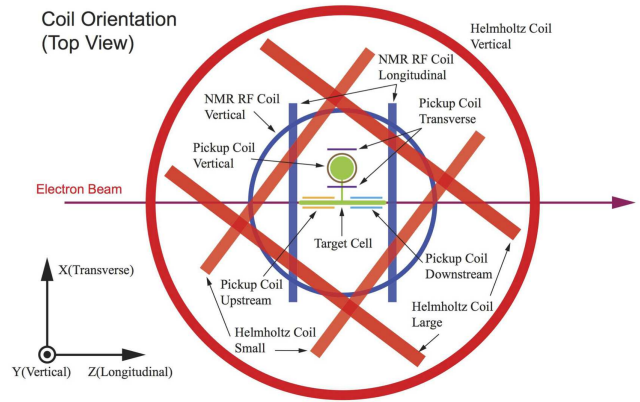}
    \caption{(Color online) The target setup.  The Helmholtz coils for the holding
             magnetic field and coils for the RF field are shown.  The pickup coils 
	     near the target cell are used for NMR measurements.  
	     Figure reproduced from~\protect\cite{HallANIM}.  
             }
    \label{fig:target-layout}
 \end{figure}

The target apparatus was composed of a number of different elements: the target 
cells, target oven, target ladder system, Helmholtz coils for the holding 
magnetic field, RF coils and polarizing lasers.  The layout of the target system 
is shown in Fig.~\ref{fig:target-layout}.  The outer circle and large straight lines 
intersecting at right angles inscribed in the large circle represent Helmholtz coils.  
The smaller vertical straight lines and circle overlapping with the Helmholtz coils 
signify the RF coils.  Pickup coils mounted near the target cell are also shown. 

Two pairs of Helmholtz coils, capable of producing magnetic fields in two 
orthogonal directions, were utilized in E06-014: longitudinal (along the direction 
of the beam), and transverse in-plane (perpendicular and horizontal to the beam).    
The field reached a magnitude of 25\,G, requiring $\sim 7$\,A of current in 
each coil.  The RF coils and pickup coils are important for the measurement of 
the target polarization, as presented in Sections~\ref{sec:epr} and~\ref{sec:nmr}. 

%===============================================================================
\subsection{Target cells} \label{sec:target-cells} 
%===============================================================================

\begin{figure}[hbt]
    \centering 
    \includegraphics[width=0.45\textwidth]{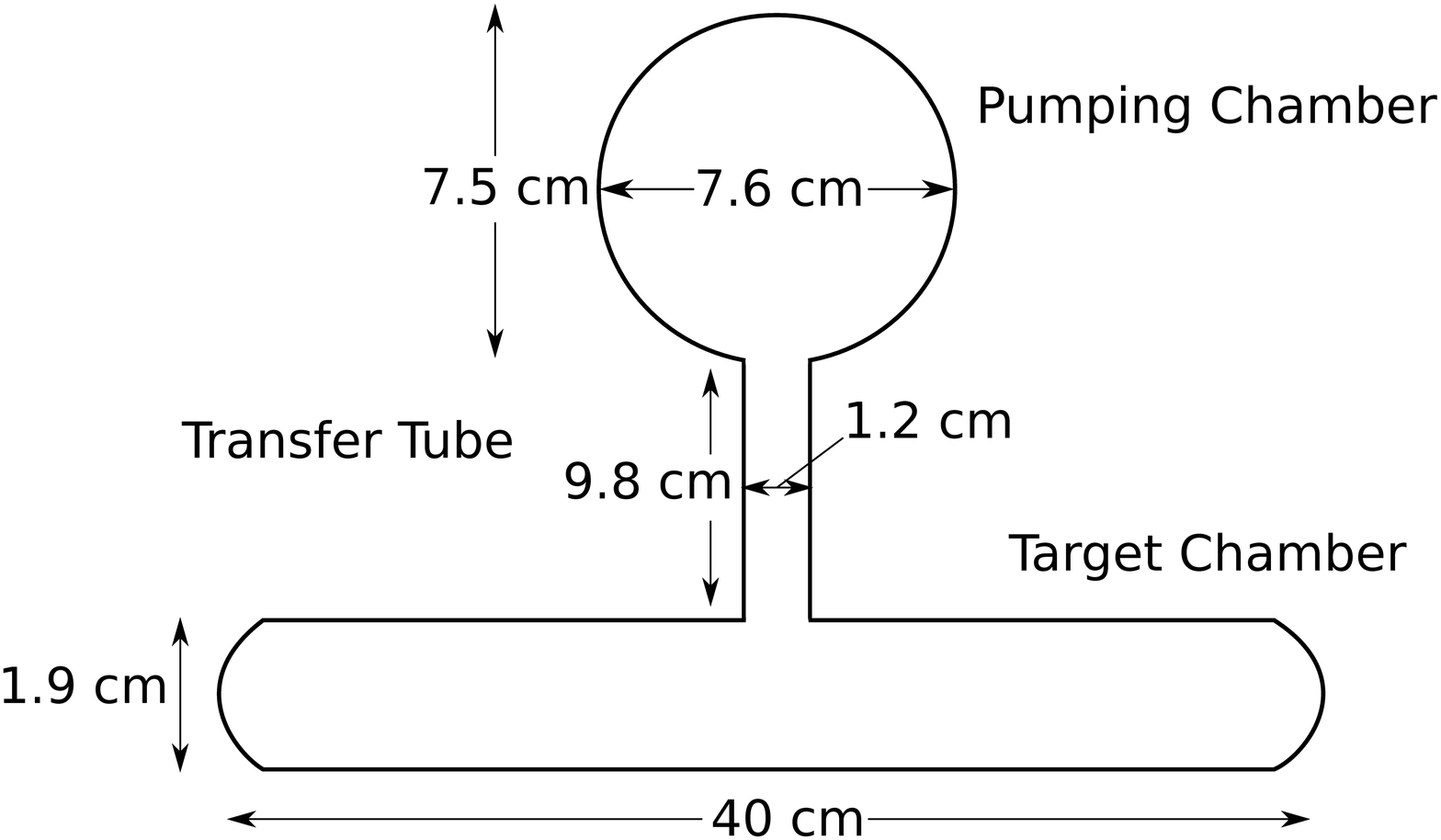}
    \caption{The production target cell used in our experiment. 
             The top spherical chamber is the pumping chamber 
             where the polarization of $^{3}$He takes place.  
             The long cylindrical chamber is the target cell,
             through which the electron beam passes longitudinally.
             The thin tube connecting these two chambers is 
             the transfer tube, which allows polarized $^{3}$He to drift
             down into the target chamber.  Drawing not to scale.  
             }
    \label{fig:target-cell}
 \end{figure}

The production target cell, named Samantha, is shown schematically
in Fig.~\ref{fig:target-cell}.  The upper chamber, called the pumping chamber, 
contained $^{3}$He, alkali metals ($^{85}$Rb and $^{39}$K at equal densities) and 
N$_2$, with number densities of $10^{20}$\,cm$^{-3}$, $10^{14}$\,cm$^{-3}$ and $10^{18}$\,cm$^{-3}$, 
respectively~\cite{CDuttaThesis}.  This chamber was heated to $\sim 265^{\circ}$C in 
order to keep the Rb and K in a gaseous state.  The polarization process took place 
in this chamber.  The polarized $^{3}$He gas (with the N$_{2}$ mixture) flowed 
through a thin transfer tube to the target chamber, thanks to the temperature gradient 
between the pumping chamber and the target cell that was kept at room temperature.  
This chamber was 40\,cm long and contained $\sim 8$\,atm of $^{3}$He and $\sim 0.13$\,atm 
of N$_2$ during the experiment.  The temperature of the cell was monitored via resistive temperature
devices (RTDs), which were placed equidistant from one another along the length
of the target chamber, along with two more placed on the pumping chamber; one at the 
top and the other at the base.  The production cell was made out of aluminoscilicate 
glass (GE-180), which was filled and characterized at the University of Virgina and the College of 
William and Mary~\cite{KAlladaThesis}.  This characterization consisted of 
measuring the polarization, gas density, glass thickness of the cell and rate of polarization.  

An additional reference cell was used~\footnote{This cell does not have a pumping chamber.},
which could be filled with H$_2$, N$_2$ or $^{3}$He.  This allowed the 
determination of the dilution factors that contribute to the cross sections and 
asymmetries.  RTDs were also mounted on the reference cell to monitor its temperature, in 
a similar configuration as was done for the target cell.  A multi-carbon foil 
(``optics'') target---as well as the reference cell filled with hydrogen gas---was 
used for the calibration of the optics for the two spectrometers.  All of these targets 
were mounted on a target ladder, which could be moved vertically up and down to select 
the target needed.  In addition to these targets, a ``no target'' position was available, 
corresponding to a hole in the target ladder.  It was used during M\o ller polarimeter 
measurements, so that the target assembly would not be damaged in the process.

%===============================================================================
\subsection{Laser system}
%===============================================================================

Our experiment utilized an upgraded laser system that had been installed for the 
immediately preceding experiment, E06-010~\cite{Qian:2011py}.  These new COMET 
lasers had a linewidth of 0.2\,nm, a factor of ten less than that of their 
predecessors (FAP lasers~\cite{XZhengThesis,AKolarkarThesis,AKelleherThesis}). 
This dramatically improved the optical pumping efficiency, since a narrower 
linewidth results in proportionately more photons exciting the 
desired atomic transitions in $^{85}$Rb, so that a higher polarization of $^{3}$He 
atoms could be attained in a shorter timeframe~\cite{CDuttaThesis}.

\begin{figure}[hbt]
    \centering
    \includegraphics[width=0.5\textwidth]{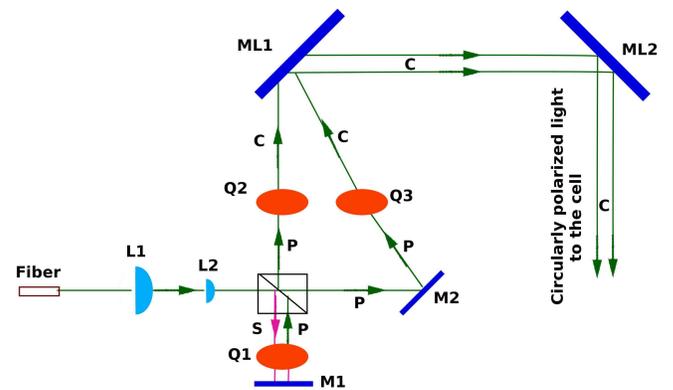}
    \caption{(Color online) The laser system used to polarize $^{85}$Rb atoms.
             The symbols labeled L1 and L2 are lenses, while symbols 
             labeled as M and ML are mirrors.  Light reflected from  
             ML2 is incident upon another mirror (not shown) which is
             attached to the oven.  Quarter wave plates are indicated by 
             Q1, Q2 and Q3.  The beam-splitting polarization cube (BSPC) is 
             represented by the rectangle with a slash through it. 
             The initial unpolarized laser light is split by the BSPC 
             into S- and P-wave components, where the P-wave component 
             has linear polarization and passes through the beam splitter.
             The S-wave component is converted into a P-wave by Q1. 
             The motorized quarter-wave plates Q2 and Q3 convert P-wave 
             light into circularly polarized light, labeled C.  
             Figure reproduced from~\cite{MPosikThesis}. 
             }
    \label{fig:laser-layout}
 \end{figure}

The laser setup is shown in Fig.~\ref{fig:laser-layout}.  It consisted of three 
COMET lasers, each with a power of 25\,W and a wavelength of 795\,nm, used to 
optically pump the $^{85}$Rb in the pumping chamber.  The lasers were installed  
in a separate laser building behind the counting house on the accelerator site 
at JLab. The fiber coming out of each COMET control unit was connected to a 
75-m-long fiber that ran from the laser building to the hall.  Then the fiber was 
connected to a 5-to-1 combiner. The output of the combiner was sent to a 
beamsplitter, yielding two linearly polarized components.  One component passed 
twice through a quarter-wave plate, after which both had the same linear polarization. 
Sending each component through another quarter wave plate converted the linear 
polarization into circular polarization. The resulting beams were then combined 
into one, with a spot size of 7.5\,cm in diameter, the size of the pumping 
chamber~\cite{CDuttaThesis}.  There were three optics lines corresponding 
to the longitudinal, transverse and vertical polarization directions.  The polarizing 
optics were set up in an antiparallel pumping configuration such that the 
target spin was always oriented opposite to the magnetic holding field~\cite{MPosikThesis}.
 
%===============================================================================
\subsection{EPR measurements} \label{sec:epr}
%===============================================================================

The target polarization was measured in an absolute sense through an electron
paramagnetic resonance (EPR) measurement, that utilized Zeeman splitting of the 
electron energy levels when an atom was placed in an external magnetic field.  
This phenomenon occurred for the $^{85}$Rb and $^{39}$K atoms, which were present 
in the pumping chamber.  The ground states of the alkali metals split into 
$2F + 1$ energy levels, where $F$ is the total angular momentum quantum number.  
Specifically, for $^{85}$Rb, the $F =$ 3 ground state split into seven sublevels 
corresponding to $m_F = -3$,\ldots, $3$.  For $^{39}$K, the $F = 2$ ground state 
split into five energy levels, given by $m_{F} = -2$,\ldots, $2$.  The splitting 
corresponded to a frequency that is proportional to the holding field.  This frequency 
was shifted due to the small effective magnetic field created by the spin-exchange 
mechanism of $^{85}$Rb--$^{39}$K and $^{39}$K--$^{3}$He, in addition to the 
polarization of the $^{3}$He nuclei.  

When the EPR transition was excited, an alkali metal (either Rb or K as chosen
by the excitation frequency) lost its polarization.  When one of the metals was 
depolarized, so was the other due to the fast spin-exchange mechanism.  Upon 
re-polarization of the Rb atoms, there was an increase in the photons emitted 
corresponding to the $P_{1/2} \rightarrow S_{1/2}$ (D1) transition (795\,nm).  
However, due to thermal mixing between the $P_{1/2}$ and $P_{3/2}$ energy states and 
occasional collisional mixing with N$_{2}$ in the cell, the $P_{3/2} \rightarrow S_{1/2}$ 
(D2) transition (780\,nm) was possible.  While the amount of D1 and D2 flourescence 
was roughly the same~\cite{CDuttaThesis}, the D1 light was suppressed due to 
a large background component corresponding to the polarizing laser light.     
Therefore, a filter was attached to a photodiode to identify the D2 light.  
During the measurement, the RF was modulated with a 100\,Hz sine wave, and 
the D2 transition was synchronized to this modulating signal and measured by 
a lock-in amplifier.  The signal from the lock-in output was proportional to the 
derivative of the EPR fluorescence curve as a function of the RF; the EPR resonance 
occured when the derivative was equal to zero~\cite{JHuangThesis,CDuttaThesis}.  
EPR measurements were performed every few days.

To determine the polarization in the target chamber, a model~\cite{RomalisThesis,Singh:2008,Ye:2009dx,Dolph:2011rc} 
was used to describe the diffusion of the $^{3}$He polarization from the pumping 
chamber to the target chamber.  The relative systematic error of the measurement 
was $\sim 4\%$, dominated by the uncertainties on the dimensionless constant 
$\kappa_{0}$~\cite{Kadlecek:2005} and the number density of the gas in the 
pumping chamber~\cite{MPosikThesis}.  

%===============================================================================
\subsection{NMR measurements} \label{sec:nmr}
%===============================================================================

Another method we used for measuring the polarization of the $^{3}$He nuclei was 
measuring the $^{3}$He nuclear magnetic resonance signal.  The magnetic moments 
of $^{3}$He nuclei aligned along the direction of an external magnetic holding field 
had their direction reversed by applying an RF field in the perpendicular direction. 
Sweeping the frequency of the RF field through the resonance of the $^{3}$He
nucleus flipped the spins of the nuclei.  This spin flip changed the field flux 
through the pick-up coils (Fig.~\ref{fig:target-layout}), inducing an electromotive force.  
The signals from the coils were pre-amplified and combined, and sent to a lock-in 
amplifier. The magnitude of the final signal was proportional to the $^{3}$He 
polarization.

The RF is swept according to the Adiabatic Fast Passage (AFP) technique~\cite{Abragam}, 
in which the sweep through the resonant frequency is done faster than the 
spin-relaxation time, but slowly enough so that the nuclear spins can follow 
the sweep of the RF field.  This minimizes the effect of these NMR measurements on 
the target polarization.

An NMR measurement is a relative measurement, so it needs to be compared against 
a known reference. A measurement of NMR on water is typically used, for which 
the polarization can be calculated exactly from statistical mechanics~\cite{Lorenzon}.  
However, in E06-014, water-cell measurements were available only for the 
longitudinal target polarization configuration, as conversion factors needed to 
account for the different positions of the water and $^{3}$He cells could not 
be measured for the transverse configuration~\cite{MPosikThesis}.  Because of this, 
the NMR measurements were calibrated against EPR measurements (Section~\ref{sec:epr})
that were done close in time relative to the NMR measurements.  
The NMR water measurements in the longitudinal configuration were used as a 
cross-check against the EPR measurements, and were found to be consistent to the 1\% level. 
NMR measurements were performed every four hours on the production $^{3}$He target.  
The systematic error on the NMR measurement was $\sim 3.9\%$ (relative), dominated 
by the uncertainties on the EPR calibration and on the computed magnetic flux through 
the pickup coils~\cite{MPosikThesis}.   

%===============================================================================
\subsection{Target performance} \label{sec:target-pol}
%===============================================================================

\begin{figure}[hbt]
    \centering 
    \includegraphics[width=0.5\textwidth]{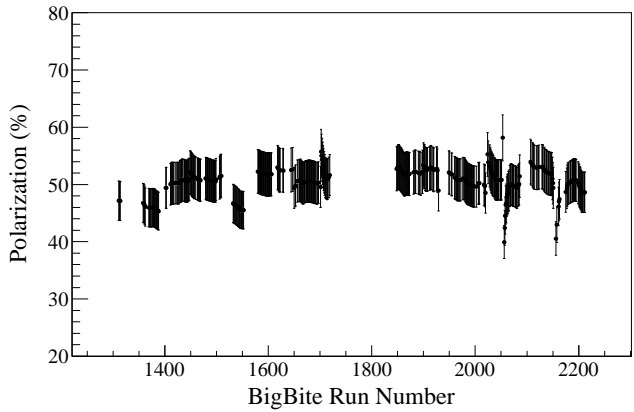}
    \caption{The $^{3}$He target polarization as a function of BigBite run number.}
    \label{fig:target-pol}
 \end{figure}

The target polarization over the course of the experiment, extracted from NMR 
measurements, is shown in Fig.~\ref{fig:target-pol}.  In the data analysis (Section~\ref{sec:asym_ana}), 
the target polarization data was utilized on a run-by-run basis.  On average, the 
target polarization achieved was 50.5\% with a relative uncertainty 
of 7.2\% (3.6\% absolute).  The dominant contribution to the uncertainty was from 
the calibration of the NMR measurements against the EPR measurements (3.9\% relative) 
and the loss of polarization due to the diffusion of polarized $^{3}$He from the pumping 
chamber to the target chamber (6\% relative).

   %===============================================================================
\section{Data Analysis} \label{sec:data-analysis} 
\subsection{Analysis procedure} 
%================================================================================

\begin{figure}[hbt]
    \centering
    \includegraphics[width=0.5\textwidth]{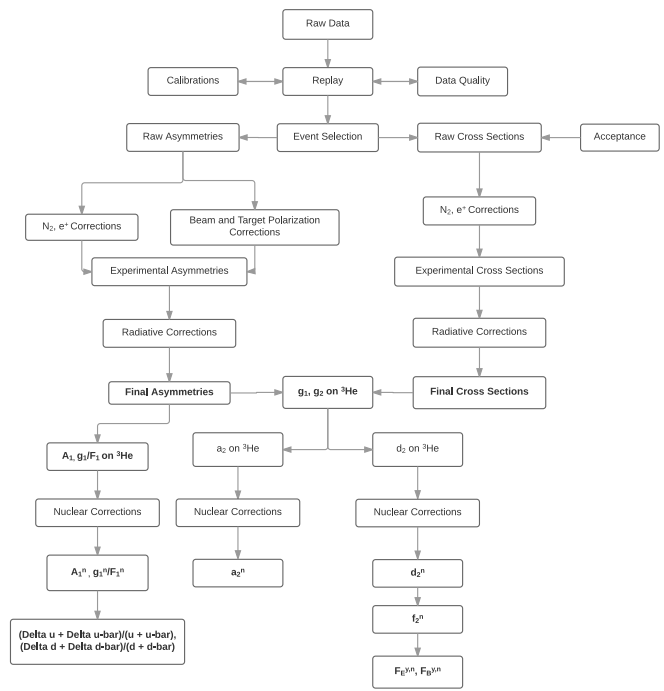}
    \caption{The data analysis procedure. }
    \label{fig:ana-flow-chart}
 \end{figure}

The analysis procedure is outlined in Fig.~\ref{fig:ana-flow-chart}, which shows 
that the raw data were first replayed, followed by the calibration and data quality checks.
Data calibrations included gain-matching ADC readings within the gas \v{C}erenkov 
and shower calorimeters to have the same responses for a given type of signal.  
Calibrations also involved optimizing the software packages that describes the optics 
of the two spectrometers.  Multi-foil carbon targets, a sieve slit collimator and
elastic $^{1}$H$(e,e')p$ data at an incident energy of $E = 1.23$\,GeV were used to
calibrate the optics software package for the LHRS~\cite{Qian:2011py} and for the 
BigBite spectrometer.  The momentum resolution achieved for the BigBite spectrometer 
was $\sim 1\%$~\cite{MPosikThesis}.  Data quality checks implied checking the calibration 
results and removing beam trips from the data.  Faulty runs (e.g., those having poor beam 
quality, detector live-times $\lesssim 80\%$, run times less than a few minutes, etc.) 
were also identified and discarded from the analysis.  

Following calibration and data quality checks, the electron sample was cleaned up
by removing events that did not generate a good trigger or had poor track
reconstruction.  Cuts were also made to remove pion tracks and events originating in the 
target window.  In the BigBite data set, geometrical cuts had to be implemented to 
remove events that rescattered from the BigBite magnet pole pieces. After all cuts 
had been applied to the data, the raw physics observables consisting of cross sections 
and asymmetries were then extracted.  Corrections were applied to account for the 
nitrogen target contamination and background due to pair-produced electrons, neither of 
which could be removed by cuts.  After these~corrections were applied, we obtained the 
experimental cross sections (Section~\ref{sec:xs-ana}) and the experimental asymmetries (Section~\ref{sec:asym_ana}).  
Applying radiative corrections yielded the final quantities for each of those, from 
which the spin structure functions $g_{1}$ and $g_{2}$ on $^{3}$He were extracted as described 
in Appendix~\ref{sec:spin-structure-funcs-app}.  The $^{3}$He results for the unpolarized 
cross sections, double-spin asymmetries and spin structure functions $g_{1}$ and $g_{2}$ 
are presented in Section~\ref{sec:res_he3}.  The Lorentz color force $d_{2}^{^{3}\text{He}}$ 
was obtained from Eq.~\ref{eqn:d2_exp}, after which nuclear corrections (Section~\ref{sec:nuclear_cor}) were 
applied to obtain $d_{2}^{n}$ (Section~\ref{sec:d2-results}).  From the $g_{1}^{^{3}\text{He}}$ 
data the $a_{2}$ matrix element on $^{3}$He (given as the third moment of $g_{1}$) was 
extracted.  Nuclear corrections, similar to those used for $d_{2}$, were applied to 
obtain $a_{2}^{n}$ (Section~\ref{sec:a2-results}).  From a twist expansion of world data 
for $\Gamma_{1}^{n}$, the twist-4 matrix element $f_{2}^{n}$ was obtained using our 
$d_{2}^{n}$ data as input (while the value of $a_{2}^{n}$ was taken from an average over 
the available model calculations, see Section~\ref{sec:res_cf}).  Additionally, $A_{1}^{^{3}\text{He}}$ and 
$g_{1}^{^{3}\text{He}}/F_{1}^{^{3}\text{He}}$ were extracted with the aid of Eqs.~\ref{eqn:a1_exp} 
and~\ref{eqn:g1f1_exp}.  Nuclear corrections were then applied to the $^{3}$He results to 
obtain the neutron quantities (Sections~\ref{sec:a1n-results} and~\ref{sec:g1f1n-results}).  
Using the $g_{1}^{n}/F_{1}^{n}$ data obtained, we then extracted the flavor-separated ratios 
$(\Delta u + \Delta \bar{u})/(u + \bar{u})$ and $(\Delta d + \Delta \bar{d})/(d + \bar{d})$ 
(Section~\ref{sec:res_flavor}).  

%===============================================================================
\subsection{Cross sections} \label{sec:xs-ana} 
\subsubsection{Extraction of raw cross sections from data} 
%===============================================================================

The unpolarized differential cross section was calculated from the data for a 
given run as follows:

 \begin{equation} \label{eqn:xs}
    \frac{d^3\sigma_{\text{raw}}}{d\Omega dE'} = \frac{t_{ps} N_{\text{cut}}}{(Q/e) n t_{LT} \varepsilon}
                                                   \frac{1}{\Delta E' \Delta \Omega \Delta Z},
 \end{equation}

\noindent where $t_{ps}$ denotes the prescale value for the T3 trigger~\footnote{A prescale factor 
restricts the number of events accepted for a given trigger.  For example, a prescale 
of 100 means that one event per every 100 will be accepted.  It was used to either 
remove certain types of events entirely, or to restrict events due to high rates.}, 
$N_{\text{cut}}$ the number of electrons that pass all cuts, $Q/e$ the number of beam 
electrons delivered to the target, $n$ the target number density in 
amagats~\footnote{1 amagat = $2.6867805 \times 10^{25}$\,m$^{-3}$.}, $t_{LT}$ the 
live-time~\footnote{The live-time is defined as the ratio of the number of triggers 
accepted by the DAQ to the total number of triggers generated for a given run~\cite{DFlayThesis}. 
This factor corrects for the high trigger rates during the experiment, which prevent the 
detectors from recording every event}, and  
$\varepsilon$ the product of all detector (cut) efficiencies.  The quantity  
$\Delta E' = \delta p/p_{0} \cdot p_0$, where $\delta p/p_{0} = (p - p_{0})/p_{0}$ is 
the scattered momentum of the electron $p$ relative to the LHRS momentum setting $p_{0}$.  
Electrons were selected according to the criterion $\left| \delta p/p_{0} \right| < 3.5$\%, 
which was based on the agreement of the Monte Carlo simulation of the spectrometer 
(see below) with the data~\cite{DFlayThesis}.  The quantity $\Delta Z$ denotes the effective 
target length seen by the spectrometer, measured in meters.  The cut on the effective 
target length was chosen such that the target windows and edge effects due to scattering 
from the magnets in the LHRS are removed.  The term $\Delta \Omega$ denotes the solid-angle 
acceptance, measured in steradians; it is defined as the product of the in-plane scattering 
angle $\Delta \theta$ and out-of-plane scattering angle $\Delta \phi$.  The cut chosen
for $\Delta \theta$ ($\Delta \phi$) was $\pm 40$\,mrad ($\pm 20$\,mrad), which amounts
to a solid angle of 3.2\,msr.  The cuts on $\delta p/p_{0}$, $\Delta Z$, $\Delta \theta$ 
and $\Delta \phi$ were informed by looking at a Monte Carlo simulation of the LHRS.    

The effective acceptance was determined with the Single-Arm Monte Carlo (SAMC)
simulation~\cite{SAMC}.  To determine how the geometrical acceptance of the LHRS deviates from
the ideal rectangular acceptance, SAMC began by generating events originating from 
the target that were uniformly distributed over the kinematical phase space.  Each event
was then transported to the focal plane using an optical model of the LHRS~\cite{LeRose}.  
As the particle encountered each magnet aperture in the LHRS, a check was performed 
to see if it successfully passed through the known apertures.  If the simulated particle 
successfully made it through all geometrical apertures, it was then reconstructed back 
to the target using the optics matrix optimized during the experiment.  The ratio $r$ of the 
number of reconstructed events to the number of generated events was used to determine 
the effective acceptance, written as 
$r = \Delta E' \Delta \Omega \Delta Z / \Delta E'_{\text{MC}} \Delta \Omega_{\text{MC}} \Delta Z_{\text{MC}}$.
The subscript MC refers to the initially generated kinematic phase space in the simulation, 
chosen to be larger than the apertures of the LHRS, so as to avoid edge effects. 

The cross sections extracted for each run of a given momentum bin were then averaged, 
weighted by their statistical errors: 

\begin{equation} \label{eqn:wavg}
   \langle \sigma \rangle = \frac{ \sum\limits_{i=0}^{n} \sigma_i \frac{1}{\delta \sigma_i^2} }
                                 { \sum\limits_{i=0}^{n} \frac{1}{\delta \sigma_i^2} }, 
\end{equation}

\noindent where $\delta \sigma_i$ is the statistical error on the cross section for 
the $i^{\text{th}}$ run. 

%===============================================================================
\subsubsection{Background corrections} 
%===============================================================================

The raw $^{3}$He cross section measured in the LHRS, $\sigma_{\text{raw}}$, 
contains contributions from electrons that were not scattered from $^{3}$He, but 
were produced in processes corresponding to electron-positron (pair) production 
(arising from $\pi^{0}$ mesons decaying predominantly to photons), or 
scattering from nitrogen nuclei.  

To remove the pair-production contributions from $\sigma_{\text{raw}}$, several 
runs were taken with the LHRS in positive polarity mode (i.e., detecting positrons) 
to measure the positron cross section, $\sigma^{e^{+}}$. 
The nitrogen electron cross section $\sigma_{\text{N}_{2}}^{e^{-}}$ was measured 
by filling the additional reference target cell (Section~\ref{sec:target-cells}) 
with nitrogen gas and exposing it to the beam.  Pair production also occurs when 
scattering from nitrogen nuclei, so a nitrogen positron cross section, $\sigma_{\text{N}_{2}}^{e^{+}}$, 
was also measured with the LHRS in positive polarity mode.  The positron cross 
section on nitrogen $\sigma_{\text{N}_{2}}^{e^{+}}$ was subtracted from 
$\sigma_{\text{N}_{2}}^{e^{-}}$ to avoid double-counting the pair-produced events 
in the measurement that were already accounted for in $\sigma^{e^{+}}$.  
In principle, one has to consider pion background contributions; however, 
given the high pion suppression in the LHRS (Section~\ref{sec:det-perf}), 
this component was found to be negligible.  Combining the measurements for $\sigma_{\text{raw}}$, 
$\sigma^{e^{+}}$, $\sigma_{\text{N}_{2}}^{e^{-}}$, $\sigma_{\text{N}_{2}}^{e^{+}}$, 
yielded the experimental $^{3}$He cross section, $\sigma_{\text{exp}}$:

\begin{eqnarray} \label{eqn:xs_bkgnd}
   \sigma_{\text{exp}}                 &=& \sigma_{\text{raw}} - \sigma^{e^{+}} 
                                        -  \sigma_{\text{N}_{2}}^{\text{dil}}  \label{eqn:exp_xs} \\
   \sigma_{\text{N}_{2}}^{\text{dil}}  &=& \frac{n_{\text{N}_{2}}}{n_{\text{N}_{2}} + n_{^{3}\text{He}}}
                                           \left(\sigma_{\text{N}_2}^{e^{-}} - \sigma_{\text{N}_{2}}^{e^{+}}\right), 
                                           \label{eqn:xs-nit-dil} 
\end{eqnarray}

\noindent where $n_{\text{N}_{2}}$ is the number density of nitrogen in the production cell
and $n_{^{3}\text{He}}$ is the number density of $^{3}$He in the production cell.

When the nitrogen reference cell was in the beam, the number density for the cell
was extracted using the measured temperature and pressure of the cell.  This number
density was used when extracting a nitrogen cross section (Eq.~\ref{eqn:xs}). 
A systematic uncertainty of 2.2\% was estimated by computing the number densities 
while varying the temperature and pressure by up to 2$^\circ$\,C and 2\,psig, 
respectively~\cite{MPosikThesis}.  In the $^{3}$He production cell, the number 
density of nitrogen $n_{\text{N}_{2}}$ was taken to be 0.113\,amg (Eq.~\ref{eqn:xs-nit-dil}).  
This value was recorded as the target was initially filled, and is accurate 
to 3\% from a pressure curve analysis. 

\begin{figure}[hbt]
    \centering 
    \includegraphics[width=0.5\textwidth]{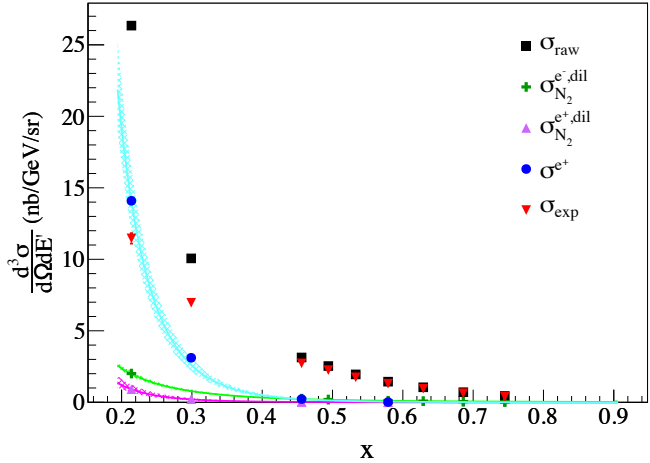}
    \caption{(Color online) Cross sections as a function of Bjorken-x at $E = 4.74$\,GeV.  
             The squares show the raw electron cross section, and the circles 
             (and fit) show the positron cross section measured on the $^{3}$He target.  
             The cross data (and fit) represent the diluted nitrogen 
             cross section measured on the N$_{2}$ target, and the up-triangle 
             data (and fit) show the diluted nitrogen cross section 
             measured in positive polarity mode on the N$_{2}$ target.  
             The down-triangle data points are the final background-subtracted data, 
             $\sigma_{\text{exp}}$.
             The error bars on the data points represent the statistical uncertainties.    
             }
    \label{fig:exp_xs_4}
 \end{figure}

\begin{figure}[hbt]
    \centering
    \includegraphics[width=0.5\textwidth]{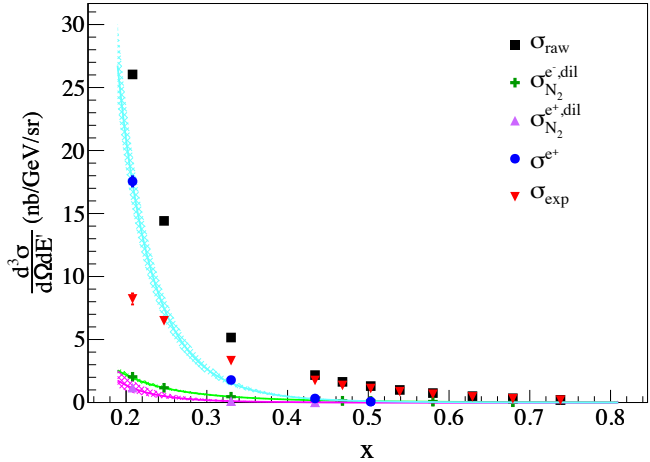}
    \caption{(Color online) Cross sections as a function of Bjorken-x at $E = 5.89$\,GeV.  
             The description of the various data sets is the same as in Fig.~\ref{fig:exp_xs_4}.    
             }
    \label{fig:exp_xs_5}
 \end{figure}

Due to time constraints and hardware problems encountered during the experiment, 
there were not enough data to map out the background contributions to the raw cross section 
for all kinematic bins.  To resolve this issue, empirical fits to the 
positron and nitrogen data (see Appendix~\ref{sec:xs-fits-app}) were used to subtract 
those contributions.  Figures~\ref{fig:exp_xs_4} and~\ref{fig:exp_xs_5} show the raw 
electron cross section, the positron and nitrogen cross sections (scaled by the ratio 
of the nitrogen number density to that of $^{3}$He in the production target cell), 
and the background-subtracted electron cross section, $\sigma_{\text{exp}}$.  
The error bars on the data points represent the statistical uncertainties.  
The largest correction was due to the positrons, at $\sim 53$\% in the lowest $x$ bin, 
and fell to a few percent for $x \gtrsim 0.5$.   

%===============================================================================
\subsubsection{Radiative corrections} \label{sec:xs_rc}
%===============================================================================

A first correction that must be done {\it before} carrying out the radiative corrections
is to subtract the elastic and quasi-elastic radiative tails, since they are long 
and affect all states of higher invariant mass $W$~\cite{Mo:1968cg}.  For these 
kinematics, the elastic tail was small and affects the lowest bins in scattered 
electron energy $E'$ at the $\lesssim$ 1\% level only. The elastic tail was computed 
following the exact formalism given by Mo and Tsai~\cite{Mo:1968cg}, and using 
elastic $^{3}$He form factors from Amroun~\cite{Amroun:1994qj}.
The $^{3}$He quasi-elastic tail, however, was much larger, at $\sim 25$--$30\%$  
in the lowest $x$ bin. The quasi-elastic radiative tail was computed by utilizing 
an appropriate model of the $^{3}$He quasi-elastic cross section~\cite{QFS} and 
applying radiative effects~\cite{Stein:1975yy}.  The tail was then subtracted from the data.  
The model was checked against existing quasi-elastic $^{3}$He 
data~\cite{Day:1979bx,Marchand:1985us,Meziani:1992xr} covering a broad range of kinematics.

After the elastic and quasi-elastic tails had been subtracted from the data, radiative
corrections were applied according to~\cite{Mo:1968cg,TsaiSLAC}, where the internal corrections
were calculated using the equivalent radiator method and the external corrections
were performed using the energy peaking approximation.  In the experiment, we 
took production data for only two beam energies of 4.74 GeV and 5.89 GeV.  
However, we needed enough data to properly calculate the integrals involved in the 
radiative correction procedure.  Therefore, we used the F1F209 cross section 
parameterization~\cite{Bosted:2012qc} to fill in the rest of the phase space for 
each data set.  The radiative corrections were as large as $\sim 50\%$ in the 
lowest measured $x$ bin, and fell off to a few percent at the large $x$ bins.

\begin{figure}[hbt]
   \centering
   \includegraphics[width=0.5\textwidth]{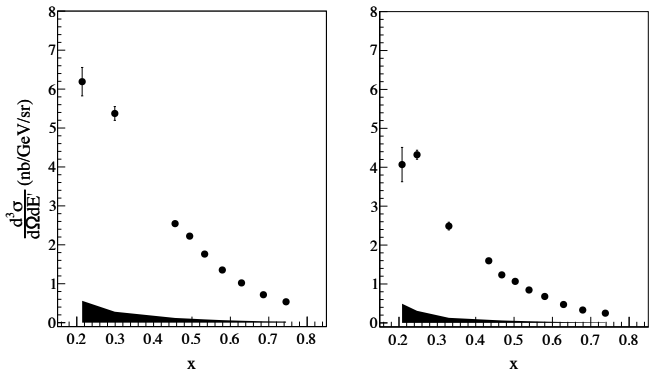}
   \caption{The final $^{3}$He unpolarized cross sections as a function of Bjorken-$x$.  
            The left (right) panel shows $E = 4.74$\,GeV (5.89\,GeV) data.  The error bars show the 
            statistical uncertainty, while the bands represent the systematic uncertainty.      
            }
   \label{fig:born-xs}
\end{figure}

The resulting final $^{3}$He cross sections for E = 4.74\,GeV and 5.89\,GeV are presented 
in Fig.~\ref{fig:born-xs}.  The data are tabulated in Tables~\ref{tab:xs_4} and~\ref{tab:xs_5}.  
The uncertainty on the final cross section arising from the radiative corrections 
was determined by varying the contributions to the radiative correction calculations,
including the radiation thicknesses in the incident and scattered electron path, 
and considering different models for the elastic and quasi-elastic tail calculations.  
Of these, the quasi-elastic tail gave the biggest uncertainty, $\sim$ 5--6\% for 
the lowest $x$ bin, falling to $\sim 1\%$ for all other bins.  A full breakdown 
of the uncertainties on the final results is given in Tables~\ref{tab:xs-syst-err-4} 
and~\ref{tab:xs-syst-err-5} in Appendix~\ref{sec:xs-syst-err-tables}.

%===============================================================================
\subsection{Asymmetries} \label{sec:asym_ana} 
\subsubsection{Extraction of raw asymmetries from data}  \label{sec:asym-ana-raw}
%===============================================================================
 
The raw double-spin asymmetries were extracted from the data recorded 
in the BigBite detectors according to: 

\begin{eqnarray} 
   A_{\parallel}^{\text{raw}} &=& \frac{N^{\downarrow \Uparrow} - N^{\uparrow \Uparrow}}{N^{\downarrow \Uparrow} + N^{\uparrow \Uparrow}} \label{eqn:raw_para_asym} \\  
   A_{\perp}^{\text{raw}}     &=& \frac{1}{\langle \cos\phi \rangle} \frac{N^{\downarrow \Rightarrow} - N^{\uparrow \Rightarrow}}{N^{\downarrow \Rightarrow} 
                                  + N^{\uparrow \Rightarrow}}, \label{eqn:raw_perp_asym}
\end{eqnarray} 

\noindent where $N$ denotes the number of electron counts after applying data quality
and PID cuts for a particular beam and target spin configuration, $\downarrow$ ($\uparrow$)
corresponds to the electron's spin polarized antiparallel (parallel) to its momentum,
$\Uparrow$ indicates that the target was polarized parallel to the electron momentum, 
and $\Rightarrow$ indicates that the target was polarized transverse to the electron momentum.  

In E06-014, there were three target-spin configurations, either parallel 
($\parallel$, referred to here as longitudinal) or transverse ($\perp$, in two 
orientations: $\Leftarrow$, 90$^{\circ}$ and $\Rightarrow$, 270$^{\circ}$) 
to the electron beam momentum.  The quantity $\phi$ is the angle between the 
electron scattering plane (defined by incident and scattered electron momenta 
$\vec{k}$ and $\vec{k}'$) and the polarization plane (defined by $\vec{k}$ and 
the target polarization $\vec{S}$).  The transverse data were normalized by 
$\left< \cos \phi \right>$, since this term is not necessarily equal to unity.  
This correction was found to be very small.  There is no $\left< \phi \right>$ 
correction for parallel asymmetry data since the electron spin is aligned along 
the target spin direction.  

For the spin asymmetry analysis, one needs to correctly classify the data according to 
the relation between the digital helicity signal and the physical helicity of 
the electrons in the beam.  M\o ller polarimetry measurements (Secton~\ref{sec:beam_pol}) 
were performed after each beam configuration change, to check the consistency of the 
electron helicity assignment in Hall A~\cite{DParnoThesis}.  To confirm the relationship 
between the digital helicity signal and the physical helicity of the electrons in the 
beam, measurements of the $^{3}$He longitudinal quasi-elastic asymmetry were conducted 
at $E = 1.23$\,GeV and $\theta = 45^{\circ}$.  The sign of the extracted raw quasi-elastic 
asymmetry was verified to be consistent with our understanding of the relationship between 
the digital helicity signal and the physical electron helicity. 

Since there were two transverse target spin configurations, care must be taken 
when combining the results for the 90$^{\circ}$ and 270$^{\circ}$ since they will 
have opposite signs relative to one another.  To determine which target configuration should carry 
which sign, one can consider the dot product of the scattered electron momentum 
vector $\vec{k'}$ and the target spin vector $\vec{S}$.  With the target polarized 
transverse to the incident electron momentum $\vec{k}$, the target spin only enters the 
cross section through the dot product of $\vec{k'}$ and $\vec{S}$~\cite{Anselmino:1994gn}.  
Their dot product is $\vec{k'} \cdot \vec{S} = E' \sin \theta \cos \phi$, where 
$\theta$ is the electron scattering angle.  For the transverse spin configurations, 
$\phi$ is nearly 0.  The positive sense of the target spin is then the direction that points 
to the side of the beamline where the scattering electron is detected 
(consistent with JLab E99-117~\cite{Zheng:2004ce}).  In this experiment, the asymmetry 
measurement was done using BigBite; therefore, when the target spin was pointing 
towards BigBite (270$^{\circ}$), the asymmetry carried a positive sign.  If the target 
spin was pointing towards the LHRS (90$^{\circ}$), it carried a negative sign.  
Using this convention, the results for 270$^{\circ}$ and 90$^{\circ}$ were averaged 
together using their statistical errors as a weight.

%===============================================================================
\subsubsection{Experimental asymmetries}
%===============================================================================

The raw asymmetry definitions shown in Eqs.~\ref{eqn:raw_para_asym} and~\ref{eqn:raw_perp_asym} 
do not account for dilution effects due to the presence of nitrogen in the target 
or the imperfect beam and target polarizations.  Therefore, the raw asymmetries must be
corrected to yield the experimental asymmetries: 

\begin{equation} \label{eqn:exp_asym}
   A_{\parallel,\perp}^{\text{exp}} = \frac{1}{D_{\text{N}_{2}}P_{b}P_{t}}A_{\parallel,\perp}^{\text{raw}},
\end{equation} 

\noindent where $P_{b}$ and $P_{t}$ denote the beam and target polarizations, 
respectively (Sections~\ref{sec:beam_pol} and~\ref{sec:target-pol}) and $D_{\text{N}_{2}}$ 
the nitrogen dilution factor.  Experimental asymmetries for one target-spin configuration
were averaged in the same way as shown in Eq.~\ref{eqn:wavg}, where the $i^{\text{th}}$ 
cross section $\sigma_i$ is replaced by the $i^{\text{th}}$ experimental asymmetry $A^{\text{exp}}_{i}$. 

The nitrogen dilution factor was determined by comparing the rates from the 
nitrogen reference cell against those from the $^{3}$He production cell:

\begin{equation}
   D_{\text{N}_{2}} = 1 - \frac{\Sigma_{\text{N}_{2}}(\text{N}_{2})}{\Sigma_{\text{total}}(^{3}\text{He})}
                   \frac{t_{ps}(\text{N}_{2})}{t_{ps}(^{3}\text{He})}
                   \frac{Q(^{3}\text{He})}{Q(\text{N}_{2})}
                   \frac{t_{LT}(^{3}\text{He})}{t_{LT}(\text{N}_{2})}
                   \frac{n_{\text{N}_{2}}(^{3}\text{He})}{n_{\text{N}_{2}}(\text{N}_{2})}, 
\end{equation} 

\noindent where $\Sigma_{\text{N}_{2}}$ and $\Sigma_{\text{total}}$ denote the 
total number of counts that pass data quality and PID cuts detected during the N$_{2}$
and $^{3}$He production target runs, while $n_{\text{N}_{2}}(\text{N}_{2})$ and $n_{\text{N}_{2}}(^{3}\text{He})$
denote the nitrogen number densities present in the two targets.  Because the nitrogen
and $^{3}$He production runs have different characteristics (e.g., scattering rates, running time, etc.),
the measured electron counts must be normalized by the total charge, given by 
$Q(\text{N}_{2})$ and $Q(^{3}\text{He})$, deposited on the two targets, the trigger 
prescale factors for the nitrogen and $^{3}$He runs, given as $t_{ps}(\text{N}_{2})$ 
and $t_{ps}(^{3}\text{He})$; and finally, the live-times for the nitrogen and $^{3}$He runs, 
given as $t_{LT}(\text{N}_{2})$ and $t_{LT}(^{3}\text{He})$. 

The nitrogen dilution factor was extracted on a run-by-run basis and the results 
were averaged, weighted by their statistical uncertainties for a given run configuration.  
The resulting dilution factor was applied bin-by-bin in $x$, and was found to be 
roughly constant at $D_{\text{N}_{2}} \approx 0.920 \pm 0.003$~\cite{MPosikThesis}.

%===============================================================================
\subsubsection{Background corrections}
%===============================================================================

\begin{figure}[hbt]
   \centering
   \includegraphics[width=0.5\textwidth]{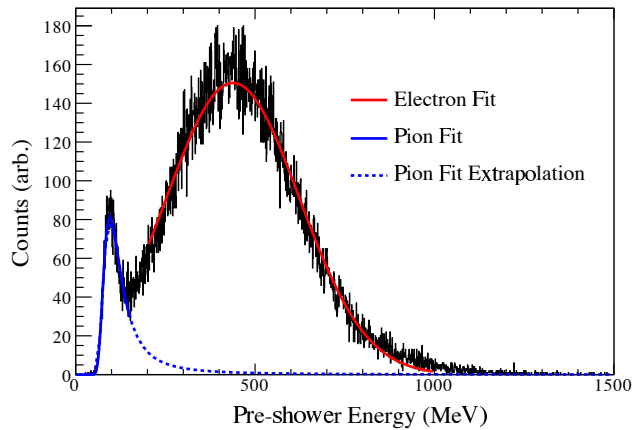}
   \caption{A signal distribution in the BigBite pre-shower, where the particle's
            total deposited energy is plotted for the $<x> = 0.425$ bin.
            Electrons (pions) are shown by the peak on the right (left).
            The curves fitted to the data are used in the analysis to determine 
            the amount of pion contamination in the electron sample, see the text. 
            }
   \label{fig:bigbite-pid-ps}
\end{figure}

As described in Section~\ref{sec:xs-ana}, the main sources of background 
contamination were charged pions and pair-produced electrons. 
To quantify the charged pion contamination in the electron sample, 
the pion peak in the pre-shower energy spectrum was fitted with a Gaussian function 
convoluted with a Landau function, and the electron peak with a Gaussian function, 
as shown in Fig.~\ref{fig:bigbite-pid-ps}.  The ratio of the pion counts to the 
electron counts was then evaluated from the integrals of the two fits above a threshold 
of 200 MeV~\cite{MPosikThesis}.  This ratio was evaluated for the $\pi^{-}$ 
($N_{\pi^{-}}/N_{e^{-}}$) and $\pi^{+}$ ($N_{\pi^{+}}/N_{e^{+}}$) mesons.  The $\pi^{+}$ 
ratio was evaluated after reversing the polarity of BigBite so 
that particles with similar trajectories could be compared.  The $N_{\pi^{-}}/N_{e^{-}}$ 
ratio was largest in the lowest $x$ bin of 0.277, at $\sim 2.7\%$, and dropped 
quickly to below 1\% by $x = 0.425$.  The $N_{\pi^{+}}/N_{e^{+}}$ ratio was larger 
and consistently $\sim 6\%$ across the whole $x$ range.  A systematic uncertainty 
of 2.5\% was assigned to the $N_{\pi}/N_{e}$ ratios, the value determined  
in the immediately preceding experiment, E06-010~\cite{Qian:2011py}, in which a 
similar fitting procedure was used and checked independently through a coincidence 
trigger between the electrons in BigBite and pions in the LHRS. In E06-010, it was 
found that these two methods were consistent to around 2--3\%~\cite{KAlladaThesis}.

\begin{figure}[hbt]
   \centering
   \includegraphics[scale=0.45]{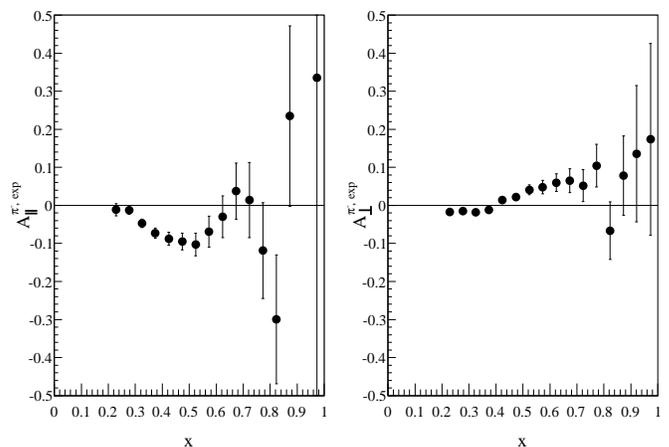} 
   \caption{The $\pi^{-}$ experimental asymmetry on $^{3}$He for $E = 5.89$\,GeV.  The left (right) 
            panel shows the data for the target polarized longitudinal (transverse) 
            to the electron beam momentum.  The error bars indicate the statistical uncertainty.}
   \label{fig:asym-pi-minus} 
\end{figure}

\begin{figure}[hbt]
   \centering
   \includegraphics[scale=0.45]{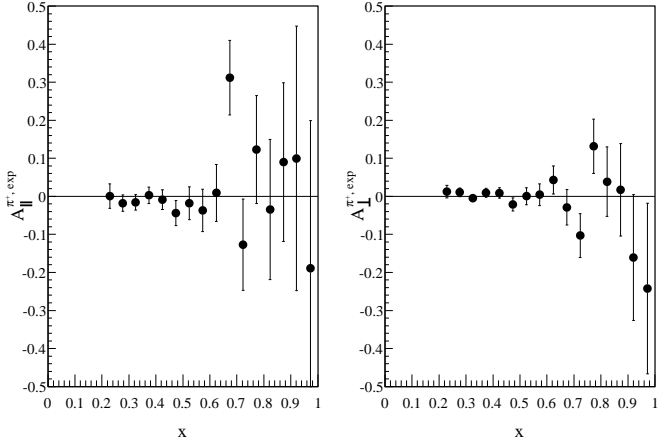} 
   \caption{The $\pi^{+}$ experimental asymmetry on $^{3}$He for $E = 5.89$\,GeV.  The left (right) 
            panel shows the data for the target polarized longitudinal (transverse) 
            to the electron beam momentum.  The error bars indicate the statistical uncertainty.}
   \label{fig:asym-pi-plus} 
\end{figure}

A spin asymmetry in the charged-pion production will affect the measured electron
asymmetries.  To study this, asymmetries in the pion sample were determined after applying  
corrections for the nitrogen dilution and the beam and target polarizations, yielding 
the $\pi^{-}$ and $\pi^{+}$ experimental asymmetries.  The results are shown in 
Figs.~\ref{fig:asym-pi-minus} and~\ref{fig:asym-pi-plus} for $E = 5.89$\,GeV.  
After scaling these measured pion asymmetries with the $N_{\pi}/N_{e}$ ratios, 
it was found that the $\pi^{-}$ ($\pi^{+}$) asymmetry contribution was less than 
5\% (3\%) of the statistical uncertainty in the asymmetries and therefore could be neglected.

\begin{figure}[hbt]
   \centering
   \includegraphics[scale=0.48]{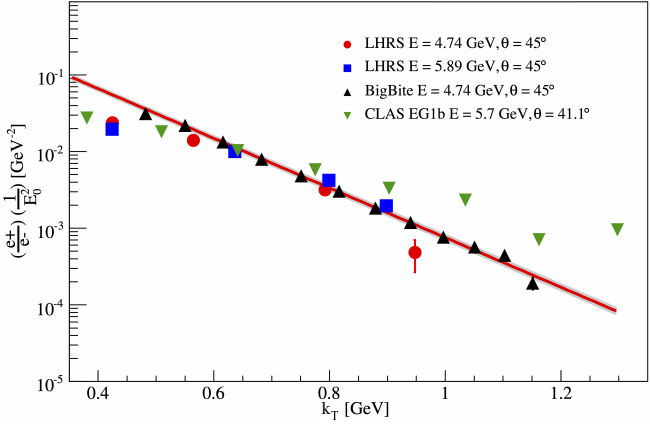} 
   \caption{(Color online) The $N_{e^{+}}/N_{e^{-}}$ data from this experiment measured
            in the LHRS and BigBite and CLAS EG1b~\protect\cite{CLASEG1b} plotted as 
            $(1/E^{2})(N_{e^{+}}/N_{e^{-}})$ versus the transverse momentum $k_{T}$.  
            Our fit function is shown by the solid curve, with its error given by the 
            band surrounding it.  See text for additional details.}
   \label{fig:e-plus-e-minus} 
\end{figure}

To quantify the contamination due to pair-produced electrons, the ratio of
positrons to electrons, $N_{e^{+}}/N_{e^{-}}$ needed to be determined. Due to time 
constraints this ratio could only be measured directly at 4.74\,GeV. To determine this 
ratio for the 5.89\,GeV run a parametrization was made of the 4.74\,GeV BigBite 
and LHRS runs, the 5.89\,GeV LHRS run together with data at 5.7\,GeV and $\theta = 41.1^{\circ}$ 
from JLab CLAS EG1b~\cite{CLASEG1b}. These data, multiplied by $(1/E^{2})(N_{e^{+}}/N_{e^{-}})$, 
were then fit to an empirical function $f(k_T) = \exp(a + b \cdot k_T)$ with $k_T = k \sin \theta$ 
being the transverse momentum. The data and the fit are shown in 
Fig.~\ref{fig:e-plus-e-minus}.  The error on the fit was propagated to be 5--6\% 
across the $x$-range of the measurements.  The lowest $x$ bin, $x = 0.23$, was then 
excluded from further analysis after the $N_{e^{+}}/N_{e^{-}}$ ratio was found to be 
in excess of 80\%. At the first bin included in the analysis, $x = 0.277$, that 
ratio was $\sim 50$\% and fell to less than 10\% by $x = 0.473$.  Beyond $x = 0.5$ 
the ratio had dropped to below 3\%~\cite{MPosikThesis}.  

\begin{figure}[hbt]
   \centering
   \includegraphics[scale=0.45]{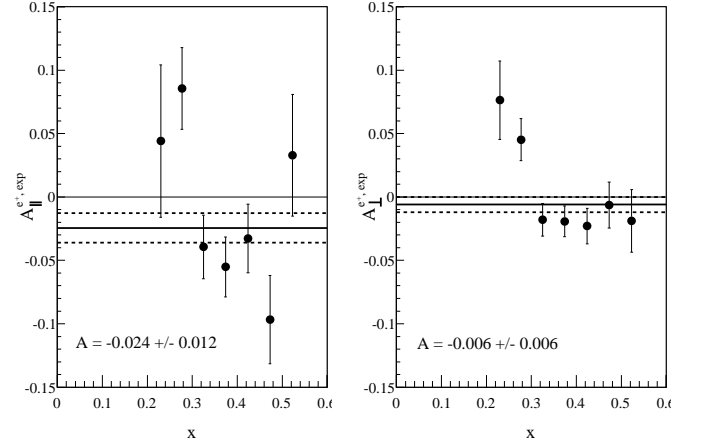} 
   \caption{The positron experimental asymmetry for $E = 4.74$\,GeV.  The left (right) 
            panel shows the data for the target polarized longitudinal (transverse) 
            to the electron beam momentum.  The error bars indicate the statistical uncertainty.
            The weighted average is indicated by the solid line, and its uncertainty is indicated
            by the surrounding dashed lines.  The numerical value of the weighted average
            with its uncertainty is also shown.}
   \label{fig:asym-positron-4} 
\end{figure}

\begin{figure}[hbt]
   \centering
   \includegraphics[scale=0.45]{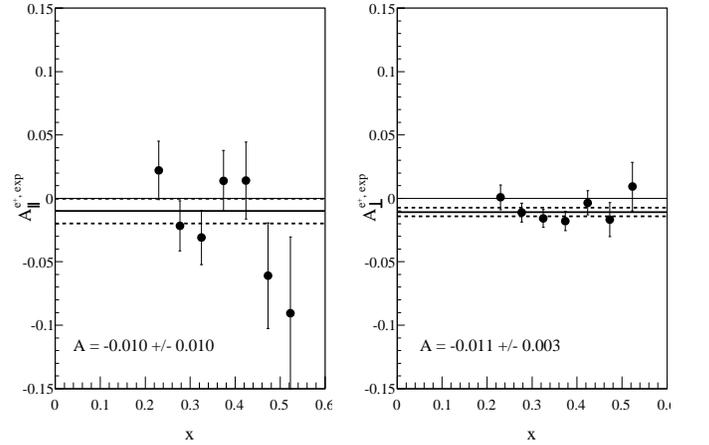} 
   \caption{The positron experimental asymmetry for $E = 5.89$\,GeV.  
            The desciption of the data and fitted line with its uncertainty 
            is the same as Fig.~\ref{fig:asym-positron-4}.  
           }
   \label{fig:asym-positron-5} 
\end{figure}

Just as with the pion contamination, pair-produced electrons can affect or dilute 
the measured electron asymmetry.  Ideally, the asymmetry should be measured by 
reversing the polarity of BigBite so that the positrons are detected with the same 
acceptance as the electrons.  However, due to the same time constraints as mentioned 
previously, this could only be completed for one target-spin orientation ($270^{\circ}$) 
at 4.74\,GeV.  Because of this, the positron asymmetries were measured bent down in 
BigBite.  The asymmetries measured in this configuration were observed to be in 
good agreement with those when the positrons were deflected upwards; however, the 
acceptance, and hence the rate, was $\sim 60$\% lower.  Therefore, for each beam 
energy and each target-spin configuration a weighted average was computed over all 
measured asymmetries.  The results are shown in Figs~\ref{fig:asym-positron-4} 
and~\ref{fig:asym-positron-5}.  The measured electron asymmetries for each beam energy 
and target spin-configuration were then corrected by the corresponding weighted 
average positron asymmetry scaled by the $N_{e^{+}}/N_{e^{-}}$ ratio.   

The effects of the charged-pion and pair-produced electron asymmetries were 
corrected for through:

\begin{equation} \label{eqn:asym_cor}
   A_{e^{-}} = \frac{A^{\text{exp}}_{e^{-}} - f_{1}A^{\text{exp}}_{\pi^{-}} - f_{3}A^{\text{exp}}_{e^{+}} + f_{2}f_{3}A^{\text{exp}}_{\pi^{+}} }{1 - f_{1} - f_{3} + f_{2}f_{3}},
\end{equation}  

\noindent where $f_{1} = N_{\pi^{-}}/N_{e^{-}}$; $f_{2} = N_{\pi^{+}}/N_{e^{+}}$; $f_{3} = N_{e^{+}}/N_{e^{-}}$;
$A^{\text{exp}}_{\pi^{\pm}}$ are the $\pi^{\pm}$ experimental asymmetries, and 
$A^{\text{exp}}_{e^{+}}$ is the positron experimental asymmetry.  Since 
the corrections for the pion experimental asymmetries were found to be negligible, 
Eq.~\ref{eqn:asym_cor} can be simplified to: 

\begin{equation} \label{eqn:asym_cor_fin}
   A_{e^{-}} \approx \frac{A^{\text{exp}}_{e^{-}}  - f_{3}A^{\text{exp}}_{e^{+}} }{1 - f_{1} - f_{3} + f_{2}f_{3}} \equiv A^{\text{cor}}. 
\end{equation}  

At this point, the bins for which $x > 0.90$ were removed from the analysis,
so as to exclude the quasi-elastic and $\Delta$ resonance contributions.

%===============================================================================
\subsubsection{False asymmetries}
%===============================================================================

When measuring a scattering asymmetry, care must be taken to ensure that the asymmetry 
was due to electron spin-dependent scattering and not to helicity-correlated changes
in the electron beam, known as {\it false asymmetries}.  One potential false 
asymmetry arises from a difference in the electron beam intensity between the 
two helicity states, resulting in an asymmetry in the deposited charge on the target. 
During the experiment, the beam charge asymmetry was limited to $\sim 100$\,ppm
through the use of a feedback loop controlled by a specialized DAQ~\cite{HAPPEXDAQ}  
and was verified by measuring the charge asymmetry using the Compton polarimeter~\cite{DParnoThesis}. 
Compared to the size of the electron asymmetry measurements, $A^{\text{cor}}$, the 
charge asymmetry was negligible.

Helicity-dependent DAQ changes can also generate false asymmetries, which can 
be observed through the measurement of the detector's live-time.  A helicity-dependent 
rate change could lead to one helicity state having a larger live-time than the other, 
resulting in an asymmetry. BigBite detector live-times were recorded for each helicity 
gate for each run of the experiment.  The helicity-dependent live-time asymmetry was 
extracted from the data and was found to be $< 100$\,ppm for the entire data set, 
which was negligible~\cite{MPosikThesis}. 

In addition to charge- and DAQ-induced false asymmetries, the analysis could also 
introduce a false asymmetry.  For example, if the data rates were high enough, it 
may be more difficult to reconstruct good tracks related to the higher-rate
helicity state as compared to the lower-rate one, resulting in an asymmetry~\cite{SRiordanThesis}.
However, the E06-014 data set was dominated by single-track events ($\sim 96\%$),
and thus the rates were not high enough for such an asymmetry to have a significant 
impact on the measured electron asymmetries. 

Potential sources of false asymmetries are limited by the 30\,Hz helicity flipping 
rate of the electron beam.  Additionally, any false asymmetry that does not change 
sign with respect to the IHWP state (Section~\ref{sec:asym-ana-raw}), such as those 
due to electronic cross-talk~\cite{GMillerThesis}, would be canceled when combining 
data from the two IHWP states.  In summary, no significant false asymmetries were observed. 

%===============================================================================
\subsubsection{Radiative corrections} \label{sec:asym_rc} 
%===============================================================================

Radiative corrections on the asymmetries were applied utilizing a similar approach 
as on the cross sections in Section~\ref{sec:xs_rc}.  We carried out the corrections 
on polarized cross section differences, $\Delta\sigma$, related to asymmetries by:

\begin{equation} \label{eqn:delta_sig}
    A_{\parallel,\perp}^{r} = \frac{\Delta\sigma_{\parallel,\perp}^{r} }{2\sigma_{0}^{r}}, 
\end{equation}

\noindent where $A_{\parallel}^{r}$ ($A_{\perp}^{r}$) indicates the longitudinal 
(perpendicular) asymmetry which includes radiative effects.  The unpolarized cross section 
is $\sigma_0^{r}$, where the $r$ indicates that radiative effects have been applied.  
We used the F1F209 parameterization~\cite{Bosted:2012qc} for the unpolarized cross section.  
The input used to fill out the integration phase space fell into three different
kinematic regions: the DIS region, the quasi-elastic region, and the $\Delta$ resonance region. 
For the DIS region the DSSV~\cite{deFlorian:2008mr} PDF parametrization was used, 
for the quasi-elastic region Bosted's nucleon form factors~\cite{Bosted:1994tm}, 
smeared by a quasi-elastic scaling function~\cite{Amaro:2004bs} was used and for 
the $\Delta$ region the MAID model~\cite{Drechsel:2007if} was used.  The $\Delta\sigma$ 
obtained after putting these models together for the three regions described the JLab E94-010 
data~\cite{Amarian:2002ar,*Slifer:2008re} reasonably well~\cite{DFlayThesis}.

In the radiative correction procedure, the quasi-elastic tail was not subtracted 
first, but rather included in the integration.  The elastic tail was found to be 
negligible and was not subtracted.  To minimize statistical fluctuations in the 
radiative corrections, the corrections were performed on a model of our data set.  
After obtaining the final $\Delta\sigma$, the corresponding asymmetry was obtained 
by inverting Eq.~\ref{eqn:delta_sig} (but using the {\it Born} $\sigma_0$ from F1F209) 
to find $A$ with the size of the radiative correction given by:

\begin{equation}
   \Delta A = A^{b} - A^{r}.  
\end{equation}

\noindent The quantity $A^{b}$ denotes the Born asymmetry and $A^{r}$ the radiated asymmetry.
Here, $A^{r}$ is the model input to the radiative corrections program.  This $\Delta A$ 
was applied to our extracted asymmetries, $A_{\parallel,\perp}^{\text{cor}}$ (see Eq.~\ref{eqn:asym_cor_fin}), 
as an additive correction.  The size of the radiative correction was found to be at 
most of the order of 45\% (10$^{-3}$ absolute) of the uncorrected asymmetry.  The radiative 
corrections in the DIS region were checked against results obtained using the formalism 
of Akushevich {\it et al.}~\cite{Akushevich:1997di}.  The results of both methods agreed 
to the $10^{-4}$ level in the asymmetry.    

\begin{figure}[hbt]
   \centering
   \includegraphics[width=0.5\textwidth]{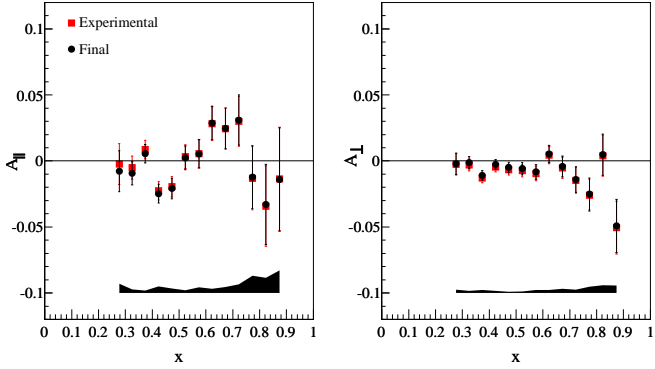} 
   \caption{(Color online) The parallel and perpendicular asymmetries on $^{3}$He for $E = 4.74$\,GeV 
            before (square) and after (circle) radiative corrections.  The error bands indicate the 
            systematic uncertainty for the final asymmetries.    
            }
   \label{fig:rad-born-asym-4}
\end{figure}

\begin{figure}[hbt]
   \centering
   \includegraphics[width=0.5\textwidth]{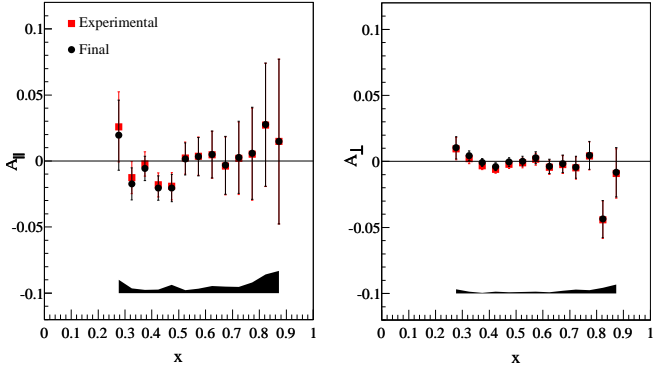} 
   \caption{(Color online) The parallel and perpendicular asymmetries on $^{3}$He for $E = 5.89$\,GeV 
            before (square) and after (circle) radiative corrections.  The error bands  
            indicate the systematic error for the final asymmetries.    
            }
   \label{fig:rad-born-asym-5}
\end{figure}

The asymmetries on $^{3}$He before and after radiative corrections for the 4.74 and 5.89\,GeV 
runs are shown in Figs.~\ref{fig:rad-born-asym-4} and~\ref{fig:rad-born-asym-5} and tabulated in 
Tables~\ref{tab:asym_4} and~\ref{tab:asym_5} in Section~\ref{sec:results}.   The systematic 
uncertainties for the radiative corrected data, were obtained by varying all input to reasonable levels. 
The inputs varied included the electron cuts, the nitrogen dilution factor, beam and target 
polarizations, pion and pair-production contamination levels and the radiative corrections. 
The latter were observed to change less than 5\% when using various input models and when 
the radiation thicknesses before and after scattering were varied by up to $\pm 10$\%. 
Tables in Appendix~\ref{sec:asym-syst-err-tables} list the various contributions to the 
systematic errors.

%===============================================================================
\subsection{From $^{3}$He to the neutron} \label{sec:nuclear_cor}
%===============================================================================

Free nucleons behave differently from those bound in nuclei, primarily due to spin 
depolarization, Fermi motion, nuclear binding, and nuclear shadowing and anti-shadowing 
effects.  Additionally, the characteristics of bound nucleons can be altered by 
the presence of non-nucleonic degrees of freedom and how far off-shell the nucleons are. 

We utilized the work of Bissey {\it et. al.}~\cite{Bissey:2001cw}, which provides
a description of the $g_{1}$ spin structure function on $^{3}$He over the range 
$10^{-4} \leq x \leq 0.8$.  It models the $^{3}$He wavefunction incorporating 
the $S$, $S'$ and $D$ states, and includes a pre-existing $\Delta$ (1232) component:  

\begin{eqnarray} \label{eqn:g1he3_nucl} 
   g_{1}^{^{3}\text{He}}(x) &=& P_{n}g_{1}^{n}(x) + 2P_{p}g_{1}^{p}(x)                                
                             - 0.014\left[ g_{1}^{p}(x) - 4 g_{1}^{n}(x)\right] \nonumber \\
                            &+& a(x) g_{1}^{n}(x) + b(x) g_{1}^{p}(x), 
\end{eqnarray} 

\noindent where $P_{p}$ and $P_{n}$ are the effective polarizations of the proton 
and neutron in $^{3}$He~\cite{Friar:1990vx}, respectively.  The third term arises 
from the $\Delta(1232)$ component in the $^{3}$He wave function~\cite{Bissey:2001cw}.  
The functions $a(x)$ and $b(x)$ describe nuclear shadowing and anti-shadowing effects. 
In the present experiment, the $x$-coverage does not drop below $x \sim 0.2$ so that 
shadowing and anti-shadowing effects can be neglected.  Therefore, Eq.~\ref{eqn:g1he3_nucl} 
becomes: 

\begin{equation} \label{eqn:g1he3_fin} 
   g_{1}^{^{3}\text{He}}(x) \approx P_{n}g_{1}^{n}(x) + 2P_{p}g_{1}^{p}(x) 
                            - 0.014\left[ g_{1}^{p}(x) - 4 g_{1}^{n}(x)\right].  
\end{equation} 

\noindent The same formula applies for $g_{2}$ data, with $g_{2}$ replacing $g_{1}$
in Eq.~\ref{eqn:g1he3_fin}. 

The nuclear corrections to our data for $d_2^{^{3}\text{He}}$ were applied to the 
{\it integral} (not bin-by-bin to the $d_{2}$ integrand), which resulted in two 
average $Q^{2}$-bins for each of the beam energies 4.74 and 5.89\,GeV.  Thus, the 
correction followed the formalism defined in Eq.~\ref{eqn:g1he3_fin}: 

\begin{equation} \label{eqn:d2n} 
   d_{2}^{n} = \frac{1}{\tilde{P}_{n}} \left( d_{2}^{^{3}\text{He}} - \tilde{P}_{p}d_{2}^{p}\right),   
\end{equation} 

\noindent for a given bin in $Q^{2}$, where the quantity $\tilde{P}_{p} = 2P_{p} - 0.014$ 
and $\tilde{P}_{n} = P_{n} + 0.056$.  For the proton an effective polarization 
$P_{p} = -0.028^{+0.009}_{-0.004}$ was used, and for the neutron $P_{n} = 0.86^{+0.036}_{-0.020}$~\cite{Zheng:2004ce}.
The matrix element $d_{2}^{p}$ was evaluated by considering various global 
analyses~\cite{deFlorian:2008mr,Bourrely:2001du,Bourrely:2007if,Leader:1997kw,deFlorian:2005mw,Gehrmann:1995ag}
to construct $g_{1}^{p}$.  The Wandzura-Wilczek relation~\cite{Wandzura:1977qf,Accardi:2009au,Accardi:2009nv} 
was used to obtain $g_{2}^{p}$, which is valid if the higher-twist effects are assumed to be small; 
this is a reasonable assumption based on the results of SLAC E155~\cite{Anthony:1999py}.  Using the 
average of the results for $g_{1}^{p}$ and $g_{2}^{p}$, the $d_{2}^{p}$ integral was evaluated over 
the same $x$-range of our experiment at $\left<Q^{2}\right> = 3.21$ and $4.32$\,GeV$^{2}$, and 
the result was inserted into Eq.~\ref{eqn:d2n}.      

To obtain the nuclear corrections needed to extract $g_{1}^{n}/F_{1}^{n}$, we first divided Eq.~\ref{eqn:g1he3_fin}
by $F_{1}^{^{3}\text{He}}$ and then rewrote $F_{1}^{^{3}\text{He}}$ in terms of $F_{2}^{^{3}\text{He}}$ 
following: 

\begin{equation} \label{eqn:f1-f2} 
   F_{1}\left(x,Q^{2}\right) = \frac{F_{2}\left(x,Q^{2}\right) \left(1 + \gamma^{2}\right)}{2x \left[ 1 + R\left(x,Q^{2}\right)\right]}, 
\end{equation}

\noindent where $R$ is taken to be target-independent~\cite{Bissey:2001cw}, which is a reasonable 
assumption for $Q^{2} > 1.5$\,GeV$^{2}$ and $x > 0.15$~\cite{Ackerstaff:1999ac}.  Solving for 
$g_{1}^{n}/F_{1}^{n}$ yielded: 

\begin{equation} \label{eqn:g1f1n} 
   \frac{g_{1}^{n}}{F_{1}^{n}} = \frac{1}{\tilde{P}_{n}} \frac{F_{2}^{^{3}\text{He}}}{F_{2}^{n}}
                                 \left( \frac{g_{1}^{^{3}\text{He}}}{F_{1}^{^{3}\text{He}}} 
                               - \tilde{P}_{p}\frac{F_{2}^{p}}{F_{2}^{^{3}\text{He}}}\frac{g_{1}^{p}}{F_{1}^{p}} \right).  
\end{equation} 

\noindent Using Eq.~\ref{eqn:g1f1n}, we extracted $g_{1}^{n}/F_{1}^{n}$ from our $^{3}$He data.  
For the unpolarized $F_{2}^{^{3}\text{He}}$ structure function, we utilized
the F1F209 parameterization~\cite{Bosted:2012qc}, which incorporates Fermi motion and EMC effects, and
for $F_{2}^{p}$ and $F_{2}^{n}$, the unpolarized PDF model CJ12~\cite{Owens:2012bv} to world 
data was used.  A fit to world $g_{1}^{p}/F_{1}^{p}$ 
data~\cite{Ackerstaff:1997ws,Abe:1998wq,Anthony:1999py,Dharmawardane:2006zd,Prok:2014ltt} 
was performed and used.  The fit used a second-order polynomial in $x$ with three free 
parameters and assumed $Q^{2}$-independence.  Any $Q^{2}$-dependence would cancel in the 
ratio of $g_{1}/F_{1}$ to leading order and next-to-leading order~\cite{Anselmino:1994gn}.  
For more details, see Appendix~\ref{sec:g1F1p-fit}.  
 
For the nuclear corrections to obtain $A_{1}^{n}$, we used the expression for $A_{1}$ in terms 
of the structure functions $g_{1}$, $g_{2}$ and $F_{1}$ (Appendix~\ref{sec:a1_app}).  
Solving for $A_{1}^{n}$ gave (cf. Eq.~\ref{eqn:g1f1n}): 

\begin{equation} \label{eqn:A1n} 
   A_{1}^{n} = \frac{1}{\tilde{P}_{n}} \frac{F_{2}^{^{3}\text{He}}}{F_{2}^{n}}
               \left( A_{1}^{^{3}\text{He}} 
             - \tilde{P}_{p}\frac{F_{2}^{p}}{F_{2}^{^{3}\text{He}}}A_{1}^{p} \right).   
\end{equation} 

\noindent The same models used in the $g_{1}/F_{1}$ analysis for $F_{2}$ on $^{3}$He, 
the proton and the neutron were used in the $A_{1}$ analysis.  Similar to the $g_{1}/F_{1}$
analysis, a $Q^{2}$-independent, second-order polynomial in $x$ was fit to world $A_{1}^{p}$ 
data~\cite{Airapetian:2006vy,Abe:1998wq,Anthony:1999py,Ashman:1987hv,Ashman:1989ig,Dharmawardane:2006zd,Alekseev:2010hc,Adeva:1998vv} 
and used in the analysis. For more details, see Appendix~\ref{sec:a1p-fit}.

Other neutron extraction methods have been studied in Ref.~\cite{Ethier:2013hna}, where the 
full convolution formalism was used at finite $Q^{2}$, including the nucleon off-shell and $\Delta$ 
contributions.  Such calculations are consistent with our extraction of $d_{2}^{n}$ and $A_{1}^{n}$ 
following Eqs.~\ref{eqn:d2n} and~\ref{eqn:A1n}, respectively.

   %===============================================================================
\section{Results} \label{sec:results}
\subsection{$^{3}$He results}  \label{sec:res_he3} 
%===============================================================================

Results for the unpolarized $e$-$^{3}$He scattering cross section (Section~\ref{sec:xs-ana}) 
for $E = 4.74$ and $5.89$\,GeV are given in Tables~\ref{tab:xs_4} and~\ref{tab:xs_5}.  All of the
contributions to the systematic uncertainty in the cross section are given in Tables~\ref{tab:xs-syst-err-4}
and~\ref{tab:xs-syst-err-5}.  The biggest contribution to the systematic uncertainty 
was the background subtraction, at a relative uncertainty of $\sim 9$\% in the 
lowest bin in $x$.  Our extracted cross section values are in good agreement with the 
F1F209 parameterization~\cite{Bosted:2012qc}.    

\begin{table}[hbt]
\centering
\caption{The final $^{3}$He unpolarized cross sections for 4.74\,GeV data.
         The uncertainties listed are statistical and systematic, respectively. 
         }
\label{tab:xs_4}
\begin{ruledtabular} 
\begin{tabular}{ccc}
% \hline
 $\left< x \right> $    & $\left< Q^{2} \right>$ (GeV$^{2}$) & $\frac{d^{3}\sigma}{d\Omega dE'}$ (nb/GeV/sr) \\ 
\hline 
  0.214 & 1.659 & 6.191 $\pm$ 0.365 $\pm$ 0.561 \\ 
  0.299 & 2.209 & 5.374 $\pm$ 0.178 $\pm$ 0.281 \\ 
  0.456 & 3.094 & 2.544 $\pm$ 0.048 $\pm$ 0.121 \\ 
  0.494 & 3.285 & 2.223 $\pm$ 0.034 $\pm$ 0.103 \\ 
  0.533 & 3.472 & 1.762 $\pm$ 0.026 $\pm$ 0.084 \\ 
  0.579 & 3.694 & 1.353 $\pm$ 0.027 $\pm$ 0.065 \\ 
  0.629 & 3.909 & 1.021 $\pm$ 0.018 $\pm$ 0.050 \\ 
  0.686 & 4.149 & 0.718 $\pm$ 0.012 $\pm$ 0.035 \\ 
  0.745 & 4.387 & 0.536 $\pm$ 0.012 $\pm$ 0.028 \\ 
% \hline 
\end{tabular}
\end{ruledtabular}
\end{table}

\begin{table}[hbt]
\centering
\caption{The final $^{3}$He unpolarized cross sections for 5.89\,GeV data.
         The uncertainties listed are statistical and systematic, respectively.  
         }
\label{tab:xs_5}
\begin{ruledtabular} 
\begin{tabular}{ccc}
% \hline
 $\left< x \right> $    & $\left< Q^{2} \right>$ (GeV$^{2}$) & $\frac{d^{3}\sigma}{d\Omega dE'}$ (nb/GeV/sr) \\ 
\hline 
  0.208 & 2.064 & 4.069 $\pm$ 0.440 $\pm$ 0.492 \\ 
  0.247 & 2.409 & 4.322 $\pm$ 0.116 $\pm$ 0.310 \\ 
  0.330 & 3.095 & 2.488 $\pm$ 0.099 $\pm$ 0.130 \\ 
  0.434 & 3.882 & 1.596 $\pm$ 0.038 $\pm$ 0.079 \\ 
  0.468 & 4.124 & 1.234 $\pm$ 0.030 $\pm$ 0.063 \\ 
  0.503 & 4.360 & 1.067 $\pm$ 0.020 $\pm$ 0.052 \\ 
  0.539 & 4.603 & 0.846 $\pm$ 0.016 $\pm$ 0.042 \\ 
  0.580 & 4.873 & 0.679 $\pm$ 0.012 $\pm$ 0.033 \\ 
  0.629 & 5.173 & 0.472 $\pm$ 0.010 $\pm$ 0.022 \\ 
  0.679 & 5.478 & 0.331 $\pm$ 0.007 $\pm$ 0.016 \\ 
  0.738 & 5.811 & 0.250 $\pm$ 0.006 $\pm$ 0.013 \\ 
% \hline
\end{tabular}
\end{ruledtabular} 
\end{table}
 
The $\vec{e}$-$^{3}\vec{\text{He}}$ electron asymmetries $A_{\parallel}^{^{3}\text{He}}$ 
and $A_{\perp}^{^{3}\text{He}}$ (Section~\ref{sec:asym_ana}), the virtual photon asymmetry $A_{1}^{^{3}\text{He}}$,
and the structure function ratio $g_{1}^{^{3}\text{He}}/F_{1}^{^{3}\text{He}}$ (Eqs.~\ref{eqn:a1_exp} and~\ref{eqn:g1f1_exp}) 
for the 4.74\,GeV and 5.89\,GeV data are given in Tables~\ref{tab:asym_4} and~\ref{tab:asym_5}.
The polarized structure functions $g_{1}^{^{3}\text{He}}$ and $g_{2}^{^{3}\text{He}}$ (Eqs.~\ref{eqn:g1_exp} and~\ref{eqn:g2_exp}) 
are given in Tables~\ref{tab:g1g2_4} and~\ref{tab:g1g2_5}.  All of the contributions to 
the systematic uncertainties are given in Appendix~\ref{sec:syst-err-tables}, with the dominant 
one being the selection of the electron sample.  

\begin{figure}[hbt]
   \centering
   \includegraphics[width=0.5\textwidth]{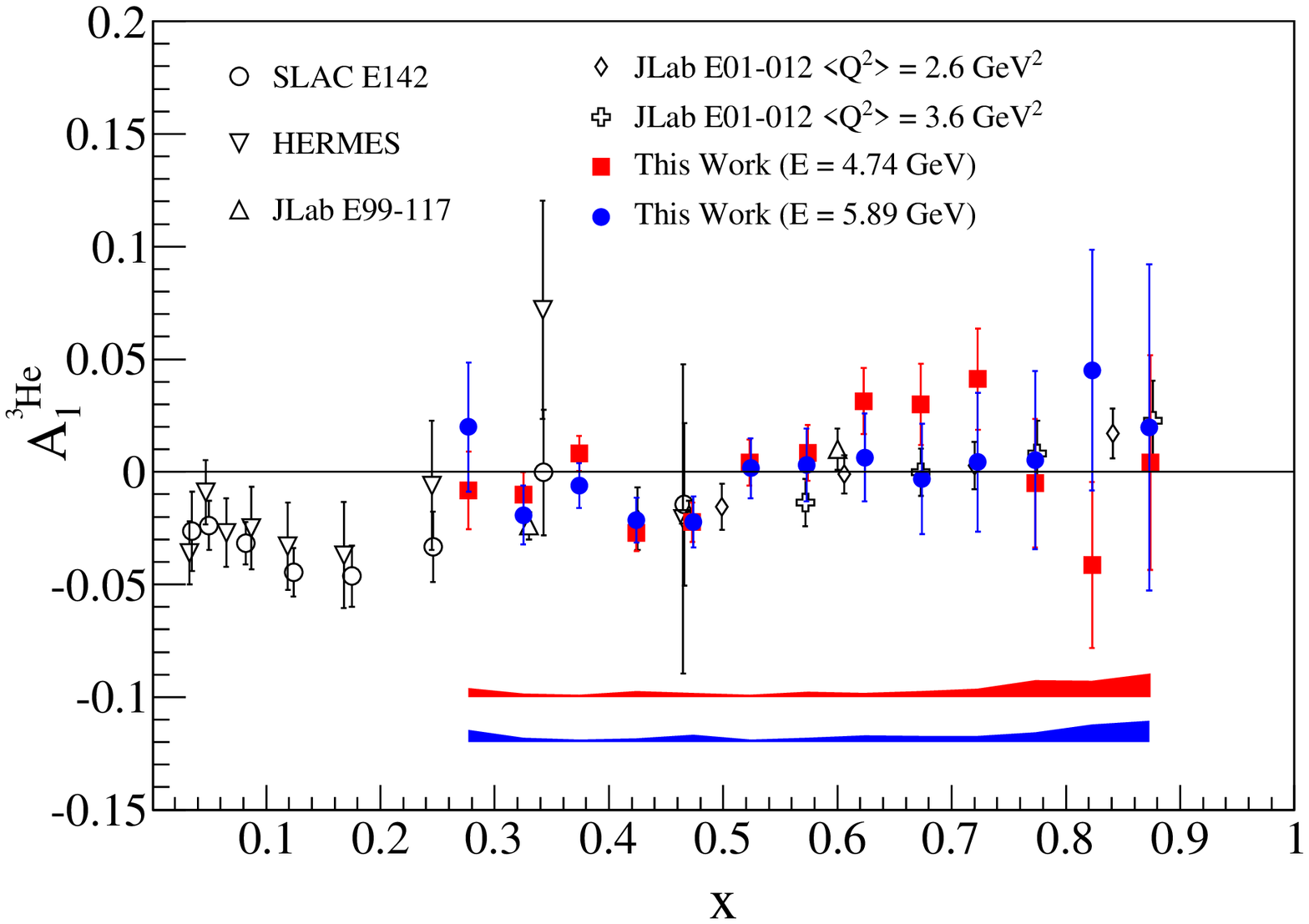}
   \caption{(Color online) Our measured result for $A_{1}^{^{3}\text{He}}$ for 4.74\,GeV (filled squares)
            and 5.89\,GeV (filled circles) data.  The error bars on our data points represent
            the statistical uncertainty.  The bands at the bottom of the plot
            indicate the systematic uncertainty for each data set, where the upper (lower) 
            band corresponds to the 4.74\,GeV (5.89\,GeV) data set.  
            The DIS data set corresponds to data for which $x < 0.519$ ($x < 0.623$) 
            for $E = 4.74$\,GeV (5.89\,GeV); the data at larger $x$ values correspond to 
            the resonance region.  Also plotted are world DIS data from SLAC E142~\protect\cite{Anthony:1996mw}, 
            HERMES~\protect\cite{Ackerstaff:1999ey} and JLab E99-117~\protect\cite{Zheng:2003un,Zheng:2004ce},  
            and resonance data from JLab E01-012~\protect\cite{Solvignon:2008hk}.  
            }
   \label{fig:a1he3_result}
\end{figure}

The asymmetry $A_{1}^{^{3}\text{He}}$ is plotted in Fig.~\ref{fig:a1he3_result}, 
compared to the world DIS data from SLAC E142~\cite{Anthony:1996mw}, HERMES~\cite{Ackerstaff:1999ey} 
and JLab E99-117~\cite{Zheng:2003un,Zheng:2004ce}.  Also plotted are resonance data from JLab E01-012~\cite{Solvignon:2008hk}.  
We find that our results reproduce the trend of existing data.  

\begin{figure}[hbt]
   \centering 
   \includegraphics[width=0.5\textwidth]{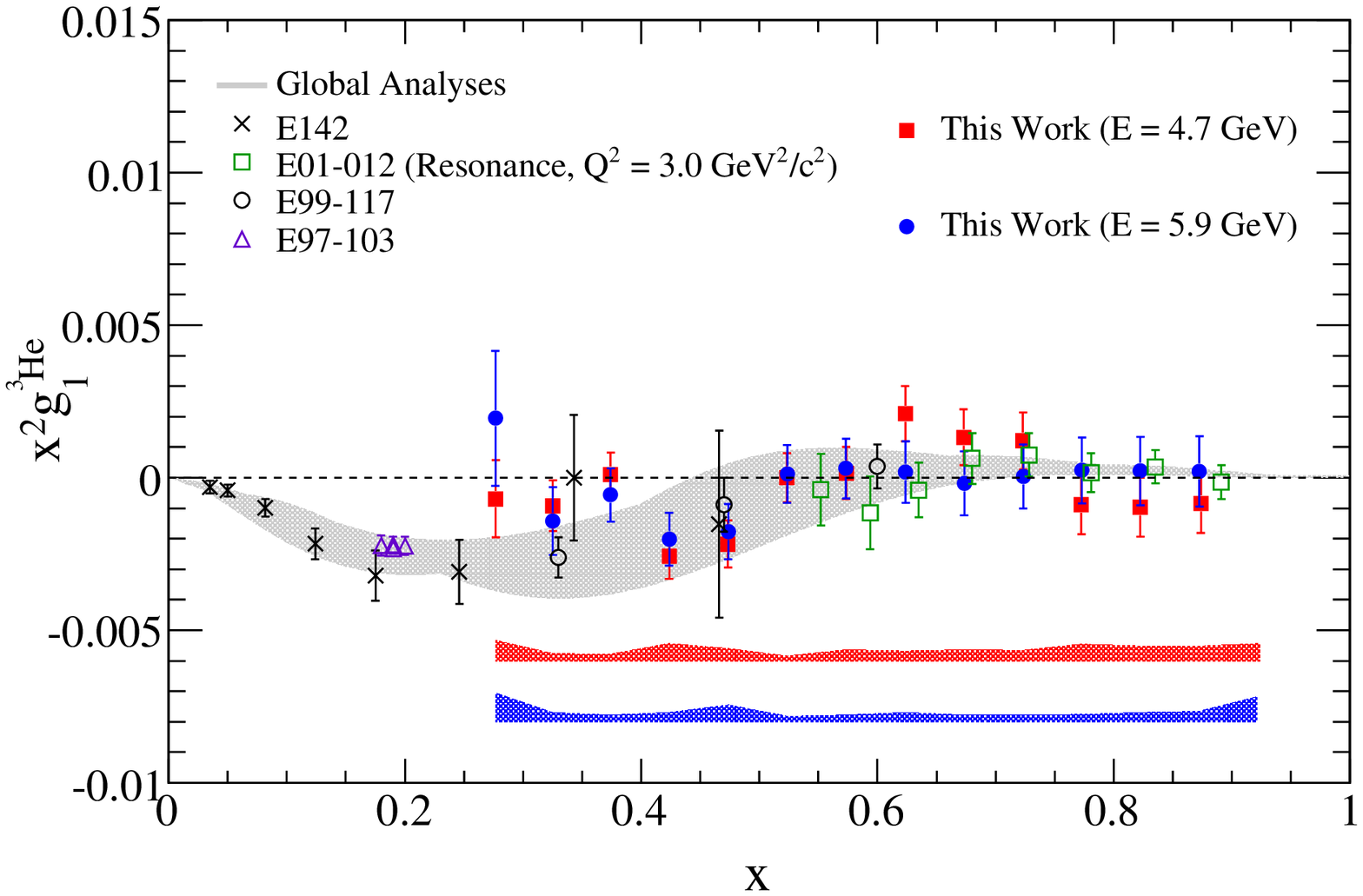}
   \caption{(Color online) Our results for $g_{1}$ on $^{3}$He for 4.74\,GeV (filled squares)
            and 5.89\,GeV (filled circles) data.  The error bars on our data points represent
            the statistical uncertainty.  The bands at the bottom of the plot
            indicate the systematic uncertainty for each data set, where the upper (lower) 
            band corresponds to the 4.74\,GeV (5.89\,GeV) data set.  
            The DIS data set corresponds to data for which $x < 0.519$ ($x < 0.623$) for 
            $E = 4.74$\,GeV (5.89\,GeV); the data at larger $x$ values correspond to the 
            resonance data.  Our data are compared to world DIS data from 
            JLab E99-117~\protect\cite{Zheng:2003un,Zheng:2004ce}, SLAC E142~\protect\cite{Anthony:1996mw}, and 
            resonance data from JLab E01-012~\protect\cite{Solvignon:2008hk} and JLab E97-103~\protect\cite{KKramerThesis,Kramer:2005qe}.  
            The gray band represents an envelope of various global 
            analyses~\protect\cite{deFlorian:2008mr,Gehrmann:1995ag,Leader:2006xc,deFlorian:2005mw,Bourrely:2001du,*Bourrely:2007if}
            for $g_{1}$ at $Q^{2} = 4.43$\,GeV$^{\text{2}}$, which was the average
            $Q^{2}$ of our data set.   
            }
   \label{fig:g1he3_result}
\end{figure}

\begin{figure}[hbt]
   \centering 
   \includegraphics[width=0.5\textwidth]{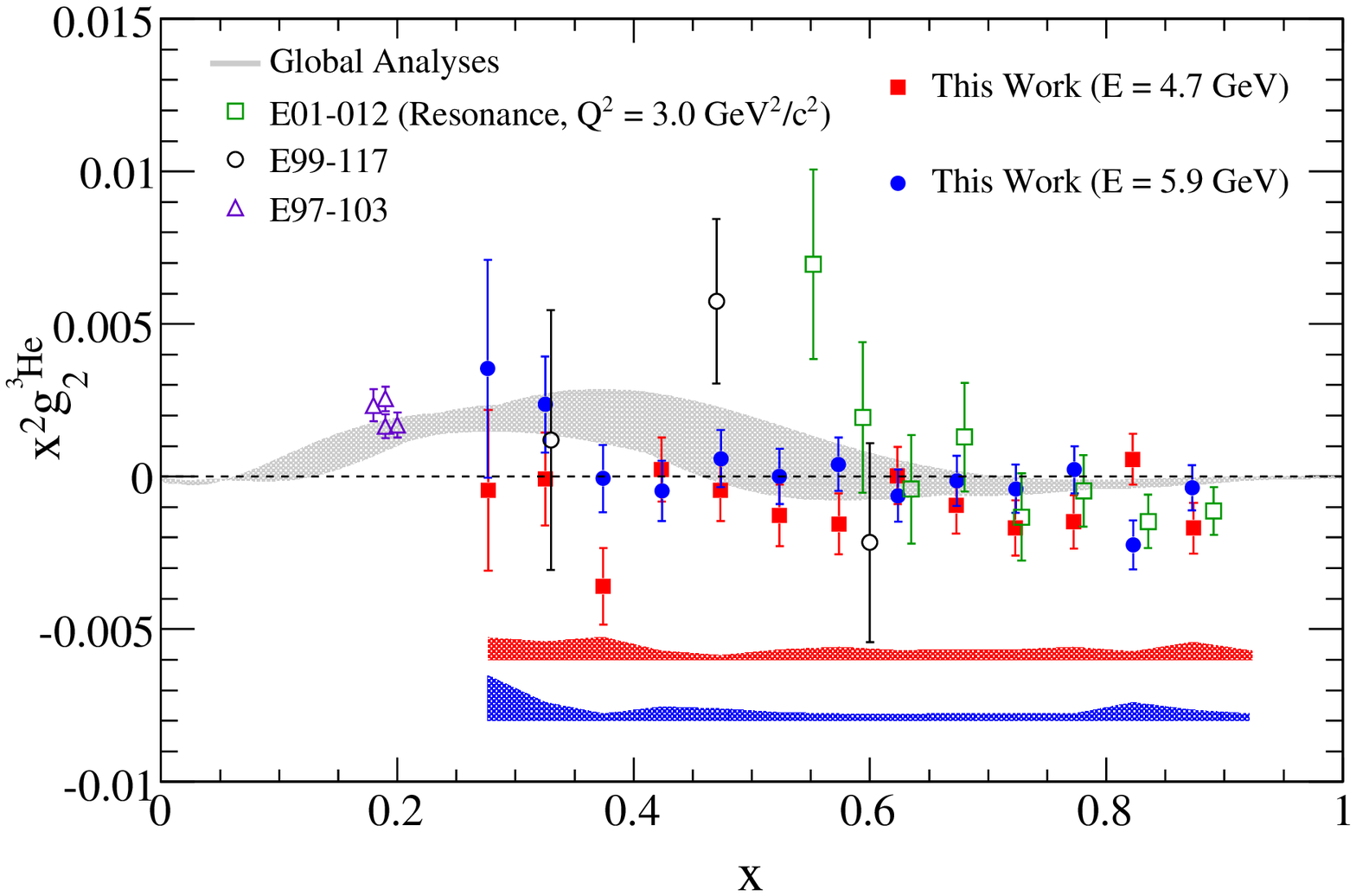}
   \caption{(Color online) Our results for $g_{2}$ on $^{3}$He for the 4.74\,GeV (filled squares)
            and 5.89\,GeV (filled circles) data.  The error bars on our data points represent
            the statistical uncertainty.  The bands at the bottom of the plot
            indicate the systematic uncertainty for each data set, where the upper (lower) 
            band corresponds to the 4.74\,GeV (5.89\,GeV) data set.  
            The DIS data set corresponds to data for which 
            $x < 0.519$ ($x < 0.623$) for $E = 4.74$\,GeV (5.89\,GeV); the data at larger $x$ 
            values correspond to the resonance data.  Our data are compared to world data from JLab E99-117~\protect\cite{Zheng:2003un,Zheng:2004ce},
            JLab E01-012~\protect\cite{Solvignon:2008hk} (resonance data) and JLab E97-103~\protect\cite{KKramerThesis,Kramer:2005qe} (resonance data).  
            The gray band represents an envelope of various global 
            analyses~\protect\cite{deFlorian:2008mr,Gehrmann:1995ag,Leader:2006xc,deFlorian:2005mw,Bourrely:2001du,*Bourrely:2007if}
            used to construct $g_{2}^{\text{WW}}$.  
            }
   \label{fig:g2he3_result}
\end{figure}

The spin-structure functions $g_{1}^{^{3}\text{He}}$ and $g_{2}^{^{3}\text{He}}$ are 
presented in Figs.~\ref{fig:g1he3_result} and~\ref{fig:g2he3_result}, in which the world 
DIS data from JLab E99-117~\cite{Zheng:2003un,Zheng:2004ce} and SLAC E142~\cite{Anthony:1996mw} 
are also shown.  Resonance data from JLab E01-012~\cite{Solvignon:2008hk} are also presented.  
The gray band represents an envelope encompassing a number of global 
analyses~\cite{deFlorian:2008mr,Gehrmann:1995ag,Leader:2006xc,deFlorian:2005mw,Bourrely:2001du,*Bourrely:2007if}. 
Our data reproduce the trend seen in existing $g_{1}^{^{3}\text{He}}$ data.  
For $g_{2}^{^{3}\text{He}}$, our data have improved on the uncertainty by about a factor of 
2 relative to the JLab E99-117 data set.    

\begin{table*}[hbt] 
\centering
\caption{Asymmetry results for $A_{\parallel}^{^{3}\text{He}}$, $A_{\perp}^{^{3}\text{He}}$,
         $A_{1}^{^{3}\text{He}}$ and $g_{1}^{^{3}\text{He}}/F_{1}^{^{3}\text{He}}$
         for the 4.74\,GeV data.  The first uncertainty is statistical, while the second is systematic.}
\label{tab:asym_4}
\begin{ruledtabular} 
\begin{tabular}{ccrrrr}
 \multicolumn{1}{c}{$\left< x \right>$}                          & 
 \multicolumn{1}{c}{$\left< Q^{2} \right>$ (GeV$^{2}$)}          & 
 \multicolumn{1}{c}{$A_{\parallel}^{^{3}\text{He}}$}             & 
 \multicolumn{1}{c}{$A_{\perp}^{^{3}\text{He}}$}                 & 
 \multicolumn{1}{c}{$A_{1}^{^{3}\text{He}}$}                     & 
 \multicolumn{1}{c}{$g_{1}^{^{3}\text{He}}/F_{1}^{^{3}\text{He}}$} \\
\hline
  0.277 & 2.038  & $-0.008 \pm  0.015 \pm  0.007$ & $-0.002 \pm 0.008 \pm 0.003$ & $-0.008 \pm 0.017 \pm 0.004$ & $-0.009 \pm 0.016 \pm 0.004$ \\
  0.325 & 2.347  & $-0.009 \pm  0.009 \pm  0.003$ & $-0.001 \pm 0.005 \pm 0.002$ & $-0.010 \pm 0.010 \pm 0.001$ & $-0.010 \pm 0.009 \pm 0.001$ \\
  0.374 & 2.639  & $ 0.005 \pm  0.007 \pm  0.002$ & $-0.011 \pm 0.004 \pm 0.002$ & $ 0.008 \pm 0.008 \pm 0.001$ & $ 0.001 \pm 0.007 \pm 0.001$ \\
  0.424 & 2.915  & $-0.025 \pm  0.007 \pm  0.005$ & $-0.003 \pm 0.004 \pm 0.002$ & $-0.027 \pm 0.008 \pm 0.003$ & $-0.026 \pm 0.007 \pm 0.002$ \\
  0.473 & 3.176  & $-0.021 \pm  0.008 \pm  0.003$ & $-0.005 \pm 0.004 \pm 0.001$ & $-0.022 \pm 0.009 \pm 0.002$ & $-0.023 \pm 0.008 \pm 0.002$ \\
  0.523 & 3.425  & $ 0.002 \pm  0.009 \pm  0.002$ & $-0.006 \pm 0.005 \pm 0.001$ & $ 0.004 \pm 0.010 \pm 0.001$ & $ 0.000 \pm 0.009 \pm 0.001$ \\
  0.574 & 3.662  & $ 0.005 \pm  0.010 \pm  0.004$ & $-0.008 \pm 0.005 \pm 0.002$ & $ 0.008 \pm 0.012 \pm 0.002$ & $ 0.002 \pm 0.011 \pm 0.002$ \\
  0.623 & 3.886  & $ 0.029 \pm  0.013 \pm  0.003$ & $ 0.005 \pm 0.007 \pm 0.002$ & $ 0.031 \pm 0.015 \pm 0.002$ & $ 0.031 \pm 0.013 \pm 0.002$ \\
  0.673 & 4.099  & $ 0.025 \pm  0.015 \pm  0.005$ & $-0.004 \pm 0.009 \pm 0.003$ & $ 0.030 \pm 0.018 \pm 0.003$ & $ 0.023 \pm 0.016 \pm 0.002$ \\
  0.723 & 4.307  & $ 0.031 \pm  0.019 \pm  0.007$ & $-0.014 \pm 0.009 \pm 0.002$ & $ 0.041 \pm 0.022 \pm 0.004$ & $ 0.026 \pm 0.020 \pm 0.003$ \\
  0.773 & 4.504  & $-0.012 \pm  0.024 \pm  0.013$ & $-0.025 \pm 0.012 \pm 0.005$ & $-0.005 \pm 0.028 \pm 0.008$ & $-0.023 \pm 0.025 \pm 0.007$ \\
  0.823 & 4.694  & $-0.033 \pm  0.030 \pm  0.012$ & $ 0.005 \pm 0.016 \pm 0.006$ & $-0.041 \pm 0.037 \pm 0.007$ & $-0.032 \pm 0.032 \pm 0.006$ \\
  0.874 & 4.876  & $-0.014 \pm  0.039 \pm  0.017$ & $-0.049 \pm 0.020 \pm 0.006$ & $ 0.004 \pm 0.048 \pm 0.010$ & $-0.035 \pm 0.041 \pm 0.009$ \\
% \hline
\end{tabular}
\end{ruledtabular} 
\end{table*}

\begin{table*}[hbt]
\centering
\caption{Asymmetry results for $A_{\parallel}^{^{3}\text{He}}$, $A_{\perp}^{^{3}\text{He}}$,
         $A_{1}^{^{3}\text{He}}$ and $g_{1}^{^{3}\text{He}}/F_{1}^{^{3}\text{He}}$
         for the 5.89\,GeV data.  The first uncertainty is statistical, while the second is systematic.}
\label{tab:asym_5}
\begin{ruledtabular} 
\begin{tabular}{ccrrrr}
 \multicolumn{1}{c}{$\left< x \right>$}                          & 
 \multicolumn{1}{c}{$\left< Q^{2} \right>$ (GeV$^{2}$)}          & 
 \multicolumn{1}{c}{$A_{\parallel}^{^{3}\text{He}}$}             & 
 \multicolumn{1}{c}{$A_{\perp}^{^{3}\text{He}}$}                 & 
 \multicolumn{1}{c}{$A_{1}^{^{3}\text{He}}$}                     & 
 \multicolumn{1}{c}{$g_{1}^{^{3}\text{He}}/F_{1}^{^{3}\text{He}}$} \\
\hline
  0.277  & 2.626 & $ 0.019 \pm 0.027 \pm 0.010$  &  $ 0.010 \pm 0.008 \pm 0.003$  & $ 0.020 \pm 0.029 \pm 0.006$ & $ 0.024 \pm 0.028 \pm 0.006$ \\
  0.325  & 3.032 & $-0.017 \pm 0.012 \pm 0.003$  &  $ 0.004 \pm 0.004 \pm 0.001$  & $-0.019 \pm 0.013 \pm 0.002$ & $-0.016 \pm 0.012 \pm 0.002$ \\
  0.374  & 3.421 & $-0.006 \pm 0.009 \pm 0.002$  &  $-0.001 \pm 0.003 \pm 0.001$  & $-0.006 \pm 0.010 \pm 0.001$ & $-0.006 \pm 0.009 \pm 0.001$ \\
  0.424  & 3.802 & $-0.020 \pm 0.009 \pm 0.003$  &  $-0.004 \pm 0.003 \pm 0.001$  & $-0.021 \pm 0.010 \pm 0.002$ & $-0.022 \pm 0.009 \pm 0.002$ \\
  0.474  & 4.169 & $-0.021 \pm 0.010 \pm 0.006$  &  $ 0.000 \pm 0.003 \pm 0.001$  & $-0.022 \pm 0.011 \pm 0.003$ & $-0.021 \pm 0.010 \pm 0.003$ \\
  0.524  & 4.514 & $ 0.002 \pm 0.012 \pm 0.002$  &  $ 0.000 \pm 0.004 \pm 0.001$  & $ 0.002 \pm 0.013 \pm 0.001$ & $ 0.002 \pm 0.012 \pm 0.001$ \\
  0.573  & 4.848 & $ 0.003 \pm 0.015 \pm 0.003$  &  $ 0.003 \pm 0.004 \pm 0.001$  & $ 0.003 \pm 0.016 \pm 0.002$ & $ 0.004 \pm 0.015 \pm 0.002$ \\
  0.624  & 5.176 & $ 0.005 \pm 0.018 \pm 0.005$  &  $-0.004 \pm 0.005 \pm 0.001$  & $ 0.006 \pm 0.020 \pm 0.003$ & $ 0.003 \pm 0.018 \pm 0.003$ \\
  0.674  & 5.486 & $-0.003 \pm 0.022 \pm 0.005$  &  $-0.002 \pm 0.007 \pm 0.002$  & $-0.003 \pm 0.024 \pm 0.003$ & $-0.004 \pm 0.022 \pm 0.002$ \\
  0.723  & 5.777 & $ 0.003 \pm 0.027 \pm 0.005$  &  $-0.004 \pm 0.008 \pm 0.003$  & $ 0.004 \pm 0.031 \pm 0.003$ & $ 0.001 \pm 0.028 \pm 0.003$ \\
  0.773  & 6.059 & $ 0.006 \pm 0.035 \pm 0.008$  &  $ 0.005 \pm 0.010 \pm 0.002$  & $ 0.005 \pm 0.039 \pm 0.004$ & $ 0.008 \pm 0.035 \pm 0.004$ \\
  0.823  & 6.325 & $ 0.028 \pm 0.047 \pm 0.014$  &  $-0.044 \pm 0.014 \pm 0.004$  & $ 0.045 \pm 0.053 \pm 0.008$ & $ 0.009 \pm 0.047 \pm 0.013$ \\
  0.873  & 6.585 & $ 0.015 \pm 0.062 \pm 0.017$  &  $-0.008 \pm 0.019 \pm 0.007$  & $ 0.020 \pm 0.072 \pm 0.009$ & $ 0.011 \pm 0.064 \pm 0.009$ \\
% \hline
\end{tabular}
\end{ruledtabular} 
\end{table*}

\begin{table}
\caption{The $g_1$ and $g_2$ spin-structure functions measured on $^{3}$He at an incident electron energy of 4.74\,GeV.
         The first uncertainty is statistical, while the second is systematic.
         }
\label{tab:g1g2_4}
\center
\begin{ruledtabular}
\begin{tabular}{crr}
\multicolumn{1}{c}{$\left<x\right>$     } & 
\multicolumn{1}{c}{$g_1^{^{3}\text{He}}$} & 
\multicolumn{1}{c}{$g_2^{^{3}\text{He}}$} \\
\hline
 0.277 & $-0.009 \pm 0.016 \pm 0.009$ & $-0.006 \pm 0.034 \pm 0.009$ \\
 0.325 & $-0.009 \pm 0.008 \pm 0.002$ & $-0.001 \pm 0.014 \pm 0.006$ \\
 0.374 & $ 0.001 \pm 0.005 \pm 0.002$ & $-0.026 \pm 0.009 \pm 0.005$ \\
 0.424 & $-0.014 \pm 0.004 \pm 0.003$ & $ 0.001 \pm 0.006 \pm 0.002$ \\
 0.473 & $-0.010 \pm 0.003 \pm 0.002$ & $-0.002 \pm 0.004 \pm 0.001$ \\
 0.523 & $ 0.000 \pm 0.003 \pm 0.000$ & $-0.005 \pm 0.004 \pm 0.001$ \\
 0.574 & $ 0.000 \pm 0.003 \pm 0.001$ & $-0.005 \pm 0.003 \pm 0.001$ \\
 0.623 & $ 0.005 \pm 0.002 \pm 0.001$ & $ 0.000 \pm 0.002 \pm 0.001$ \\
 0.673 & $ 0.003 \pm 0.002 \pm 0.001$ & $-0.002 \pm 0.002 \pm 0.001$ \\
 0.723 & $ 0.002 \pm 0.002 \pm 0.001$ & $-0.003 \pm 0.002 \pm 0.001$ \\
 0.773 & $-0.001 \pm 0.002 \pm 0.001$ & $-0.002 \pm 0.001 \pm 0.001$ \\
 0.823 & $-0.001 \pm 0.001 \pm 0.001$ & $ 0.001 \pm 0.001 \pm 0.000$ \\
 0.874 & $-0.001 \pm 0.001 \pm 0.001$ & $-0.002 \pm 0.001 \pm 0.001$ \\
\end{tabular}
\end{ruledtabular}
\end{table}

\begin{table}
\center
\caption{The $g_1$ and $g_2$ spin-structure functions measured on $^{3}$He at an incident electron energy of 5.89\,GeV.
         The first uncertainty is statistical, while the second is systematic.
         } 
\label{tab:g1g2_5}
\begin{ruledtabular}
\begin{tabular}{crr}
\multicolumn{1}{c}{$\left<x\right>$     } & 
\multicolumn{1}{c}{$g_1^{^{3}\text{He}}$} & 
\multicolumn{1}{c}{$g_2^{^{3}\text{He}}$} \\
\hline
 0.277 & $ 0.026 \pm 0.029 \pm 0.012$ & $ 0.046 \pm 0.047 \pm 0.019$ \\
 0.325 & $-0.013 \pm 0.010 \pm 0.003$ & $ 0.022 \pm 0.015 \pm 0.006$ \\
 0.374 & $-0.004 \pm 0.006 \pm 0.002$ & $ 0.000 \pm 0.008 \pm 0.002$ \\
 0.424 & $-0.011 \pm 0.005 \pm 0.002$ & $-0.002 \pm 0.006 \pm 0.002$ \\
 0.474 & $-0.008 \pm 0.004 \pm 0.002$ & $ 0.003 \pm 0.004 \pm 0.002$ \\
 0.524 & $ 0.000 \pm 0.003 \pm 0.001$ & $ 0.000 \pm 0.003 \pm 0.001$ \\
 0.573 & $ 0.001 \pm 0.003 \pm 0.001$ & $ 0.001 \pm 0.003 \pm 0.001$ \\
 0.624 & $ 0.000 \pm 0.002 \pm 0.001$ & $-0.002 \pm 0.002 \pm 0.000$ \\
 0.674 & $ 0.000 \pm 0.002 \pm 0.000$ & $ 0.000 \pm 0.002 \pm 0.000$ \\
 0.723 & $ 0.000 \pm 0.002 \pm 0.000$ & $-0.001 \pm 0.002 \pm 0.000$ \\
 0.773 & $ 0.000 \pm 0.002 \pm 0.000$ & $ 0.000 \pm 0.001 \pm 0.000$ \\
 0.823 & $ 0.000 \pm 0.002 \pm 0.000$ & $-0.003 \pm 0.001 \pm 0.001$ \\
 0.873 & $ 0.000 \pm 0.002 \pm 0.000$ & $ 0.000 \pm 0.001 \pm 0.000$ \\
\end{tabular}
\end{ruledtabular}
\end{table}
 
%===============================================================================
\subsection{Neutron results}   \label{sec:res_n} 
\subsubsection{The matrix element $d_{2}^{n}$} \label{sec:d2-results} 
%===============================================================================

\begin{figure}[hbt]
   \centering 
   \includegraphics[width=0.5\textwidth]{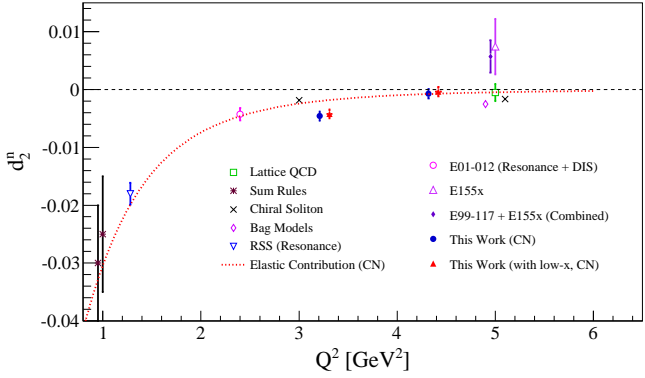}
   \caption{(Color online) Our measured $d_{2}^{n}$ data as a function of $Q^{2}$
            compared to the world data from SLAC E155x~\protect\cite{Anthony:2002hy}, JLab E99-117 
            and SLAC E155x~\protect\cite{Zheng:2004ce}, JLab RSS~\protect\cite{Slifer:2008xu} and JLab E01-012~\protect\cite{Solvignon:2013yun}.
            Also shown are various theoretical calculations, including a QCD sum rule approach~\protect\cite{Stein:1994zk,Balitsky:1989jb},
            a chiral soliton model~\protect\cite{Weigel:1996jh,*Weigel:2000gx} and a bag model~\protect\cite{Song:1996ea}.  
            Additionally, a lattice QCD~\protect\cite{Gockeler:2005vw} calculation is shown.  The 
            elastic contribution to $d_{2}^{n}$ is given by the dashed curve, evaluated 
            using the CN moments.  Figure reproduced from Ref.~\protect\cite{Posik:2014usi}.
           }
   \label{fig:d2n_result}
\end{figure}

Results for the $d_{2}^{n}$ matrix element (Eq.~\ref{eqn:d2_exp} and Section~\ref{sec:nuclear_cor}), 
first published in Ref.~\cite{Posik:2014usi}, are shown in Fig.~\ref{fig:d2n_result} 
and tabulated in Table~\ref{tab:d2n_result}.  The matrix element was extracted using 
the Cornwall-Norton (CN) moments~\cite{Melnitchouk:2005zr}. Since our measurement 
did not cover the full $x$ range, a low-$x$ ($x<0.25$) and a high-$x$ ($x>0.90$) 
contribution needed to be evaluated from other sources.  For the low-$x$ region, 
a third-order polynomial fit to the world data on 
$x^{2}g_{1}^{n}$~\cite{Anthony:1996mw,Abe:1998wq,Abe:1997qk,Kramer:2005qe} and 
$x^{2}g_{2}^{n}$~\cite{Kramer:2005qe,Anthony:1999py,Anthony:2002hy} was used to evaluate 
$d_{2}^{n}$.  This contribution is relatively small, considering the $x^{2}$-weighting.  
The large-$x$ contribution comes from the elastic peak, which was modeled using 
the Riordan~\cite{Riordan:2010id} and Kelly~\cite{Kelly:2004hm} parameterizations.  
The contribution from the range $0.90 < x < 0.99$ was considered to be negligible 
when taking into account the size of our $g_{1}$ and $g_{2}$ data at $x \sim 0.90$.  
Nuclear corrections were applied to our $^{3}$He data as described in Section~\ref{sec:nuclear_cor}.  
Adding the low-$x$ and high-$x$ (elastic) contributions to our measured result 
gave the full $d_{2}^{n}$ integral.  Target mass corrections (TMCs) were checked 
by extracting $d_{2}^{n}$ using the Nachtmann moments~\cite{Nachtmann:1973mr,*Accardi:2008pc,*Sidorov:2006fi,*Matsuda:1979ad,*Wandzura:1977ce}; 
the difference between the CN and Nachtmann approaches was found to be small relative 
to the statistical uncertainties.  An overview of the systematic uncertainties is given in 
Table~\ref{tab:d2n_sys}.  The largest contribution to the systematic uncertainty comes 
from the unmeasured low-$x$ region.   

\begin{figure}[hbt]
   \centering 
   \includegraphics[width=0.5\textwidth]{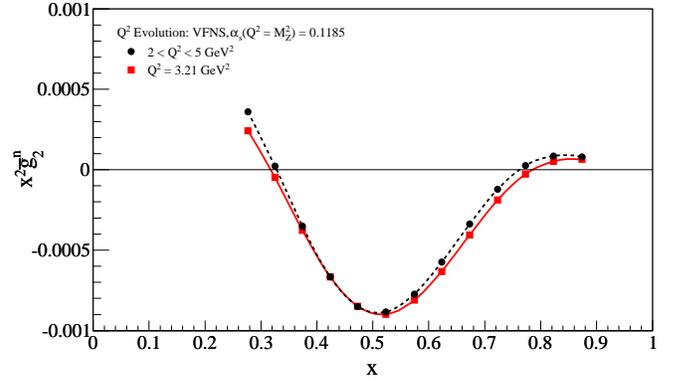}
   \caption{(Color online) The $d_{2}^{n}$ integrand for 4.74\,GeV kinematics.  
            For each measured $\left< x \right>$ bin, the model~\protect\cite{Braun:2011aw} was evolved~\protect\cite{Braun:2001qx,Botje:2010ay} 
            to the measured (average) $Q^{2}$ value indicated by the circles (squares). 
            For more details, see the text. 
           }
   \label{fig:d2n_q2-evol_4-pass}
\end{figure}

\begin{figure}[hbt]
   \centering 
   \includegraphics[width=0.5\textwidth]{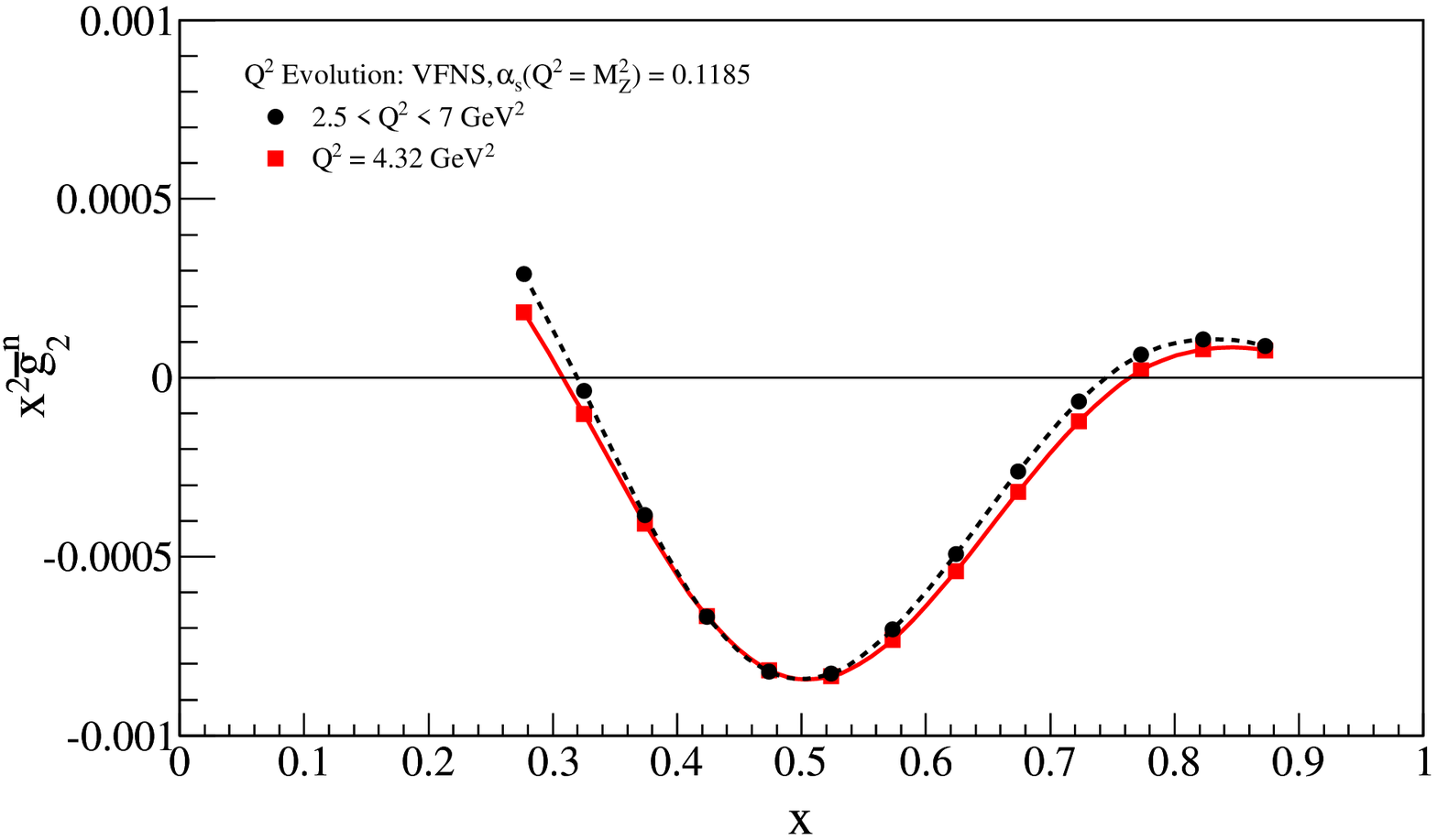}
   \caption{(Color online) The $d_{2}^{n}$ integrand for 5.89\,GeV kinematics.  
            For each measured $\left< x \right>$ bin, the model~\protect\cite{Braun:2011aw} was evolved~\protect\cite{Braun:2001qx,Botje:2010ay} 
            to the measured (average) $Q^{2}$ value indicated by the circles (squares). 
            For more details, see the text.
           }
   \label{fig:d2n_q2-evol_5-pass}
\end{figure}

Our unpolarized cross section and double-spin asymmetry data were obtained at various $Q^{2}$ 
values; since the $d_{2}^{n}$ integral is typically carried out at constant $Q^{2}$, we considered 
what the effect of evolving $\bar{g}_{2}^{n}$ (the twist-3 part of $g_{2}^{n}$) would have on 
the $d_{2}^{n}$ value.  To do this, we utilized the $\bar{g}_{2}^{n}$ model from Ref.~\cite{Braun:2011aw} 
along with the $Q^{2}$ evolution description for $g_{2}^{n}$ from Ref.~\cite{Braun:2001qx}, which uses 
flavor-nonsinglet evolution equations and utilizes large-$N_{c}$ and large-$x$ ($x \gtrsim 0.1$) approximations.  
The $Q^{2}$-evolution calculations were performed using QCDNum~\cite{Botje:2010ay} in the variable-flavor 
number scheme (VFNS) and with $\alpha_{s}\left(Q^{2} = M_{Z}^{2}\right) = 0.1185$~\cite{Agashe:2014kda}.
For each of our measured $x$ bins at a given beam energy, the model was evolved from its initial 
value at $Q^{2} = 1$\,GeV$^{2}$ to the measured $Q^{2}$-value $\left( Q_{m}^{2} \right)$ for 
that particular $x$-bin and also to the $\left< Q^{2} \right>$ value for the given beam energy, 
see Figs.~\ref{fig:d2n_q2-evol_4-pass} and~\ref{fig:d2n_q2-evol_5-pass}.  We then evaluated: 

\begin{equation}
   \Delta d_{2}^{n} = \vert d_{2}^{n}\left( \left< Q^{2} \right> \right) - d_{2}^{n}\left( Q_{m}^{2} \right) \vert
\end{equation}  

\noindent where $\left< Q^{2} \right> = 3.21$\,GeV$^{2}$ (4.32\,GeV$^{2}$) for 
$E = 4.74$\,GeV (5.89\,GeV).  The $d_{2}^{n}$ integral was evaluated according to 
Eq.~\ref{eqn:d2-sf-intro} for our measured $\left< x \right>$ bins corresponding to 
$0.277 \leq x \leq 0.874$ ($0.277 \leq x \leq 0.873$) for $E = 4.74$\,GeV (5.89\,GeV).  
We found $\Delta d_{2}^{n} = 0.00008$ (0.00007) for the $E = 4.74$\,GeV (5.89\,GeV), 
which is a factor of 6 (5) smaller than the systematic uncertainty on our measured $d_{2}^{n}$ 
for $E = 4.74$\,GeV (5.89\,GeV).  The difference between the constant $Q^{2}$ evaluation of 
$d_{2}^{n}$ compared to the varying $Q^{2}$ approach was taken as an estimate of the 
systematic uncertainty due to not performing the $Q^{2}$ evolution on our data. 

\begin{table}
\centering
\caption{The $d_2^n$ results with statistical and systematic uncertainties.  The last 
         uncertainty represents that due to neglecting the $Q^{2}$-evolution of the $d_{2}^{n}$ integrand.}
\label{tab:d2n_result}
\begin{ruledtabular} 
\begin{tabular}{c c}
$\left< Q^2 \right>$ $\left(\text{GeV}^{2}\right)$ & $d_2^n$ $\left(\times 10^{-5}\right)$  \\ \hline
3.21 &  $-421.0 \pm 79.0 \pm 82.0 \pm 8.0$  \\   
4.32 &  $-35.0  \pm 83.0 \pm 69.0 \pm 7.0$  \\   
\end{tabular}
\end{ruledtabular} 
\end{table}

Our result at $\left<Q^{2}\right> = 3.21$\,GeV$^{2}$ is small and negative, while 
the result at $\left<Q^{2}\right> = 4.32$\,GeV$^{2}$ is consistent with zero (Table~\ref{tab:d2n_result}).  
The trend of our measurements appears to be in agreement with the lattice QCD 
calculation~\cite{Gockeler:2005vw} at $Q^{2} = 5$\,GeV$^{2}$.  Our $d_{2}^{n}$ 
extraction is also consistent with bag~\cite{Song:1996ea,Stratmann:1993aw,Ji:1993sv} 
and chiral soliton~\cite{Weigel:1996jh,*Weigel:2000gx} models, as shown in Fig.~\ref{fig:d2n_result}. 

The Jefferson Lab Angular Momentum (JAM) collaboration has published new results 
from their global QCD analysis~\cite{Sato:2016tuz}.  Their work utilizes an 
iterative Monte Carlo technique that aims to reduce the influence of unphysical 
fitting parameters and lower the impact of parameter initial values on the results. 
JAM has included our data in their global analysis by fitting directly our DIS data 
for $A_{\parallel}^{^{3}\text{He}}$ and $A_{\perp}^{^{3}\text{He}}$.  Extracting 
a pure twist-3 $d_{2}^{n}$ without higher-twist contributions from resonances at the 
same $Q^{2}$ values as our results, they find a sizable effect on their (extrapolated)
prediction at $Q^{2} = 1$\,GeV$^{2}$, reducing it from $0.005 \pm 0.005$ to 
$-0.001 \pm 0.001$.  They also found their $d_{2}^{n}$ value to be consistent 
with lattice calculations~\cite{Gockeler:2005vw} when extrapolating to $Q^{2} = 5$\,GeV$^{2}$.  

%===============================================================================
\subsubsection{The matrix element $a_{2}^{n}$} \label{sec:a2-results} 
%===============================================================================

Following a similar procedure to that discussed for $d_{2}$, we can extract the $a_{2}$ 
matrix element from our $g_{1}$ data according to the third CN moment of $g_{1}$: 

\begin{equation}
   a_{2}\left( Q^{2} \right) = \int_{0}^{1} x^{2} g_{1} \left( x,Q^{2} \right) dx.  
\end{equation}

\begin{figure}[!ht]
   \centering
   \includegraphics[width=0.5\textwidth]{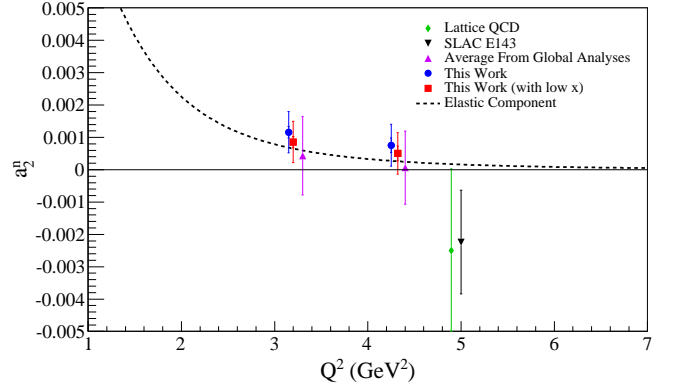}
   \caption{(Color online) Our extracted $a_{2}^{n}$ measurement compared to SLAC E143~\protect\cite{Abe:1998wq}
            and a lattice QCD calculation~\protect\cite{Gockeler:2005vw}, both of which are at $Q^{2} = 5$\,GeV$^{2}$.  
            The lattice calculation is offset in $Q^{2}$ for clarity.  The up-triangles  represent the 
            average over global 
            analyses~\protect\cite{deFlorian:2008mr,Bourrely:2001du,Bourrely:2007if,Gehrmann:1995ag,deFlorian:2005mw,Leader:2006xc},
            which are offset in $Q^{2}$ for clarity. 
            Our measurements shown as the circles (squares) exclude (include) the unmeasured low-$x$ region.
            The inner error bar on our data is the statistical error and the outer error bar is the 
            in-quadrature sum of the statistical and systematic uncertainties.  
            The results excluding the low-$x$ contribution are offset in $Q^{2}$ for clarity. 
            The elastic contribution is computed by using the Riordan~\protect\cite{Riordan:2010id} 
            and Kelly~\protect\cite{Kelly:2004hm} parameterizations for $G_{E}^{n}$ and $G_{M}^{n}$, respectively.}
   \label{fig:a2n-result}
\end{figure}

\noindent The low-$x$, high-$x$ and measured regions were treated in the same way 
as was done for the $d_{2}$ analysis, using the same model inputs.  Our $g_{1}$ data 
were not evolved to a constant $Q^{2}$, as our investigation into the $Q^{2}$ evolution
of our $g_{1}$ data revealed that the $Q^{2}$ dependence was negligible~\cite{MPosikThesis}.  
Our results for $a_{2}^{n}$ are shown in Fig.~\ref{fig:a2n-result}, where the inner error bars are 
the statistical uncertainties and the outer error bars represent the in-quadrature sum of 
statistical and systematic uncertainties.  The circle (square) data points exclude (include) 
the unmeasured low-$x$ region.  Both data points extracted are positive,
and the elastic contribution is sizable.  The SLAC E143~\cite{Abe:1998wq} data are 
also plotted, where their uncertainty is the in-quadrature sum of the statistical and systematic 
contributions.  The up-triangles represent the average over global 
analyses~\cite{deFlorian:2008mr,Bourrely:2001du,Bourrely:2007if,Gehrmann:1995ag,deFlorian:2005mw,Leader:2006xc}.
The lattice calculation shown is from G\"{o}ckeler {\it et al.}~\cite{Gockeler:2005vw}, 
where the error bar is statistical with a 15\% systematic uncertainty added in quadrature.  
The systematic uncertainty arises from their extrapolation of their result to the chiral limit~\cite{Gockeler:2005vw}.
Our results are tabulated in Table~\ref{tab:a2n-result}, broken down into the low-$x$,
measured and high-$x$ regions.  The last column shows the full extraction.  The 
systematic uncertainties on $a_{2}^{n}$ are dominated by that in $a_{2}^{^{3}\text{He}}$ 
and that due to the parameterization of $a_{2}^{p}$.  The uncertainties for $a_{2}^{n}$ are 
summarized in Appendix~\ref{sec:a2-syst-err}. 

\begin{table*}[!ht]
\caption{The extracted $a_{2}^{n}$ over the full $x$-range, decomposed into
         the low-$x$, measured and high-$x$ components.  The column labeled ``full''
         is the sum of all three regions.  The uncertainties are only listed for the full
         extraction, where the first quantity is the statistical uncertainty and the second 
         is the systematic uncertainty.  All uncertainties are absolute.}
\label{tab:a2n-result}
\center
\begin{ruledtabular}
\begin{tabular}{ccccc}
$\left<Q^2\right>$ $\left(\text{GeV}^{\text{2}}\right)$ 
& Low-$x$  $\left(\times 10^{-4}\right)$ 
& Measured $\left(\times 10^{-4}\right)$ 
& High-$x$ $\left(\times 10^{-4}\right)$ 
& Full     $\left(\times 10^{-4}\right)$ \\
\hline
3.21 & $-3.056$ & 5.078 & 6.530 & 8.552 $\pm$ 1.761 $\pm$ 6.125 \\ 
4.32 & $-3.056$ & 5.499 & 2.601 & 5.044 $\pm$ 2.270 $\pm$ 6.042 \\
\end{tabular}
\end{ruledtabular}
\end{table*}

%===============================================================================
\subsubsection{Color force extraction} \label{sec:res_cf}
%===============================================================================

In order to decompose the Lorentz color force into its electric and magnetic 
components, one needs to first extract the twist-4 matrix element, $f_{2}^{n}$.  
This is accomplished by considering our measured $d_{2}^{n}$ value along with
the $a_{2}^{n}$ matrix element. 

The $a_{2}^{n}$ matrix element was evaluated for various global 
analyses~\cite{deFlorian:2008mr,Bourrely:2001du,Bourrely:2007if,Gehrmann:1995ag,deFlorian:2005mw,Leader:2006xc} 
over the range $0.02 < x < 0.90$, and the average of the results at each 
$\left<Q^{2}\right>$-bin is taken as the value of $a_{2}^{n}$.  These results are 
consistent with our extracted values (Fig.~\ref{fig:a2n-result}).  The elastic 
contribution was added to the integral in a similar fashion as was done for $d_{2}^{n}$. 

With $d_{2}^{n}$ and $a_{2}^{n}$ in hand, the twist-4 matrix element $f_{2}^{n}$ 
was evaluated following the analysis presented in~\cite{Meziani:2004ne,Ji:1997gs}.  Results
for $\Gamma_{1}^{n}$ (Eq.~\ref{eqn:gamma1}) from HERMES~\cite{Airapetian:2002wd}, SMC~\cite{Adeva:1998vv}, 
JLab RSS~\cite{Slifer:2008xu} and E94-010~\cite{Slifer:2008re}, and the SLAC experiments 
E142~\cite{Anthony:1996mw}, E143~\cite{Abe:1998wq} and E154~\cite{Abe:1997cx} were used 
in the updated extraction analysis.  The data sets chosen for this analysis were from 
those experiments that published $g_{1}^{n}$ data at constant $Q^{2}$. 
Since $\Gamma_{1}^{n}$ is an integral over all $x$ ($0 \leq x \leq 1$), the 
unmeasured low-$x$ and high-$x$ regions need to be accounted for in a consistent fashion.  
The method we implemented for all data sets is that shown in Ref.~\cite{Meziani:2004ne}, 
with the exception of the HERMES and JLab data, which had already used such an extrapolation.  
The extrapolation calls for fitting the $g_{1}^{n}$ data to an appropriate function over the appropriate 
$x$-range.  For the low-$x$ region, the fit function was a constant $f(x) = A$, with 
$A$ being a free parameter.  The fit was performed over the range $x_{\text{min}} < x < x'$, where $x'$
is the lowest measured $x$ bin for a given experiment.  The lower bound $x_{\text{min}}$ 
is defined by $W = \sqrt{1000}$\,GeV.  The uncertainty in this low-$x$ extrapolation 
was estimated by taking the difference between our fit of $f(x) = A$ with 
that of a simple Regge parameterization where $f(x) = Ax^{-1/2}$~\cite{Heimann:1973hq,Ellis:1988mn}.  
For the high-$x$ region, the fit function was $f(x) = A(1 - x)^{3}$, with $A$ being a 
free parameter, over the range $x' < x < x_{\text{max}}$. The quantity $x'$ is the 
highest $x$ bin for which there were data available and $x_{\text{max}}$ is defined by the pion 
production threshold, $W = 1.12$\,GeV.  The fit functions for the low- and high-$x$ 
regions were chosen based on the trend of the data in the last three or two bins 
in each case, respectively.   

The elastic contribution to $\Gamma_{1}^{n}$ was evaluated using the 
Riordan~\cite{Riordan:2010id} and Kelly~\cite{Kelly:2004hm} parameterizations, 
and was added to all of the world data.  The uncertainty on the elastic contribution 
was estimated as the difference between using the Riordan (Kelly) parameterization 
for $G_{E}^{n}$ ($G_{M}^{n}$) compared to using Galster~\cite{Galster:1971kv} (dipole) 
for $G_{E}^{n}$ ($G_{M}^{n}$).  The resulting $\Gamma_{1}^{n}$ data from this analysis 
are presented in Appendix~\ref{sec:gamma1-world-app}.  

In order to extract the higher-twist contribution, the twist-2 contribution must
first be removed. Using the OPE~\cite{Wilson:1969zs}, $\Gamma_{1}^{n}$ was expanded as 
an inverse power series in $Q^{2}$, revealing the higher-twist components, that 
was accessed by subtracting the leading twist (twist-2) contribution. 
The twist-2 contribution $\mu_{2}^{n}$ was calculated using Eq.~\ref{eqn:mu_2}, 
where $\alpha_{s}$ was parameterized according to~\cite{Bethke:2009jm} and normalized to 
$\alpha_{s}\left(1\text{ GeV}^{2}\right) = 0.45 \pm 0.05$ for $\Lambda_{\text{QCD}} = 315$\,MeV. 
We used $N_{f} = 3$ and $N_{\text{loop}} = 3$~\cite{Ji:1997gs}, $g_{A} = 1.2723 \pm 0.0023$~\cite{Agashe:2014kda}, 
and $a_{8} = 0.587 \pm 0.016$~\cite{Agashe:2014kda}.  Note that the values for $g_{A}$ and $a_{8}$ have been 
updated relative to those used in Ref.~\cite{Posik:2014usi}. At large $Q^{2}$, the higher-twist
contributions should be small due to the $Q^{-2}$ suppression; therefore, 
$\Gamma_{1}^{n}\left(Q^{2} \right) = \mu_{2}\left(Q^{2}\right)$.  Because of this,
the axial charge $\Delta\Sigma$ was extracted (Eq.~\ref{eqn:mu_2}) using the highest 
$Q^{2}$-measurements from SLAC E154~\cite{Abe:1997cx} $(Q^{2} = 5\text{ GeV}^{2})$, 
SMC~\cite{Adeva:1998vv} $(Q^{2} = 10\text{ GeV}^{2})$, and HERMES~\cite{Airapetian:2002wd} 
$(Q^{2} = 6.5\text{ GeV}^{2})$.  Statistically averaging the results of these experiments 
yielded $Q^{2} = 5.77$\,GeV$^{2}$ and $\Gamma_{1}^{n} = -0.03851 \pm 0.00535$, resulting in  
$\Delta \Sigma = 0.375 \pm 0.052$.  This calculation is in excellent agreement with~\cite{Accardi:2012qut} 
and is consistent with Ref.~\cite{deFlorian:2008mr}, but at odds with Ref.~\cite{Nocera:2014gqa}.  
In the latter case, we suspect this disagreement may be due in part to differing approaches in 
the low-$x$ extrapolation of the world data in the various global analyses; additionally, Ref.~\cite{Nocera:2014gqa} 
is dominated by proton data (and does not include JLab neutron data~\cite{Zheng:2003un,Zheng:2004ce,Solvignon:2013yun,Slifer:2008re}), 
which may be biasing the extraction of $\Delta\Sigma$, though it is clear that there is 
a need for more neutron data in general.   
 
As described in Appendix~\ref{sec:gamma1-world-app}, a fit to $\Gamma_{1}^{n} - \mu_{2}$
as a function of $Q^{2}$ allows the extraction of $f_{2}^{n}$ after inserting the average 
$a_{2}^{n}$ (see beginning of Section~\ref{sec:res_cf}) and $d_{2}^{n}$ from the present experiment. 
The extracted $f_{2}^{n}$ values are given in Table~\ref{tab:f2n-result}.  This result differs from that 
in Ref.~\cite{Posik:2014usi} because we have improved our analysis, where we have updated the values for $g_{A}$ 
and $a_{8}$ used in the evaluation of the twist-2 term $\mu_{2}$.  We also now include uncertainties on 
the low-$x$ and elastic terms in the $\Gamma_{1}^{n}$ analysis mentioned above; additionally, our uncertainty 
on $f_{2}^{n}$ has changed as we now consider the full error matrix of our fit function, accounting for 
correlations between the fit parameters $A$ and $B$.  Our results reported here are larger than those in 
Ref.~\cite{Posik:2014usi} by about 23\% (25\%) for $E = 4.74$\,GeV ($5.89$\,GeV), and the systematic uncertainty 
has been reduced by a factor of 1.5. 

\begin{figure}[hbt]
   \centering
   \includegraphics[width=0.52\textwidth]{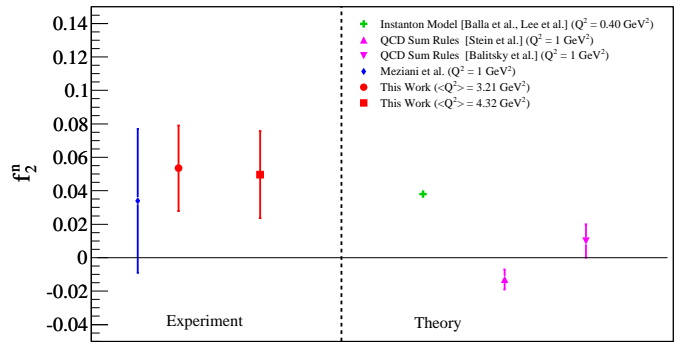}
   \caption{(Color online) Our extracted result for $f_{2}^{n}$ as compared to that from an instanton 
            model~\protect\cite{Lee:2001ug,Balla:1997hf} and QCD sum rules~\protect\cite{Stein:1994zk,Stein:1995si,Balitsky:1989jb}.  
            The result from the analysis of Meziani {\it et al.}~\protect\cite{Meziani:2004ne} is also shown.   
            For the present data, the inner error bars represent the statistical uncertainties (smaller than the markers), 
            while the outer error bars represent the statistical and systematic uncertainties added in quadrature.
            }
   \label{fig:f2n}
\end{figure}

Our result for $f_{2}^{n}$ is compared to that from an instanton model~\cite{Lee:2001ug,Balla:1997hf} and QCD 
sum rule calculations from E.~Stein {\it et al.}~\cite{Stein:1994zk,Stein:1995si} and Balitsky {\it et al.}~\cite{Balitsky:1989jb}, 
shown in Fig.~\ref{fig:f2n}.  We find good agreement with the instanton model and reasonable agreement
with the QCD sum rule result from Balitsky {\it et al}.  Currently, there are no lattice QCD calculations of $f_{2}^{n}$,
and it would be interesting to see what a lattice approach would yield. One can compare our result to that of Meziani 
{\it et al.}~\cite{Meziani:2004ne}, who found $f_{2}^{n} = 0.034 \pm 0.043$ normalized to $Q^{2} = 1$\,GeV$^{2}$ 
using a similar data set in their extraction of $\Gamma_{1}^{n}$.  The main difference between 
these analyses was that they fit world neutron data to obtain $a_{2}^{n}$ and used the results of 
SLAC E155x~\cite{Anthony:2002hy} for $d_{2}^{n}$, both at $Q^{2} = 5$\,GeV$^{2}$.  
In contrast, this work used the measured $d_{2}^{n}$, and the $a_{2}^{n}$ matrix element was obtained at 
the necessary $Q^{2}$ from an average over the global 
analyses~\cite{deFlorian:2008mr,Bourrely:2001du,Bourrely:2007if,Gehrmann:1995ag,deFlorian:2005mw,Leader:2006xc}. 
The statistical uncertainty on our extracted $f_{2}^{n}$ arises from $d_{2}^{n}$, while the systematic uncertainty 
contains contributions from the fit to extract $f_{2}^{n}$, and from the $a_{2}^{n}$ and $d_{2}^{n}$ systematic uncertainties, 
but is dominated by the first.   

\begin{table}
\center
\caption{Extracted $f_2^n$ results.  The uncertainties given are the statistical and systematic uncertainties, respectively.}
\label{tab:f2n-result}
\begin{ruledtabular} 
\begin{tabular}{c c}
$\left<Q^2\right>$ $\left(\text{GeV}^{2}\right)$ & $f_2^n$ $\left(\times 10^{-3}\right)$ \\
\hline
3.21 & $53.45 \pm 0.79 \pm 25.55$ \\
4.32 & $49.68 \pm 0.83 \pm 25.99$ \\
 
\end{tabular}
\end{ruledtabular} 
\end{table}

With the matrix elements $d_{2}^{n}$ and $f_{2}^{n}$ evaluated, the Lorentz 
color force was decomposed into its electric and magnetic components via~\cite{Burkardt:2008ps}:

\begin{eqnarray}
   F_{E}^{y,n} = - \frac{M_{n}^{2}}{6}\left( 2d_{2}^{n} + f_{2}^{n} \right) \label{eqn:f_e}    \\  
   F_{B}^{y,n} = - \frac{M_{n}^{2}}{6}\left( 4d_{2}^{n} - f_{2}^{n} \right) \label{eqn:f_b},  
\end{eqnarray}

\begin{figure}[hbt]
   \centering
   \includegraphics[width=0.52\textwidth]{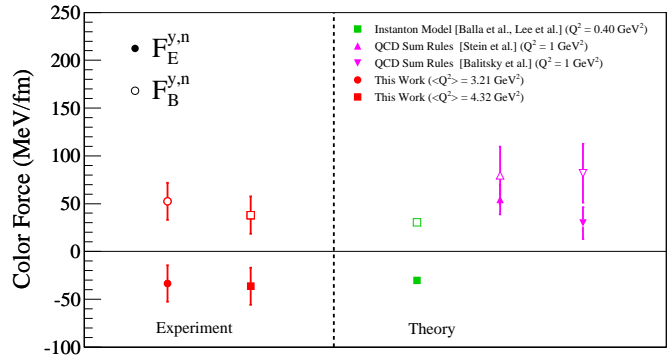}
   \caption{(Color online) Our extracted result for $F_{E}^{y,n}$ (filled markers) and $F_{B}^{y,n}$ (open markers) as 
            compared to an instanton model~\protect\cite{Lee:2001ug,Balla:1997hf} and QCD sum 
            rules~\protect\cite{Stein:1994zk,Stein:1995si,Balitsky:1989jb}.  For our data points, the inner 
            error bars represent the statistical uncertainties (smaller than the markers), while the 
            outer error bars represent the statistical and systematic uncertainties added in quadrature.  
            }
   \label{fig:color-forces}
\end{figure}

\noindent The results for the electric $\left(F_{E}^{y,n}\right)$ and magnetic $\left(F_{B}^{y,n}\right)$ 
color forces averaged over the volume of the neutron are shown in Table~\ref{tab:color-forces-result}
and in Fig.~\ref{fig:color-forces}, where we compare to an instanton model~\cite{Lee:2001ug,Balla:1997hf} 
and QCD sum rules from Stein {\it et al.}~\cite{Stein:1994zk,Stein:1995si} and 
Balitsky {\it et al.}~\cite{Balitsky:1989jb}.  In the figure, filled markers represent 
$F_{E}^{y,n}$, while open markers indicate $F_{B}^{y,n}$.  We find that the electric and magnetic 
color forces are approximately equal and opposite.  The electric color force component 
$F_{E}^{y,n}$ is in agreement with the instanton model, while the magnetic component $F_{B}^{y,n}$ 
is consistent with the instanton model and QCD sum rules.  However, those calculations 
were performed at $Q^{2} = 0.4$ and 1\,GeV$^{\text{2}}$, respectively.  Note that 
the values for $F_{E}^{y,n}$ and $F_{B}^{y,n}$ reported here differ from those presented
in Ref.~\cite{Posik:2014usi} because we have re-evaluated the color forces using 
the updated $f_{2}^{n}$ values given in Table~\ref{tab:f2n-result}.  The central
values for $F_{E}^{y,n}$ at $E = 4.74$\,GeV ($E = 5.89$\,GeV) have increased in magnitude by 
28\% (25\%), while for $F_{B}^{y,n}$ the central values have increased in magnitude by 16\% (24\%).  
The systematic uncertainties for both $F_{E}^{y,n}$ and $F_{B}^{y,n}$ have been reduced 
by a factor of 1.5.  

\begin{table}
\center
\caption{Extracted magnetic and electric Lorentz color force components.  The uncertainties given are the statistical and 
         systematic uncertainty, respectively.}
\label{tab:color-forces-result}
\begin{ruledtabular} 
\begin{tabular}{c c c}
$\left<Q^2\right>$ $\left(\text{GeV}^2\right)$ 
& $F_{E}^{y,n}$ $\left(\text{MeV}/\text{fm}\right)$ 
& $F_{B}^{y,n}$ $\left(\text{MeV}/\text{fm}\right)$  \\
\hline
3.21 & $-33.53 \pm 1.32 \pm  19.07$ & $52.35 \pm 2.43 \pm 19.18$ \\ 
4.32 & $-36.48 \pm 1.38 \pm  19.38$ & $38.04 \pm 2.55 \pm 19.46$ \\ 

\end{tabular}
\end{ruledtabular} 
\end{table}

%===============================================================================
\subsubsection{The virtual photon-nucleon asymmetry $A_{1}^{n}$} \label{sec:a1n-results} 
%===============================================================================

The asymmetry $A_{1}^{n}$ (Section~\ref{sec:nuclear_cor}) was extracted for our 
DIS data at the two beam energies for $\left< Q^{2} \right> = 2.59$\,GeV$^{2}$ ($E = 4.74$\,GeV) and 
$\left< Q^{2} \right> = 3.67$\,GeV$^{2}$ ($E = 5.89$\,GeV); the results are given in 
Tables~\ref{tab:n_dis_4} and~\ref{tab:n_dis_5}.  These results were averaged using 
the statistical uncertainty as the weight, while the systematic uncertainties were averaged 
using equal weights.  The averaged results, first published in~\cite{Parno:2014xzb}, 
are given in Table~\ref{tab:n_dis_avg} and plotted in Fig.~\ref{fig:A1n_dis-result}.  
The systematic uncertainties are outlined in tables given in Appendix~\ref{sec:a1-g1f1-syst-err}.  
The biggest contribution to the uncertainty is from the effective proton polarization.  
Our result is consistent with the trend seen in current DIS data from SLAC E142~\cite{Anthony:1996mw} 
and E154~\cite{Abe:1997cx}, HERMES~\cite{Ackerstaff:1997ws} and JLab E99-117~\cite{Zheng:2003un,Zheng:2004ce}. 
Although this experiment was optimized for the measurement of $d_{2}^{n}$, we obtained 
a data set with uncertainties that are competitive with the previous JLab data from E99-117.  
Our extraction provides a proof-of-principle measurement in an open-geometry detector 
(BigBite) yielding data with competitive uncertainties compared to the majority of the 
world data in the mid-$x$ range, and showing a zero-crossing at $x \approx 0.5$.  
Our data tend to follow the trend of the pQCD-based parameterization that includes orbital 
angular momentum~\cite{Avakian:2007xa}, possibly indicating the importance of orbital 
angular momentum in the spin of the nucleon.  Our result also shows good agreement with 
the NJL-type model from Clo\"{e}t {\it et al.}~\cite{Cloet:2005pp}.  
Dyson-Schwinger Equation treatment predictions~\cite{Roberts:2013mja} are presented 
at $x = 1$ (Fig.~\ref{fig:A1n_dis-result}).  

\begin{figure}[hbt]
   \centering 
   \includegraphics[width=0.5\textwidth]{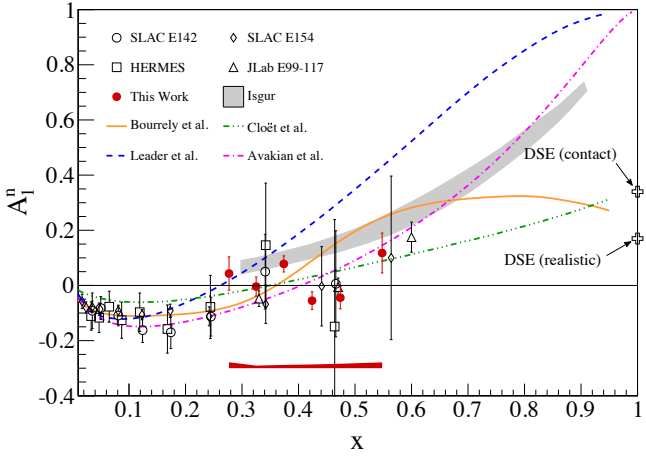}
   \caption{(Color online) Our measured $A_{1}^{n}$ results compared to world 
            data~\protect\cite{Anthony:1996mw,Abe:1997qk,Zheng:2003un,Zheng:2004ce,Ackerstaff:1997ws}
            and a pQCD-inspired global analysis (dashed)~\protect\cite{Leader:1997kw}, a statistical quark 
            model (solid)~\protect\cite{Bourrely:2015kla}, 
            a relativistic CQM model (gray band)~\protect\cite{Isgur:1998yb} and a pQCD-based parameterization 
            including orbital angular momentum (dash-dotted)~\protect\cite{Avakian:2007xa}.
            An NJL-type model (dash triple-dotted) is also shown~\protect\cite{Cloet:2005pp}.
            Dyson-Schwinger Equation treatment predictions~\protect\cite{Roberts:2013mja} are presented at 
            $x = 1$. The band at the bottom of the plot indicates the systematic uncertainty 
            for the present data. 
            }
   \label{fig:A1n_dis-result}
\end{figure}

\begin{table}[hbt] 
\centering
\caption{Results for $A_{1}^{n}$ and $g_{1}^{n}/F_{1}^{n}$ for $E = 4.74$\,GeV.  
         The uncertainties given are the statistical and systematic uncertainties, respectively.
         }
\label{tab:n_dis_4}
\begin{ruledtabular} 
\begin{tabular}{crr}
 \multicolumn{1}{c}{$\left< x \right>$   } & 
 \multicolumn{1}{c}{$A_{1}^{n}$          } & 
 \multicolumn{1}{c}{$g_{1}^{n}/F_{1}^{n}$} \\
\hline
 0.277 & $ 0.012 \pm 0.071 \pm 0.008$ & $ 0.007 \pm 0.068 \pm 0.010$ \\
 0.325 & $ 0.011 \pm 0.043 \pm 0.009$ & $ 0.008 \pm 0.041 \pm 0.008$ \\
 0.374 & $ 0.102 \pm 0.037 \pm 0.014$ & $ 0.065 \pm 0.034 \pm 0.011$ \\
 0.424 & $-0.064 \pm 0.040 \pm 0.014$ & $-0.066 \pm 0.038 \pm 0.013$ \\
 0.473 & $-0.044 \pm 0.051 \pm 0.015$ & $-0.058 \pm 0.047 \pm 0.014$ \\
\end{tabular}
\end{ruledtabular} 
\end{table}

\begin{table}[hbt] 
\centering
\caption{Results for $A_{1}^{n}$ and $g_{1}^{n}/F_{1}^{n}$
         for $E = 5.89$\,GeV.  
         The uncertainties given are the statistical and systematic uncertainties, respectively.
         }
\label{tab:n_dis_5}
\begin{ruledtabular} 
\begin{tabular}{crr}
 \multicolumn{1}{c}{$\left< x \right>$   } & 
 \multicolumn{1}{c}{$A_{1}^{n}$          } & 
 \multicolumn{1}{c}{$g_{1}^{n}/F_{1}^{n}$} \\
\hline
 0.277 & $ 0.127 \pm 0.116 \pm 0.035$ & $ 0.143 \pm 0.112 \pm 0.014$ \\
 0.325 & $-0.031 \pm 0.058 \pm 0.009$ & $-0.019 \pm 0.056 \pm 0.009$ \\
 0.374 & $ 0.035 \pm 0.049 \pm 0.010$ & $ 0.031 \pm 0.046 \pm 0.009$ \\
 0.424 & $-0.039 \pm 0.053 \pm 0.013$ & $-0.049 \pm 0.050 \pm 0.012$ \\
 0.474 & $-0.044 \pm 0.066 \pm 0.017$ & $-0.044 \pm 0.062 \pm 0.015$ \\
 0.524 & $ 0.109 \pm 0.088 \pm 0.019$ & $ 0.098 \pm 0.082 \pm 0.017$ \\
 0.573 & $ 0.135 \pm 0.126 \pm 0.023$ & $ 0.132 \pm 0.116 \pm 0.021$ \\
\end{tabular}
\end{ruledtabular} 
\end{table}

\begin{table}[hbt] 
\centering
\caption{Results for $A_{1}^{n}$ and $g_{1}^{n}/F_{1}^{n}$
         averaged over our $E = 4.74$ and $5.89$\,GeV results 
         for $\left< Q^{2} \right> = 3.08$\,GeV$^{2}$.  
         The uncertainties given are the statistical and systematic uncertainties, respectively.
         }
\label{tab:n_dis_avg}
\begin{ruledtabular} 
\begin{tabular}{crr}
 \multicolumn{1}{c}{$\left< x \right>$   } & 
 \multicolumn{1}{c}{$A_{1}^{n}$          } & 
 \multicolumn{1}{c}{$g_{1}^{n}/F_{1}^{n}$} \\
\hline
 0.277 & $ 0.043 \pm 0.060 \pm 0.022$ & $ 0.044 \pm 0.058 \pm 0.012$ \\
 0.325 & $-0.004 \pm 0.035 \pm 0.009$ & $-0.002 \pm 0.033 \pm 0.009$ \\
 0.374 & $ 0.078 \pm 0.029 \pm 0.012$ & $ 0.053 \pm 0.028 \pm 0.010$ \\
 0.424 & $-0.055 \pm 0.032 \pm 0.014$ & $-0.060 \pm 0.030 \pm 0.012$ \\
 0.474 & $-0.044 \pm 0.040 \pm 0.016$ & $-0.053 \pm 0.037 \pm 0.015$ \\
 0.548 & $ 0.118 \pm 0.072 \pm 0.021$ & $ 0.110 \pm 0.067 \pm 0.019$ \\
\end{tabular}
\end{ruledtabular} 
\end{table}
 
%===============================================================================
\subsubsection{The structure function ratio $g_{1}^{n}/F_{1}^{n}$} \label{sec:g1f1n-results} 
%===============================================================================

Similar to the $A_{1}^{n}$ analysis (Section~\ref{sec:a1n-results}), the $g_{1}^{n}/F_{1}^{n}$ 
ratio was extracted for our DIS data at each beam energy for $\left< Q^{2} \right> = 2.59$\,GeV$^{2}$ ($E = 4.74$\,GeV) 
and $\left< Q^{2} \right> = 3.67$\,GeV$^{2}$ ($E = 5.89$\,GeV).  These data are 
given in Tables~\ref{tab:n_dis_4} and~\ref{tab:n_dis_5}.  The results were averaged 
using the statistical uncertainty as the weight, and systematic uncertainties 
were averaged using equal weights.  The averaged results, first published in~\cite{Parno:2014xzb}, 
are given in Table~\ref{tab:n_dis_avg} and plotted in Fig.~\ref{fig:g1F1n_dis-result}.  
The systematic uncertainties are given in tables presented in Appendix~\ref{sec:a1-g1f1-syst-err}.  
The biggest contribution to the uncertainty is from the effective proton polarization 
and from $A_{1}^{p}$.  Our results are comparable to the JLab E99-117 data~\cite{Zheng:2003un,Zheng:2004ce}
in reach and precision, and are consistent with the trend seen in the DIS data from 
SLAC E143~\cite{Anthony:1996mw} and E155~\cite{Anthony:2000fn}  as shown in Fig.~\ref{fig:g1F1n_dis-result}. 

\begin{figure}[hbt]
   \centering 
   \includegraphics[width=0.5\textwidth]{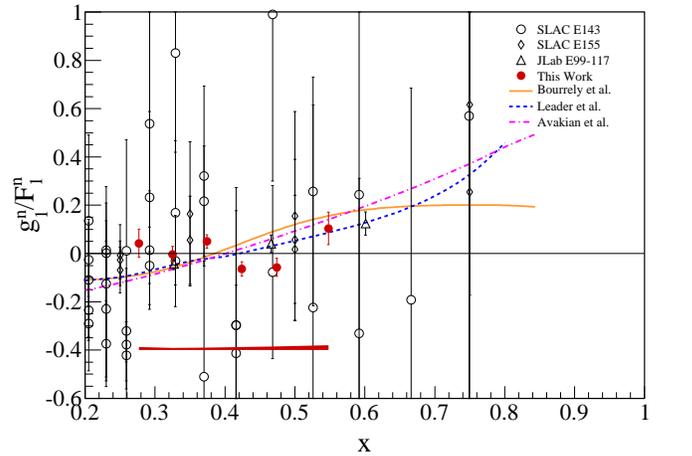}
   \caption{(Color online) Our measured $g_{1}^{n}/F_{1}^{n}$ results, compared to world 
            data~\protect\cite{Anthony:1996mw,Abe:1997cx,Zheng:2003un,Zheng:2004ce}
            and an NLO QCD global analysis~\protect\cite{Leader:2006xc} (dashed)
            and pQCD-inspired fit~\protect\cite{Avakian:2007xa} (dash-dotted), and a 
            statistical quark model~\protect\cite{Bourrely:2001du,*Bourrely:2007if} (solid).  
            The band at the bottom of the plot indicates the systematic uncertainty for our data set. 
            }
   \label{fig:g1F1n_dis-result}
\end{figure}

%===============================================================================
\subsection{Flavor decomposition via the quark-parton model} \label{sec:res_flavor} 
%===============================================================================

\begin{figure}[hbt]
   \centering
   \includegraphics[width=0.5\textwidth]{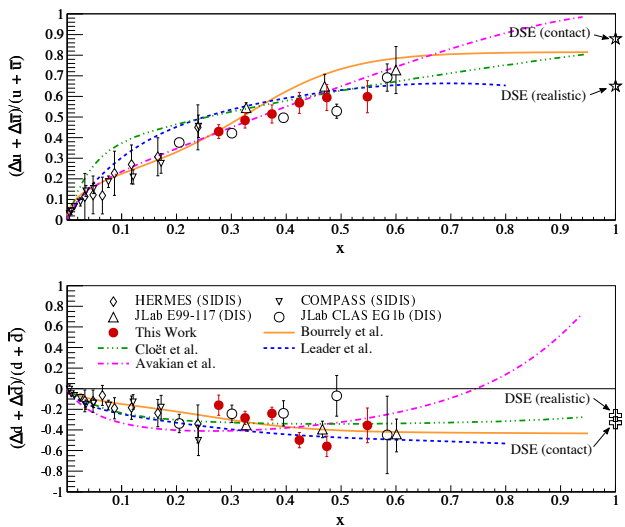}
   \caption{(Color online) Our combined $E = 4.74$ and $5.89$\,GeV data for the up and down quark ratios.  
            Our results are compared to existing data~\protect\cite{Zheng:2004ce,Dharmawardane:2006zd,Alekseev:2010ub,Airapetian:2004zf}, 
            where the error bars on all data sets are the in-quadrature sum of the statistical 
            and systematic uncertainties. Also presented is a statistical quark model 
            (solid)~\protect\cite{Bourrely:2015kla}, an NLO QCD global analysis from 
            Leader {\it et al.} (dashed)~\protect\cite{Leader:2006xc}, and a pQCD analysis 
            including orbital angular momentum from Avakian {\it et al.} (dash-dotted)~\protect\cite{Avakian:2007xa}.  
            An NJL-type model from Clo\"{e}t {\it et al.} (dash triple-dotted)~\protect\cite{Cloet:2005pp} is also shown.  
            DSE predictions~\protect\cite{Roberts:2013mja} are shown at $x = 1$.
            }
   \label{fig:flavor-sep-dis-avg}
\end{figure}

Under the quark-parton model~\cite{Feynman:1969ej}, if one assumes that the strange quark
distributions $s(x)$, $\bar{s}(x)$, $\Delta s(x)$ and $\Delta \bar{s}(x)$ are negligible
for $x > 0.3$, and neglecting any $Q^{2}$-dependence in the ratio of structure functions, 
the polarized-to-unpolarized quark ratios were extracted through Eqs.~\ref{eqn:up}
and~\ref{eqn:down}.  We utilized the $R^{du}$ ratio from the CJ12~\cite{Owens:2012bv} model.  
Using our fit to the $g_{1}^{p}/F_{1}^{p}$ world data sets, we obtained at leading order the 
quantities $\left( \Delta u + \Delta \bar{u}\right)/\left( u + \bar{u}\right)$ and 
$\left( \Delta d + \Delta \bar{d}\right)/\left( d + \bar{d}\right)$ for $E = 4.74$\,GeV 
(5.89\,GeV) where $\left< Q^{2} \right> = 2.59$\,GeV$^{2}$ (3.67\,GeV$^{2}$).  The results are tabulated
in Tables~\ref{tab:u-d-4} and~\ref{tab:u-d-5}.  Averaging the two data sets, we
obtained the values given in Table~\ref{tab:u-d-dis-avg} at $\left< Q^{2} \right> = 3.08$\,GeV$^{2}$.  
These averaged results, first published in~\cite{Parno:2014xzb}, are compared to various 
world data~\cite{Zheng:2004ce,Dharmawardane:2006zd,Alekseev:2010ub,Airapetian:2004zf} 
and theoretical calculations~\cite{Avakian:2007xa,Leader:2006xc,Roberts:2013mja,Bourrely:2015kla} 
in Fig.~\ref{fig:flavor-sep-dis-avg}.  The uncertainty due to neglecting the 
strange contribution was determined by computing Eqs.~\ref{eqn:up} and~\ref{eqn:down} 
with the strange component {\it included}~\cite{DFlayThesis}:  

\begin{eqnarray}
   \frac{\Delta u + \Delta \bar{u}}{u + \bar{u}} &=& \left( \frac{\Delta u + \Delta \bar{u}}{u + \bar{u}}\right)_{s,\bar{s}=0} \nonumber \\
                                                 &+& \frac{s + \bar{s}}{u}\left[ \frac{4}{15}\frac{g_{1}^{p}}{F_{1}^{p}} 
                                                            - \frac{1}{15}\frac{g_{1}^{n}}{F_{1}^{n}}
                                                  - \frac{1}{5}\frac{\Delta s + \Delta \bar{s}}{s + \bar{s}}\right] \label{eqn:up-str-incl} \\  
   \frac{\Delta d + \Delta \bar{d}}{d + \bar{d}} &=& \left( \frac{\Delta u + \Delta \bar{u}}{u + \bar{u}}\right)_{s,\bar{s}=0} \nonumber \\ 
                                                 &+& \frac{s + \bar{s}}{d}\left[ \frac{4}{15}\frac{g_{1}^{n}}{F_{1}^{n}} 
                                                            - \frac{1}{15}\frac{g_{1}^{n}}{F_{1}^{n}}
                                                  - \frac{1}{5}\frac{\Delta s + \Delta \bar{s}}{s + \bar{s}}\right], \label{eqn:down-str-incl}  
\end{eqnarray}

\noindent where the terms $(\ldots)_{s,\bar{s}=0}$ are defined in Eqs.~\ref{eqn:up} and~\ref{eqn:down}.   
The second term in Eqs.~\ref{eqn:up-str-incl} and~\ref{eqn:down-str-incl} is the strange contribution, 
which was evaluated using various 
parameterizations~\cite{deFlorian:2005mw,Leader:2006xc,deFlorian:2008mr,Bourrely:2001du,Bourrely:2007if,Owens:2012bv}
and taking the maximum difference between calculations using all possible model combinations 
as the uncertainty.  It was found to be sizable in the lowest $x$-bins for the down-quark results,
but was small for $x \geq 0.424$.  For the up-quark results, the strange uncertainty was small.   
The uncertainty due to neglecting the strange contribution is included in our reported uncertainties. 

The extracted up-and down-quark ratios agree with the general trend of the world data
within our uncertainties.  Our analysis supports the notion that the down-quark 
ratio stays negative into the large $x$ region, with no clear indication that it turns 
positive towards $x \simeq 0.6$, as predicted by the Avakian {\it et al.} calculation~\cite{Avakian:2007xa}. 

The largest contribution to our systematic uncertainties on the up-quark ratio is from our fit to 
the $g_{1}^{p}/F_{1}^{p}$ data, while for the down-quark ratio, the largest contributions are due 
to the $g_{1}^{p}/F_{1}^{p}$ fit and the $d/u$ ratio.  An overview of the systematic uncertainties 
is given in Appendix~\ref{sec:flavor-syst-err}.        

\begin{table}[!ht]
\centering
\caption{Results for $(\Delta u + \Delta \bar{u})/(u + \bar{u})$ and
         $(\Delta d + \Delta \bar{d})/(d + \bar{d})$ at $E = 4.74$\,GeV.
         The uncertainties given are the statistical and systematic uncertainties, respectively.
         }
\label{tab:u-d-4}
\begin{ruledtabular}
\begin{tabular}{crr}
 \multicolumn{1}{c}{$\left< x \right>$                         } & 
 \multicolumn{1}{c}{$(\Delta u + \Delta \bar{u})/(u + \bar{u})$} & 
 \multicolumn{1}{c}{$(\Delta d + \Delta \bar{d})/(d + \bar{d})$} \\
\hline
0.277 &  $0.437 \pm 0.013 \pm 0.031$ & $-0.219 \pm 0.110 \pm 0.028$ \\
0.325 &  $0.482 \pm 0.008 \pm 0.036$ & $-0.267 \pm 0.069 \pm 0.032$ \\
0.374 &  $0.513 \pm 0.006 \pm 0.043$ & $-0.218 \pm 0.060 \pm 0.038$ \\
0.424 &  $0.570 \pm 0.006 \pm 0.050$ & $-0.508 \pm 0.068 \pm 0.051$ \\
0.473 &  $0.596 \pm 0.007 \pm 0.063$ & $-0.566 \pm 0.088 \pm 0.069$ \\
\end{tabular}
\end{ruledtabular}
\end{table}

\begin{table}[!ht]
\centering
\caption{Results for $(\Delta u + \Delta \bar{u})/(u + \bar{u})$ and
         $(\Delta d + \Delta \bar{d})/(d + \bar{d})$ at $E = 5.89$\,GeV.
         The uncertainties given are the statistical and systematic uncertainties, respectively.
         }
\label{tab:u-d-5}
\begin{ruledtabular}
\begin{tabular}{crr}
 \multicolumn{1}{c}{$\left< x \right>$                         } & 
 \multicolumn{1}{c}{$(\Delta u + \Delta \bar{u})/(u + \bar{u})$} & 
 \multicolumn{1}{c}{$(\Delta d + \Delta \bar{d})/(d + \bar{d})$} \\
\hline
0.277 &  $0.410 \pm 0.022 \pm 0.032$ & $ 0.001 \pm 0.182 \pm 0.027$ \\
0.325 &  $0.487 \pm 0.010 \pm 0.037$ & $-0.314 \pm 0.094 \pm 0.033$ \\
0.374 &  $0.518 \pm 0.008 \pm 0.045$ & $-0.281 \pm 0.081 \pm 0.040$ \\
0.424 &  $0.567 \pm 0.008 \pm 0.051$ & $-0.482 \pm 0.090 \pm 0.051$ \\
0.474 &  $0.593 \pm 0.009 \pm 0.063$ & $-0.547 \pm 0.116 \pm 0.071$ \\
0.524 &  $0.594 \pm 0.012 \pm 0.070$ & $-0.352 \pm 0.165 \pm 0.083$ \\
0.573 &  $0.606 \pm 0.015 \pm 0.085$ & $-0.365 \pm 0.250 \pm 0.111$ \\
\end{tabular}
\end{ruledtabular}
\end{table}

\begin{table}[!ht]
\centering
\caption{Results for $(\Delta u + \Delta \bar{u})/(u + \bar{u})$ and
         $(\Delta d + \Delta \bar{d})/(d + \bar{d})$ averaged over the two beam
         energies, for $\left< Q^{2} \right> = 3.08$\,GeV$^{2}$.
         The uncertainties given are the statistical and systematic uncertainties, respectively.
         }
\label{tab:u-d-dis-avg}
\begin{ruledtabular}
\begin{tabular}{crr}
 \multicolumn{1}{c}{$\left< x \right>$                         } & 
 \multicolumn{1}{c}{$(\Delta u + \Delta \bar{u})/(u + \bar{u})$} & 
 \multicolumn{1}{c}{$(\Delta d + \Delta \bar{d})/(d + \bar{d})$} \\
\hline
0.277 &  $0.430 \pm 0.011 \pm 0.031$ & $-0.160 \pm 0.094 \pm 0.028$ \\
0.325 &  $0.484 \pm 0.006 \pm 0.037$ & $-0.283 \pm 0.055 \pm 0.032$ \\
0.374 &  $0.515 \pm 0.005 \pm 0.044$ & $-0.241 \pm 0.048 \pm 0.039$ \\
0.424 &  $0.569 \pm 0.005 \pm 0.051$ & $-0.499 \pm 0.054 \pm 0.051$ \\
0.474 &  $0.595 \pm 0.006 \pm 0.063$ & $-0.559 \pm 0.070 \pm 0.070$ \\
0.548 &  $0.598 \pm 0.009 \pm 0.077$ & $-0.356 \pm 0.138 \pm 0.097$ \\
\end{tabular}
\end{ruledtabular}
\end{table}

   %===============================================================================
\section{Conclusions} \label{sec:conclusion} 
%===============================================================================

Scattering a longitudinally-polarized electron beam of energies of $E = 4.74$\,GeV 
and $5.89$\,GeV from a polarized $^{3}$He target in two orientations, longitudinal 
and transverse (with respect to the electron beam momentum), we measured the 
unpolarized electron-scattering cross section with the LHRS and the electron double-spin
asymmetries with the BigBite spectrometer, with both spectrometers set at 45$^{\circ}$ with respect 
to the beamline.  Combining the unpolarized cross sections and double-spin 
asymmetries allowed the extraction of the twist-3 matrix element $d_{2}^{n}$.
This quantity was extracted at two $\left<Q^{2}\right>$ values of 3.21\,GeV$^{2}$ 
and 4.32\,GeV$^{2}$.  The result at the lower $\left<Q^{2}\right>$ value is small and negative,
while that for the higher $\left<Q^{2}\right>$ value is consistent with zero.   
The data indicate a trend towards the Lattice QCD~\cite{Gockeler:2005vw} calculation 
at $Q^{2} = 5$\,GeV$^{2}$.  The extracted $d_{2}^{n}$ values are also
consistent with predictions of chiral soliton~\cite{Weigel:1996jh,*Weigel:2000gx} 
and bag~\cite{Song:1996ea,Stratmann:1993aw,Ji:1993sv} models.  The size of the 
present $d_{2}^{n}$ results predicts the twist-3 contribution is small.

Utilizing our measured $d_{2}^{n}$ twist-3 matrix element combined with global analyses 
for $a_{2}^{n}$ and world data on $\Gamma_{1}^{n}$, we have extracted the twist-4 matrix 
element $f_{2}^{n}$.  This matrix element is observed to be larger than $d_{2}^{n}$ in magnitude, 
resulting in approximately equal and opposite Lorentz color magnetic and electric forces in 
the neutron.  The results for $f_{2}^{n}$ are consistent with instanton model calculations~\cite{Balla:1997hf,Lee:2001ug}
and a QCD sum rule calculation~\cite{Stein:1994zk}.  The extracted values for $F_{E}^{y,n}$ 
and $F_{B}^{y,n}$ are in agreement with the instanton model calculations of~\cite{Balla:1997hf,Lee:2001ug}, 
while the value for $F_{B}^{y,n}$ is additionally in accordance with QCD sum-rule calculations 
from~\cite{Stein:1994zk,Balitsky:1989jb}.  We look forward to Lattice QCD calculations of 
the $f_{2}^{n}$ matrix element, not yet available at this time due to their difficulty, 
before the new wave of planned experiments in the 12\,GeV upgrade era of Jefferson Lab 
to measure $d_2$ and $f_2$ with more precision.

Following a similar analysis procedure to the one used for $d_{2}^{n}$, we extracted the $a_{2}^{n}$
matrix element using our $g_{1}^{^{3}\text{He}}$ data.  Our $a_{2}$ results are positive 
for both $\left< Q^{2} \right> = 3.21$ and $4.32$\,GeV$^{2}$, with an improved precision 
compared to the currently published data and lattice calculations. 

The extracted virtual photon asymmetry $A_{1}^{n}$ is consistent with the current world
data, especially JLab E99-117~\cite{Zheng:2003un,Zheng:2004ce}, and shows good agreement with pQCD
calculations that incorporate quark orbital angular momentum~\cite{Avakian:2007xa}.
This suggests that orbital angular momentum may play an important role in the spin of
the nucleon.

The measured structure function ratio of $g_{1}^{n}/F_{1}^{n}$ shows a similar trend
to JLab E99-117~\cite{Zheng:2003un,Zheng:2004ce}, adding higher precision data to 
the world data set.

Fitting the world $g_{1}/F_{1}$ data on the proton allowed the extraction of $(\Delta u + \Delta \bar{u})/(u + \bar{u})$ 
and $(\Delta d + \Delta \bar{d})/(d + \bar{d})$ when using the CJ12~\cite{Owens:2012bv} model for $d/u$.  
The extracted up-quark ratio is consistent with existing measurements and models, and its uncertainty 
is dominated by our fit to the world $g_{1}^{p}/F_{1}^{p}$ data.  The down-quark 
ratio is observed to remain negative into the large-$x$ region, with no clear indication of a change 
to positive values in the range of $x \simeq 0.75$ as predicted in Ref.~\cite{Avakian:2007xa}.  
The down-quark ratio is very sensitive to the $d/u$ ratio, as is evident in the systematic errors.  
Better precision on $d/u$ from the projected experiment~\cite{E1210103} at Jefferson Lab in the 
12\,GeV upgrade era will help to constrain the $(\Delta d + \Delta \bar{d})/(d + \bar{d})$ ratio. 

A future experiment~\cite{E1206120} proposed at Jefferson Lab calls for an even higher precision
measurement of $d_{2}^{n}$ at four central $Q^{2}$-bins in a range from 2\,GeV$^{2}$ to 7\,GeV$^{2}$.  
While our data provide a good understanding of $d_{2}^{n}$ at $Q^{2} = 3.21$ and $4.32$\,GeV$^{2}$,
those data will provide a direct comparison to the Lattice QCD calculation at $Q^{2} = 5$\,GeV$^{2}$. 
There are also two dedicated $A_{1}^{n}$ experiments approved to run at Jefferson 
Lab~\cite{E1206122,E1206110} that aim to extend DIS $A_{1}^{n}$ measurements to larger $x$ ($\sim 0.77$) 
in addition to studying the $Q^{2}$-evolution of the asymmetry.  These measurements are important for 
broadening our insight in the large-$x$ spin structure of the nucleon, as suggested by the results
presented in this paper.

   %===============================================================================
\section*{Acknowledgements}
%===============================================================================

We would like to thank the Jefferson Lab Accelerator Division and Hall A staff 
for their efforts which resulted in the successful completion of the experiment.  
We would also like to thank M.~Burkardt, L.~Gamberg, W.~Melnitchouk, and J.~Soffer 
for theoretical support and useful discussions.  This material is based upon work 
supported by the U.S. Department of Energy (DOE) Office of Science under award 
numbers DE-FG02-94ER40844 and DE-FG02-87ER40315.  Jefferson Lab is operated by 
the Jefferson Science Associates, LLC, under DOE grant DE-AC05-060R23177.

   \begin{appendix} 
      %===============================================================================
\appendix{}
\section{Deep inelastic electron scattering formalism}  \label{sec:dis_app}
\subsection{Structure functions and cross sections}
%===============================================================================

In electron scattering, the electrons are accelerated to high energies and scatter from 
a nuclear or nucleon target.  In practice, the target is typically fixed.  The 
electron interacts with the target by exchanging a virtual photon with the target 
object, transferring its energy and momentum to the target.  An advantage 
of lepton scattering is that the interaction at the leptonic vertex  
is solely described by quantum electrodynamics, which simplifies the 
mathematics.  The electromagnetic nature of the interaction also results in the 
process being a ``clean'' probe into the structure of the nucleon, where the QCD 
physics is contained entirely in the description of the nucleon and is not convoluted 
with the leptonic probe.

To describe the process more quantitatively, consider Fig.~\ref{fig:e-N-scat}.
The incident and scattered electrons have the four-momenta $k = (E,\vec{k})$
and $k' = (E',\vec{k}')$, respectively.  The target has a four-momentum
of $p = (E_T,\vec{p})$.  The virtual photon exchanged between the
incident electron and the target has the four-momentum $q = (\nu,\vec{q})$.
If the incident electron has enough energy, the target can break up into a number
of distinct hadrons; otherwise, the target will remain intact.  In the
latter case, the recoiling target object would have a four-momentum $p'$
in the final hadronic state.  Electron scattering data are presented and discussed
in terms of a number of Lorentz-invariant variables, namely $\nu$, $y$, $Q^{2}$, $W$ and $x$.
Since the four-momentum at each vertex is conserved, we begin by defining $q$ 
in terms of the incoming and outgoing electron four-momenta:

\begin{figure}[ht]
   \centering
   \includegraphics[width=0.5\textwidth]{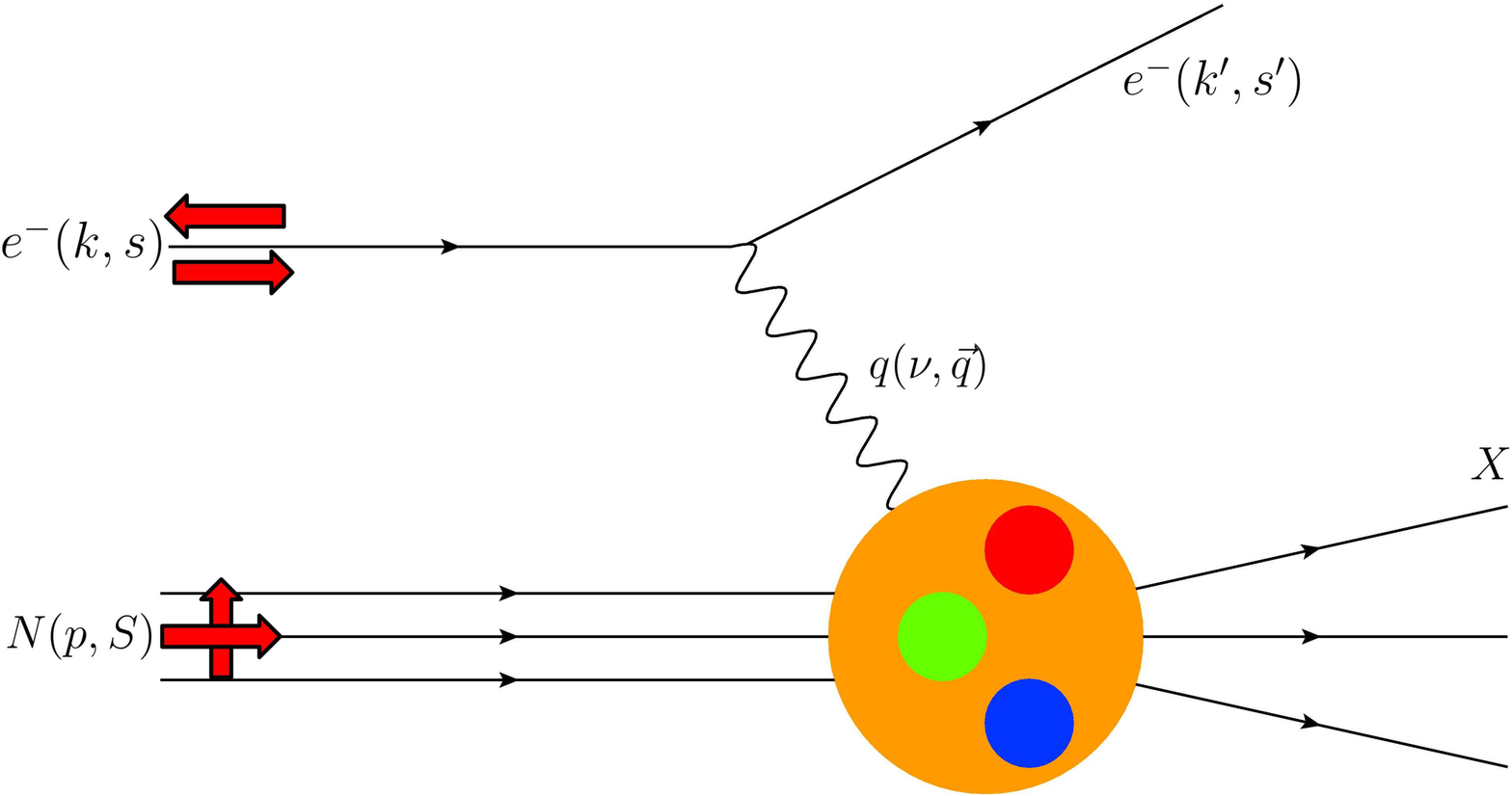}
   \caption{(Color online) A cartoon describing inclusive polarized electron-nucleon scattering.
            The large arrows indicate possible spin orientations of the incident electron and 
            nucleon.  The quantities $s$ and $S$ indicate the spin four-vectors of the 
            electron and nucleon, respectively.  The other kinematic variables
            are described in the text.}
   \label{fig:e-N-scat}
\end{figure}

\begin{equation}
   q = k - k' = (E-E',\vec{k}-\vec{k}') = (\nu,\vec{q}), \label{eqn:v-photon}
\end{equation}

\noindent where $\nu$ can be defined in an invariant form:

\begin{equation}
   \nu \equiv p \cdot q/M,  \label{eqn:nu}
\end{equation}

\noindent with $M$ being the mass of the nucleon.  In the target rest frame, $p = (M,\vec{0})$,
so Eq.~\ref{eqn:nu} reduces to $\nu = E - E'$, and is known as the {\it electron energy loss}.
The fractional energy loss, $y$, can be defined in an invariant form:  

\begin{equation}
   y \equiv \frac{p \cdot q}{q \cdot k}, 
\end{equation}

\noindent which simplifies to the non-invariant form $y = (E-E')/E$. 

The four-momentum transfer squared, $q^2$, always evaluates to less than or equal to zero.  
For convenience, we define a positive quantity $Q^{2}$:

\begin{equation}
   Q^{2} \equiv -q^{2} = 4EE'\sin^{2}\left(\theta/2\right), \label{eqn:Qsq}
\end{equation}

\noindent where $\theta$ is the scattering angle of the electron in the laboratory frame
and we have neglected the electron mass.

Shifting our focus to the hadronic side of Fig.~\ref{fig:e-N-scat}, there are 
two possibilities for the final state: there is one object (i.e., the target remains intact) 
or several, determined by the energy with which the target is probed.  Furthermore, the 
measured interaction may be described by two general terms: {\it exclusive} or 
{\it inclusive} scattering.  In the case of exclusive scattering, the scattered 
electron {\it and} all final-state hadrons are detected; there is also the case of 
detecting an electron and at least one particle in the final state, which is 
called {\it semi-inclusive}.  For inclusive scattering, only the scattered electron is 
detected in the final state. 

Inclusive scattering can be represented as $eN \rightarrow eX$, where $e$ is the electron, $N$
is the target nucleon and $X$ is the final (unmeasured) hadronic state.  In the context
of an unmeasured final hadronic state (which could consist of any of the
multitude of particle states for a given combination of $\nu$ and $Q^{2}$ values), 
we can define the invariant mass of the system, $W$:

\begin{equation}
   W^{2} \equiv \left( q + p \right)^2 = M^2 + 2M\nu - Q^{2}. \label{eqn:w} 
\end{equation}

Finally, we come to the variable $x$.  It is defined in terms of the other invariants $\nu$
and $Q^{2}$ as:

\begin{equation}
   x \equiv \frac{Q^{2}}{2p \cdot q} = \frac{Q^{2}}{2M\nu}. \label{eqn:x-bj}
\end{equation}

\noindent  The simplest interpretation of $x$ comes in the infinite momentum frame, where
the nucleon is traveling with a large momentum along $\vec{q}$.  In this frame, the active
quark in the interaction (struck by the virtual photon) carries the momentum fraction $x$
of the nucleon momentum in the leading-order DIS process~\cite{HalzenMartin}.

The DIS region is characterized by $W > 2$\,GeV, where $\nu$ and $Q^{2}$ become 
large enough so that the quarks can be resolved inside the nucleon.  In this case, 
the electron is scattering from an asymptotically free quark (or anti-quark) in the nucleon.

Consider scattering unpolarized electrons from pointlike, unpolarized spin-1/2
particles that are infinitely heavy with a charge of $+1$.  In this case, energy
conservation dictates $E' = E$ and the cross section is given by:

\begin{equation} \label{eqn:mott-xs}
   \left( \frac{d\sigma}{d\Omega}\right)_{\text{Mott}} = \frac{\alpha^{2}\cos^{2}\left( \theta/2\right)}{4E^{2}\sin^{4}\left(\theta/2\right)}, 
\end{equation}

\noindent with $\theta$ being the scattering angle of the electron.  This quantity
is known as the {\it Mott cross section}.  However, since the nucleon is a composite
object and is not infinitely massive, the cross section is more complicated than
that seen in Eq.~\ref{eqn:mott-xs}, and is given by:

\begin{eqnarray} \label{eqn:unpol-xs-f-sf}
   \frac{d^{2}\sigma^{\text{unpol.}}}{d\Omega dE'} &=& \left( \frac{d\sigma}{d\Omega}\right)_{\text{Mott}}
                                                \Bigg[ \frac{1}{\nu}F_{2}\left(x,Q^{2}\right) \nonumber \\
                                            &+& \frac{2 \tan^{2}\left(\theta/2\right)}{M}F_{1}\left(x,Q^{2}\right)\Bigg],    
\end{eqnarray}

\noindent where $F_{1}$ and $F_{2}$ are the unpolarized structure functions, 
and are related to one another through Eq.~\ref{eqn:f1-f2}. 
 
For experiments that use targets that are not nucleons ($A \neq 1$),
there are two conventions for expressing the quantities $F_{1}$ and $F_{2}$.
The first is {\it per nucleon}, written as $F_{1}/A$ and $F_{2}/A$.  The second
is {\it per nucleus}, where the structure functions are reported without dividing
by $A$.  The latter representation is used in this paper.

When both the incident electron beam and target are polarized, one can 
access the spin structure functions $g_{1}$ and $g_{2}$.  A full discussion
may be found in Ref.~\cite{Anselmino:1994gn}.  The polarized cross section
difference for when the target spin (double arrows) and electron spin (single arrows) 
are polarized along the direction of the electron momentum is given as: 
 
 \begin{eqnarray}
    \frac{d^{2}\sigma^{\downarrow,\Uparrow}}{d\Omega dE'} -\frac{d^{2}\sigma^{\uparrow,\Uparrow}}{d\Omega dE'} &=& 
    \frac{4\alpha^{2}}{Q^{2}}\frac{E'}{\nu E} 
    \big[ \left( E + E'\cos\theta \right)g_{1}\left(x,Q^{2} \right) \nonumber \\ 
    &-& 2Mx g_{2}\left(x,Q^{2} \right) \big]. 
 \end{eqnarray}

\noindent When the target spin is perpendicular to the electron spin, 
the cross section difference is written as:

\begin{eqnarray}
   \frac{d^{2}\sigma^{\downarrow,\Rightarrow}}{d\Omega dE'} -\frac{d^{2}\sigma^{\uparrow,\Rightarrow}}{d\Omega dE'} &=& 
   \frac{4\alpha^{2}}{Q^{2}}\frac{E'^{2}}{\nu E}\sin\theta 
   \big[ g_{1}\left(x,Q^{2}\right) \nonumber \\ 
   &+& \frac{2ME}{\nu}g_{2}\left(x,Q^{2} \right) \big]. 
\end{eqnarray}

%===============================================================================
\subsection{Bjorken scaling, the quark-parton model and scaling violation}
%===============================================================================

When probing an object of finite size, the measurement will depend upon the
spatial resolution of the probe; in the case of electron scattering, this is $Q^{2}$,
the negative of the momentum transferred to the target squared.  If we increase 
$Q^{2}$ so that we can resolve the internal structure of the nucleon, the quarks 
will become visible.  At this point, inelastic electron-nucleon scattering 
may be seen as elastic scattering from a single quark, while the other 
quarks remain undisturbed.  Considering that quarks are pointlike particles,
increasing the resolution $Q^{2}$ will no longer affect the interaction.

In the limit where $Q^{2}$ $\rightarrow \infty$ and $\nu \rightarrow \infty$,
with $x = Q^{2}/(2M\nu)$ fixed (the Bjorken limit), the phenomenon where experimental
observables lose their $Q^{2}$-dependence is known as {\it Bjorken scaling}~\cite{Bjorken:1969ja}.
As a result, the structure functions depend upon a single variable $x$.  Furthermore,
the $F_{2}$ structure function can be related to the $F_{1}$ structure function
by the {\it Callan-Gross relation}~\cite{Callan:1969uq}:

\begin{equation} \label{eqn:callan-gross}
   F_2(x) = 2x F_1(x).  
\end{equation}

To connect the quark behavior to the structure functions, we turn to the quark-parton 
model (QPM).  In this model, deep inelastic scattering of electrons from nucleons is 
described as the incoherent scattering of electrons from free partons (quarks and anti-quarks)
inside the nucleon~\cite{Feynman:1969ej}, via the exchange of a virtual photon.
Therefore, the nucleon structure functions $F_{1}$ and $g_{1}$ can be written in terms of the parton
distribution functions (PDFs)~\cite{Feynman:1969ej,HalzenMartin}: 

\begin{eqnarray} 
   F_{1}\left(x\right) &=& \frac{1}{2} \sum_{i} e_{i}^{2} q_{i}\left(x\right)         \label{eqn:f1_qpm} \\ 
   g_{1}\left(x\right) &=& \frac{1}{2} \sum_{i} e_{i}^{2} \Delta q_{i}\left(x\right), \label{eqn:g1_qpm} 
\end{eqnarray}

\noindent with $q = q^{\uparrow}(x) + q^{\downarrow}(x)$ and 
$\Delta q = q^{\uparrow}(x) - q^{\downarrow}(x)$, where $\uparrow$ ($\downarrow$) indicates
quark spin parallel (antiparallel) to the parent nucleon spin. 

The scaling behavior of the structure functions is only exact in the limit of infinite $Q^{2}$ and $\nu$.  
At finite values of $Q^{2}$ and $\nu$, it is only an approximation.  In reality, 
the quarks participating in the interaction with the electron may radiate gluons 
before or after scattering.  Such processes result in an infinite cross section, 
and can only be treated properly when all other processes of the same order are 
considered.  These gluonic radiative corrections result in the cross section 
acquiring a logarithmic $Q^{2}$ dependence, which can be computed exactly 
in pQCD under the formalism of the Dokshitzer-Gribov-Lipatov-Altarelli-Parisi (DGLAP)
evolution equations~\cite{Dokshitzer:1977sg,*Gribov:1972ri,*Altarelli:1977zs}.  As a 
result, the $Q^{2}$-dependence manifests itself in the structure functions. 

As a result of the scaling violation, we re-cast the PDFs and the structure functions
in terms of both $x$ and $Q^{2}$.  In particular, the definition of the PDF is now 
$q^{\uparrow (\downarrow)}(x,Q^{2})$: this is the probability of finding a quark 
$q$ with its spin parallel (antiparallel) to its parent nucleon with 
momentum fraction $x$ {\it when viewed at an energy scale} $Q = \sqrt{Q^{2}}$.

The physical interpretation tied to scaling violation is that structure functions
at low $Q^{2}$ are dominated by three {\it valence} quarks ``dressed'' by sea
quarks (manifesting as $q$-$\bar{q}$ pairs) and gluons.  As $Q^{2}$ is increased,
the resolving power increases, allowing for sensitivity to the ``bare'' quarks  and gluons
which make up the nucleon.

%===============================================================================
\subsection{The resonance region} \label{sec:res_app} 
%===============================================================================

Due to the kinematics of our experiment, about half of our data set corresponds
to the DIS regime, while the other corresponds to the {\it resonance} region.
With this in mind, we give a brief description of the resonance region.  

When $\nu$ and $Q^{2}$ have values such that $1.2 < W < 2$\,GeV, we explore 
the substructure of the nucleon.  In this energy range, the quarks that make up 
the nucleon collectively absorb the energy of the virtual photon, leading to unstable
excited states of the nucleon called {\it nucleon resonances}.  The most prominent 
resonance occurs at $W = 1.232$\,GeV, and is known as the $\Delta$ resonance.  
Higher resonances are also possible at $W > 1.4$\,GeV, but are difficult to discern 
from one another, as their peaks and tails tend to overlap.

%===============================================================================
\subsection{The virtual photon-nucleon asymmetry $A_{1}$} \label{sec:a1_app}
%===============================================================================

Let us consider a scattering interaction in which the nucleon is longitudinally polarized, while the
virtual photon is circularly polarized with a helicity of $\pm 1$.  As a result, two possible helicity-dependent 
cross sections for a given nucleon polarization arise, denoted $\sigma_{3/2}$
and $\sigma_{1/2}$.  The subscripts denote the projection of the total spin of the virtual
photon-nucleon system along the direction of the virtual photon momentum~\cite{Abe:1998wq,XZhengThesis}.
When the virtual photon spin is parallel (antiparallel) to the nucleon spin, they add to 3/2 (1/2). 
From these two cross sections, the $A_{1}$ asymmetry is formed as:

\begin{equation} \label{eqn:a1_def} 
   A_{1} \equiv \frac{\sigma_{1/2} - \sigma_{3/2}}{\sigma_{1/2} + \sigma_{3/2}}. 
\end{equation}

\noindent In terms of the structure functions, $A_{1}$ may be written as~\cite{Anselmino:1994gn}: 

\begin{equation}
   A_{1}\left(x,Q^{2}\right) = \frac{g_{1}\left(x,Q^{2}\right) - \gamma^{2}g_{2}\left(x,Q^{2}\right)}{F_{1}\left(x,Q^{2}\right)}. \label{eqn:a1-sf}  
\end{equation} 

\noindent At large $Q^{2}$, $A_{1} \approx g_{1}/F_{1}$.  This can be seen by
observing that $\gamma^{2} \rightarrow 0$ as $Q^{2}$ gets increasingly large.  
A conceptual argument on the quark level is as follows: if the spin of the 
virtual photon is antiparallel to that of the quark, then the virtual
photon can be absorbed and the quark spin is flipped; however, if the spins are
{\it parallel}, then the absorption of the virtual photon is forbidden, since the
total projection of the spins along $\vec{q}$ is 3/2 and the quark is a spin-1/2
particle.  The mathematical form of the approximation can be illustrated using
this physical interpretation: for the case where the spins of the nucleon and 
virtual photon are parallel ($\sigma_{3/2}$), then the quark that {\it can} 
absorb the virtual photon has its spin antiparallel to the nucleon spin.
This translates to: $\sigma_{3/2} \sim \sum\limits_{i} e_{i}^2 q_{i}^{\downarrow}(x)$.
A similar argument may be made for the $\sigma_{1/2}$ case where only quarks with
spins parallel to the parent nucleon can absorb virtual photons. Thus, we have:
$\sigma_{1/2} \sim \sum\limits_{i} e_{i}^2 q_{i}^{\uparrow}(x)$.  Rewriting
$A_{1}$ in terms of these approximations, we obtain:

\begin{equation}
   A_{1} \sim \frac{\sum\limits_{i} e_{i}^2 \left[ q_{i}^{\uparrow}(x) - q_{i}^{\downarrow}(x)\right]}{
                    \sum\limits_{i} e_{i}^2 \left[ q_{i}^{\uparrow}(x) + q_{i}^{\downarrow}(x)\right]} 
           =  \frac{\sum\limits_{i} e_{i}^2 \Delta q_{i}(x)}{\sum\limits_{i} e_{i}^2 q_{i}(x)} = \frac{g_{1}(x)}{F_{1}(x)}, 
\end{equation}

\noindent where the last term arises from the quark-parton model description
of the $F_{1}$ and $g_{1}$ structure functions (Eqs.~\ref{eqn:f1_qpm} and~\ref{eqn:g1_qpm}).   

The $A_{1}$ asymmetry is a ratio of structure functions ($\approx g_{1}/F_{1}$),
and as a result there is very little $Q^{2}$ dependence.  This is because
$g_{1}$ and $F_{1}$ follow the same $Q^{2}$ evolution described by the DGLAP
equations~\cite{Dokshitzer:1977sg,*Gribov:1972ri,*Altarelli:1977zs} which tends 
to cancel in the ratio, leading to $A_{1}$ being roughly $Q^{2}$-independent.

%===============================================================================
\subsection{Electron asymmetries} \label{sec:electron_asym_app} 
%===============================================================================

The virtual photon-nucleon asymmetry $A_{1}$ is defined in terms of a ratio of the 
difference in virtual photon cross sections to their sum.  Due to the difficulty 
associated with aligning the virtual photon spin along the direction of the nucleon 
spin, another approach is utilized to measure $A_{1}$ that consists of aligning the 
{\it incident electron} spin relative to the direction of the nucleon spin.  The 
extraction of the {\it electron asymmetries} allows the determination of $A_{1}$.  
After some algebra, the electron asymmetries (Section~\ref{sec:g2-intro}) may be
written as~\cite{Melnitchouk:2005zr}:

 \begin{eqnarray}
    A_{\parallel} &=& D\left( A_1 + \eta A_2 \right) \label{eqn:a_para_a1a2} \\ 
    A_{\perp}     &=& d\left( A_2 - \eta A_1 \right). \label{eqn:a_perp_a1a2}  
 \end{eqnarray}

\noindent The asymmetry $A_{2}$ is defined as $A_{2} \equiv 2\sigma_{LT}/\left( \sigma_{1/2} + \sigma_{3/2}\right)$,
where $\sigma_{LT}$ is the cross section describing the interference between virtual photons with longitudinal
and transverse polarizations.  In the QPM, there is no clear interpretation for $\sigma_{LT}$, 
and in turn $A_{2}$, as there is for $\sigma_{1/2}$, $\sigma_{3/2}$ and $A_{1}$~\cite{Drechsel:2000ct}.  
From Eqs.~\ref{eqn:a_para_a1a2} and~\ref{eqn:a_perp_a1a2}, one can write $A_{1}$ and $A_{2}$ in terms of the double-spin 
asymmetries $A_{\parallel}$ and $A_{\perp}$.  Similar equations may be obtained for $g_{1}/F_{1}$
and $g_{2}/F_{1}$~\cite{Melnitchouk:2005zr}.  The equations for $A_{1}$ and $g_{1}/F_{1}$ 
are shown in Section~\ref{sec:a1-intro}.  

%===============================================================================
\subsection{Spin structure functions} \label{sec:spin-structure-funcs-app} 
%===============================================================================

The spin structure functions $g_{1}$ and $g_{2}$ may be obtained from the measured
unpolarized cross section $\sigma_{0}$ and the double-spin asymmetries $A_{\parallel}$
and $A_{\perp}$ through: 

\begin{eqnarray}
  g_1 &=& \frac{MQ^2}{4{\alpha}^2}\frac{2y}{\left( 1 - y\right) \left( 2 - y\right)}\sigma_0 
          \Big[ A_{\parallel}  \\ 
      & & + \tan \left(\theta/2 \right)A_{\perp}\Big] \nonumber  \label{eqn:g1_exp} \\
  g_2 &=& \frac{MQ^2}{4{\alpha}^2}\frac{y^2}{\left( 1 - y\right) \left( 2 - y\right)}\sigma_0 \times  \\ 
      & & \left[ - A_{\parallel} + \frac{1 + \left( 1 - y \right)\cos \theta}
          {\left( 1 - y \right)\sin \theta}A_{\perp}\right] .  \nonumber \label{eqn:g2_exp}
\end{eqnarray}

      %===============================================================================
\section{Quark-gluon correlations}  \label{sec:ope_app} 
\subsection{The operator product expansion and twist}
%===============================================================================

The operator product expansion (OPE) allows the separation of the perturbative
and non-perturbative components in structure functions at finite $Q^{2}$.  This concept
is illustrated in the product of two {\it local} quark (or gluon) operators 
$\mathcal{O}_{a}(d)\mathcal{O}_{b}$ separated by a distance $d$ in the limit of 
$d \rightarrow 0$:

\begin{equation} \label{eqn:ope_space}
   \lim_{d \rightarrow 0} \mathcal{O}_{a}(d)\mathcal{O}_b(0) = \sum\limits_{k}c_{abk}(d) \mathcal{O}_{k}(0), 
\end{equation}

\noindent where the coefficient functions $c_{abk}$ are the Wilson coefficients
and contain the perturbative part, which can be computed using perturbation theory 
since non-perturbative effects occur at distances much larger than $d$~\cite{Manohar:1992tz}.  
The non-perturbative components manifest in $\mathcal{O}_{k}(0)$, and contribute to 
the cross section on the order of $x^{-n}(Q/M)^{D-2-n}$.  The exponents $n$ and $D$ 
are the spin and (mass) dimension of the operator, respectively.  The quantity 
$Q = \sqrt{Q^{2}}$.  The {\it twist} $\tau$ of the operator is defined by:

\begin{equation} \label{eqn:twist} 
   \tau \equiv D - n. 
\end{equation}

\noindent At large $Q^{2}$, $\tau = 2$ terms dominate in the OPE; at low $Q^{2}$,
higher-twist $(\tau > 2)$ operators become important.  

%===============================================================================
\subsection{Cornwall-Norton moments and Nachtmann moments}
%===============================================================================

Using the OPE, an infinite set of sum rules may be derived under a twist expansion
of the spin structure functions $g_{1}$ and $g_{2}$~\cite{Anselmino:1994gn}. 
Such expansions of $g_{1}$ and $g_{2}$ are known as the Cornwall-Norton (CN) 
moments~\cite{Melnitchouk:2005zr}: 

\begin{eqnarray}
   \int_{0}^{1} x^{n-1} g_{1}\left(x,Q^{2}\right) dx  &=& \frac{1}{2}a_{n-1}, \text{  } n = 1, 3, 5,\ldots \label{eqn:g1_exp}   \\ 
   \int_{0}^{1} x^{n-1} g_{2}\left(x,Q^{2}\right) dx  &=& \frac{n-1}{2n}\left( d_{n-1} - a_{n-1}\right), \label{eqn:g2_exp} \\ 
                                                      & & n = 3, 5, 7,\ldots, \nonumber  
\end{eqnarray} 

\noindent where $n$ indicates the $n^{\text{th}}$ moment.  In Eqs~\ref{eqn:g1_exp} and~\ref{eqn:g2_exp}, 
only twist-2 and twist-3 contributions are considered. The quantities
$a_{n-1}$ and $d_{n-1}$ represent the twist-2 and twist-3 matrix elements, 
respectively~\footnote{The convention used by~\cite{Melnitchouk:2005zr} is such that the matrix 
elements are labeled according to $n$ as opposed to $n-1$, as defined in this paper.}.
The expansions are only over odd integers, which is a result of the symmetry 
of the structure functions under charge conjugation~\cite{KSliferThesis}. 

The twist-3 matrix elements $d_{n-1}$ may be accessed by combining Eqs.~\ref{eqn:g1_exp} 
and~\ref{eqn:g2_exp}.  One obtains~\cite{Melnitchouk:2005zr}: 

\begin{equation}
   \int_{0}^{1} x^{n-1} \left[ g_{1}\left(x,Q^{2}\right) + \frac{n}{n-1} g_{2}\left(x,Q^{2}\right) \right] dx 
   = \frac{d_{n-1}}{2}, \quad n \geq 3. 
\end{equation} 

\noindent Choosing $n = 3$ yields the equation for $d_{2}$ (cf. Eq.~\ref{eqn:d2-sf-intro}).

The study of higher twist in structure functions has traditionally been done through
the formalism of the CN moments; however, the exact relation of the CN moments
to the dynamical higher-twist contributions has come into question in recent
analyses~\cite{Melnitchouk:2005zr,Dong:2008zza}.  It is argued that the CN moments 
are only valid when the terms connected to the finite mass of the nucleon are neglected.  
Such terms are known as {\it target mass corrections}.  These corrections are related
to twist-2 operators, and are of order $\mathcal{O}(M^{2}/Q^{2})$.  Analogous 
to the CN moments, the Nachtmann moments $M_{1}$ and $M_{2}$ can be used to separate
the higher-twist contribution from the target mass 
corrections~\cite{Nachtmann:1973mr,Wandzura:1977ce,Georgi:1976ve,Matsuda:1979ad,Sidorov:2006fi,Accardi:2008pc}. 
They are defined as~\cite{Melnitchouk:2005zr,Piccione:1997zh,Slifer:2008xu}:

\begin{eqnarray}
   M_{1}^{n}\left(Q^{2}\right) &\equiv& \frac{1}{2}a_{n-1} = \frac{1}{2}\tilde{a}_{n}E_{1}^{n} \\ 
                           &=&      \int_{0}^{1} \frac{\xi^{n+1}}{x^{2}} \Bigg[ \frac{x}{\xi} 
                                    - \frac{n^{2}}{(n+2)^{2}} \frac{M^{2}x\xi}{Q^{2}}g_{1}\left(x,Q^{2}\right)   \nonumber \\
                                    &-& \frac{4n}{n+2}\frac{M^{2}x^{2}}{Q^{2}}g_{2}\left(x,Q^{2}\right)\Bigg] dx, \nonumber \\
                                    & & n = 1,3,\ldots \nonumber \\ 
   M_{2}^{n}\left(Q^{2}\right) &\equiv& \frac{1}{2}d_{n-1} = \frac{1}{2}\tilde{d}_{n}E_{2}^{n} \\ 
                               &=& \int_{0}^{1} \frac{\xi^{n+1}}{x^{2}} \Bigg[ \frac{x}{\xi}g_{1}\left(x,Q^{2}\right) \nonumber \\ 
                               &+& \left(\frac{n}{n-1}\frac{x^{2}}{\xi^{2}} 
                               -   \frac{n}{n+1}\frac{M^{2}x^{2}}{Q^{2}}\right)g_{2}\left(x,Q^{2}\right)\Bigg] dx, \nonumber \\
                               & & n = 3,5,\ldots. \nonumber 
\end{eqnarray}  

We have performed the analysis to extract $d_{2}^{n}$ according to the CN moments,
and checked that result against what was obtained from using the Nachtmann moments.  
The difference in results between the two approaches was found to be negligible,
on the order of $10^{-5}$ in absolute value.

      %===============================================================================
\section{Fits to data}  \label{sec:fits-app} 
\subsection{Cross section fits} \label{sec:xs-fits-app}
%===============================================================================

As discussed in Section~\ref{sec:xs-ana}, due to time constraints not all cross section 
measurements for studying pair production on $^{3}$He and N$_{2}$ were carried out.
To resolve the absence of those data, the data that were collected were fit to 
a function of the form:

\begin{equation}
   f \left( E_{p} \right) = \frac{1}{E_{p}^{2}}e^{\left( a_{0} + a_{1} E_{p} \right)}, 
\end{equation}

\noindent where the scattered electron energy $E_{p}$ is in GeV.  This was done for
$\sigma^{e^{+}}$, $\sigma_{\text{N}_{2}}^{e^{-}}$ and $\sigma_{\text{N}}^{e^{+}}$.  
The fits were performed in ROOT~\cite{ROOT}, and the extracted parameters and their 
uncertainties were obtained from the Minuit minimization package~\cite{James:1975dr}.
The fits were done separately for the data sets corresponding to each beam energy.  
The systematic uncertainties on the fits was obtained by varying the parameters 
within their errors and observing the change on the fit (see Figs~\ref{fig:exp_xs_4} 
and~\ref{fig:exp_xs_5}).  The parameters together with their uncertainties are 
listed in Table~\ref{tab:neg-nitro-xs-fit-pars} for the nitrogen cross section 
when scattered electrons were detected, $\sigma_{\text{N}_{\text{2}}}^{e^{-}}$ 
(LHRS set to negative polarity mode), and in Table~\ref{tab:pos-nitro-xs-fit-pars} 
when positrons were detected, $\sigma_{\text{N}_{\text{2}}}^{e^{+}}$ (LHRS in positive polarity mode). 
Table~\ref{tab:positron-xs-fit-pars} gives the fit parameters and their uncertainties 
for the unpolarized $^{3}$He cross section where positrons were detected, $\sigma^{e^{+}}$ 
(LHRS in positive polarity mode).

\begin{table}[hbt]
\centering
\caption{Fit parameters for the nitrogen cross section (negative polarity), $\sigma_{\text{N}_{\text{2}}}^{e^{-}}$, 
         for $E = 4.74$\,GeV and $E = 5.89$\,GeV.}
\label{tab:neg-nitro-xs-fit-pars}
\begin{ruledtabular} 
\begin{tabular}{crr}
% \hline
 \multicolumn{1}{c}{Par.}               &
 \multicolumn{1}{c}{$E = 4.74$\,GeV }   &
 \multicolumn{1}{c}{$E = 5.89$\,GeV }   \\
 \hline
 $a_{0}$    & $ 1.465\text{E+}01 \pm 4.919\text{E-}02$  & $ 1.480\text{E+}01 \pm 5.647\text{E-}02$ \\ 
 $a_{1}$    & $-1.825\text{E-}03 \pm 4.770\text{E-}05$  & $-2.123\text{E-}03 \pm 5.607\text{E-}05$ \\ 
% \hline
\end{tabular}
\end{ruledtabular} 
\end{table}

\begin{table}[hbt]
\centering
\caption{Fit parameters for the nitrogen cross section (positive polarity), $\sigma_{\text{N}_{\text{2}}}^{e^{+}}$, 
         for $E = 4.74$\,GeV and $E = 5.89$\,GeV.}
\label{tab:pos-nitro-xs-fit-pars}
\begin{ruledtabular} 
\begin{tabular}{crr}
% \hline
 \multicolumn{1}{c}{Par.}               &
 \multicolumn{1}{c}{$E = 4.74$\,GeV }   &
 \multicolumn{1}{c}{$E = 5.89$\,GeV }  \\ 
 \hline
 $a_{0}$    & $ 1.559\text{E+}01 \pm 1.604\text{E-}01$ & $ 1.614\text{E+}01 \pm 2.120\text{E-}01 $\\ 
 $a_{1}$    & $-4.699\text{E-}03 \pm 2.255\text{E-}04$ & $-5.232\text{E-}03 \pm 3.126\text{E-}04 $\\ 
% \hline
\end{tabular}
\end{ruledtabular} 
\end{table}

\begin{table}[hbt]
\centering
\caption{Fit parameters for the positron cross section on $^{3}$He, $\sigma^{e^{+}}$, 
         for $E = 4.74$\,GeV and $E = 5.89$\,GeV.}
\label{tab:positron-xs-fit-pars}
\begin{ruledtabular} 
\begin{tabular}{crr}
% \hline
 \multicolumn{1}{c}{Par.}               &
 \multicolumn{1}{c}{$E = 4.74$\,GeV }   &
 \multicolumn{1}{c}{$E = 5.89$\,GeV }  \\ 
 \hline
 $a_{0}$    & $ 1.887\text{E+}01 \pm 7.998\text{E-}02$ & $ 1.896\text{E+}01 \pm 6.899\text{E-}02$ \\ 
 $a_{1}$    & $-5.620\text{E-}03 \pm 1.194\text{E-}04$ & $-5.421\text{E-}03 \pm 9.286\text{E-}05$ \\ 
% \hline
\end{tabular}
\end{ruledtabular} 
\end{table}

%===============================================================================
\subsection{$g_{1}^{p}/F_{1}^{p}$ fit} \label{sec:g1F1p-fit}
%===============================================================================

To carry out the analysis to obtain $g_{1}^{n}/F_{1}^{n}$ from our $g_{1}^{^{3}\text{He}}/F_{1}^{^{3}\text{He}}$ 
data, a parameterization of the $g_{1}^{p}/F_{1}^{p}$ data was needed.  We fit 
the world data to a three-parameter, $Q^{2}$-independent function given by:

\begin{equation}
   f(x) = p_{0} + p_{1}x + p_{2}x^{2}.   \label{eqn:asym-fit-func}
\end{equation}

\begin{figure}[!ht]
   \centering
   \includegraphics[width=0.5\textwidth]{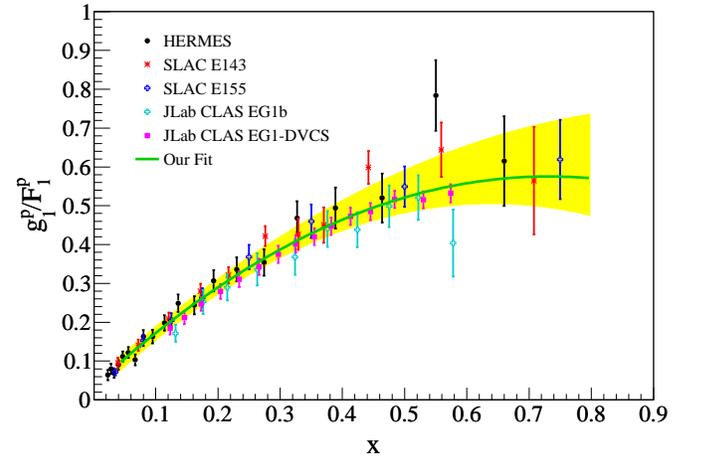}
   \caption{(Color online) Our fit to world $g_{1}^{p}/F_{1}^{p}$ data.  The error bars on the data
            are the in-quadrature sums of their statistical and systematic
            uncertainties.  The band indicates the error on the fit.}
   \label{fig:g1p-fit}
\end{figure}

\noindent The assumption of $Q^{2}$-independence is reasonable as the $Q^{2}$-evolution 
in $g_{1}$ and $F_{1}$ partially cancels in the ratio to leading order 
and next-to-leading order in $Q^{2}$~\cite{Anselmino:1994gn}.  Additionally,
the world data (which are at differing $Q^{2}$) show roughly the same behavior.
The world data considered were from HERMES~\cite{Ackerstaff:1997ws}, SLAC E143~\cite{Abe:1998wq} 
and E155~\cite{Anthony:1999py}, along with CLAS EG1b~\cite{Dharmawardane:2006zd}
and CLAS EG1-DVCS~\cite{Prok:2014ltt}.  
First, all of these data were rebinned in $x$, in new bins formed based on a 
statistical-error-weighted average; the systematic errors of all data contributing 
to a new bin were averaged with equal weights to yield its systematic error.  
The fit result, with $\chi^{2}/\text{ndf} = 0.91$, is shown in Fig.~\ref{fig:g1p-fit}
The fit parameters were found to be:

\begin{eqnarray}
   p_{0} &=&  0.035 \pm 0.008 \nonumber  \\ 
   p_{1} &=&  1.478 \pm 0.077            \\ 
   p_{2} &=& -1.010 \pm 0.138. \nonumber 
\end{eqnarray}

\noindent The fit was performed in ROOT~\cite{ROOT}, and the extracted parameters and
their uncertainties were obtained from the Minuit minimization package~\cite{James:1975dr}. 
The band indicates the uncertainty on the fit which was taken as the spread in the 
$g_{1}^{p}/F_{1}^{p}$ data, serving as a conservative estimate. 

%===============================================================================
\subsection{$A_{1}^{p}$ fit}  \label{sec:a1p-fit} 
%===============================================================================

\begin{figure}[!ht]
   \centering
   \includegraphics[width=0.5\textwidth]{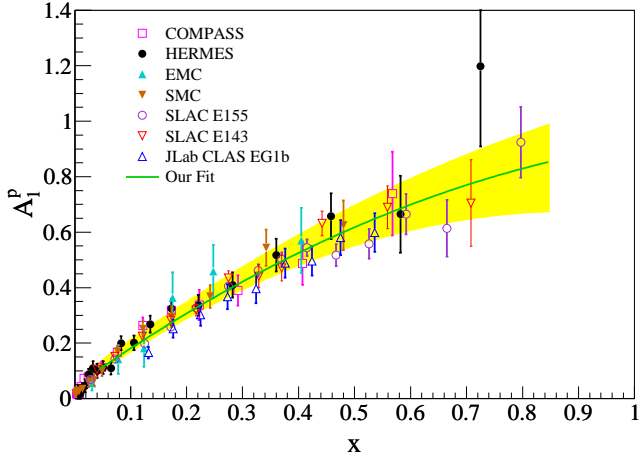}
   \caption{(Color online) Our fit to world $A_{1}^{p}$ data.  The error bars on the data
            are the in-quadrature sums of their statistical and systematic
            uncertainties.  The band indicates the uncertainty on the fit.}
   \label{fig:a1p-fit}
\end{figure}

To obtain $A_{1}^{n}$ from our $A_{1}^{^{3}\text{He}}$ data, we followed a similar 
procedure to the $g_{1}/F_{1}$ analysis. 
The data used in the fit include measurements from SMC~\cite{Adeva:1998vv}, 
HERMES~\cite{Airapetian:2006vy}, EMC~\cite{Ashman:1987hv,Ashman:1989ig}, 
SLAC E143~\cite{Abe:1998wq} and E155~\cite{Anthony:1999py}, COMPASS~\cite{Alekseev:2010hc} 
and CLAS EG1b~\cite{Dharmawardane:2006zd}.  The fit is shown in Fig.~\ref{fig:a1p-fit}, 
where we obtained $\chi^{2}/\text{ndf} = 1.11$.  The fit parameters were found to be: 

\begin{eqnarray}
   p_{0} &=&  0.044 \pm 0.007  \nonumber  \\ 
   p_{1} &=&  1.423 \pm 0.078             \\ 
   p_{2} &=& -0.552 \pm 0.158. \nonumber 
\end{eqnarray}

\noindent The fit was performed in ROOT~\cite{ROOT}, and the extracted parameters and 
their uncertainties were obtained from the Minuit minimization package~\cite{James:1975dr}. 
The band in Fig.~\ref{fig:a1p-fit} gives the fit uncertainty, computed in
the same fashion as was done for the $g_{1}^{p}/F_{1}^{p}$ fit.

      %===============================================================================
\section{World $\Gamma_{1}^{n}$ Data} \label{sec:gamma1-world-app}
%===============================================================================

\begin{figure}[!ht]
   \includegraphics[width=0.5\textwidth]{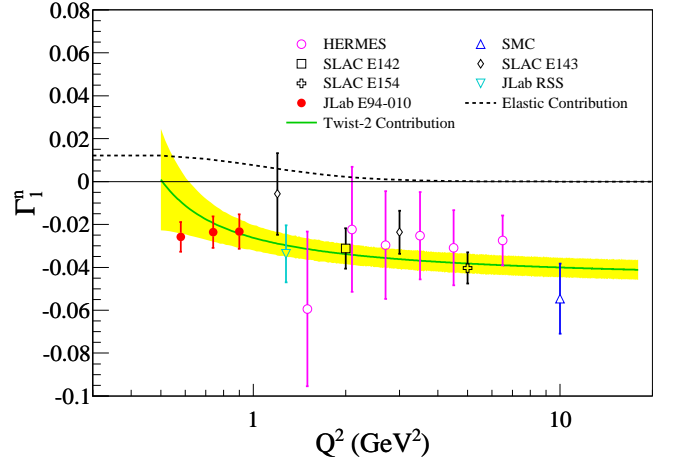}
   \caption{(Color online) $\Gamma_{1}^{n}$ for the world data from SLAC E142~\protect\cite{Anthony:1996mw},
            E143~\protect\cite{Abe:1998wq} and E154~\protect\cite{Abe:1997cx}, SMC~\protect\cite{Adeva:1998vv},
            HERMES~\protect\cite{Airapetian:2002wd}, and JLab RSS~\protect\cite{Slifer:2008xu} and E94-010~\protect\cite{Slifer:2008re}.
            The uncertainties on the world data are the in-quadrature sum of statistical and systematic uncertainties.
            The elastic contribution is given by the dashed curve, and has been added to the data.
            The twist-2 contribution is indicated by the solid curve and its uncertainty is given by the band,
            which is dominated by the uncertainty in $\alpha_{s}$.
   }
   \label{fig:gamma1}
\end{figure}

The world $\Gamma_{1}^{n}$ data from SLAC E142~\cite{Anthony:1996mw}, E143~\cite{Abe:1998wq} 
and E154~\cite{Abe:1997cx}, SMC~\cite{Adeva:1998vv}, HERMES~\cite{Airapetian:2002wd}, 
and JLab RSS~\cite{Slifer:2008xu} and E94-010~\cite{Slifer:2008re} used in our 
$f_{2}^{n}$ analysis are shown in Fig.~\ref{fig:gamma1}.  The twist-2 contribution, 
$\mu_{2}^{n}$, given by the solid curve; its uncertainty is indicated by the band.
The elastic contribution is also shown, given as the dashed curve.  For more
details, see Section~\ref{sec:res_cf}. 

\begin{figure}[!ht]
   \includegraphics[width=0.5\textwidth]{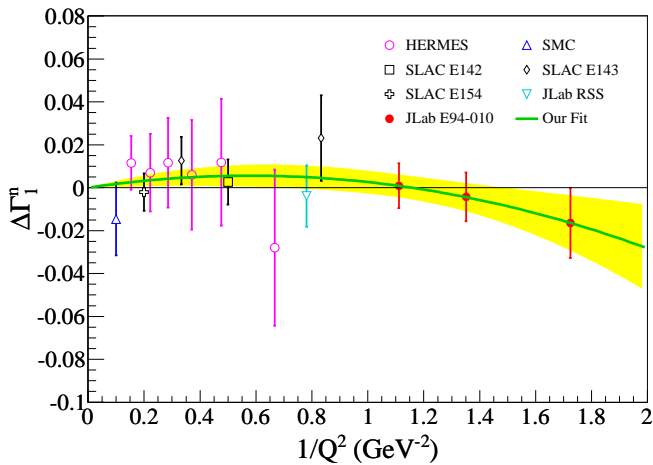}
   \caption{(Color online) $\Delta\Gamma_{1}^{n} \equiv \Gamma_{1}^{n} - \mu_{2}^{n}$ for the world data
            shown in Fig.~\ref{fig:gamma1}.  The uncertainties on the world data are the in-quadrature 
            sum of statistical and systematic uncertainties.  
            Our fit is indicated by the solid curve and its uncertainty is given by the yellow band. 
   }
   \label{fig:delta-gamma1} 
\end{figure}

Subtracting $\mu_{2}^{n}$ from $\Gamma_{1}^{n}$ yields the higher-twist 
contribution, $\Delta \Gamma_{1}^{n} \equiv \Gamma_{1}^{n} - \mu_{2}^{n}$, 
shown in Fig.~\ref{fig:delta-gamma1}.  Fitting these data to the function

\begin{equation} \label{eqn:f2_fit}
   f\left( \frac{1}{Q^{2}} \right)  = \frac{A}{Q^{2}} + \frac{B}{Q^{4}}
\end{equation}

\noindent where $A = \left(M_{n}^{2}/9\right)\left( a_{2}^{n} + 4d_{2}^{n} + 4f_{2}^{n}\right)$ and 
$B = \mu_{6}$, a higher-twist ($\tau > 4$) term, are free parameters.  Using the 
result for the fit parameter $A$, we extract $f_{2}^{n}$ after inserting the average 
$a_{2}^{n}$ from global analyses and our measured $d_{2}^{n}$ (see Section~\ref{sec:res_cf}) 
into Eq.~\ref{eqn:f2_fit}.  The values of the fit parameters $A$ and $B$ were found to be:  

\begin{eqnarray}
   A &=&  1.936\text{E-}2 \nonumber \\ 
   B &=& -1.675\text{E-}2,          
\end{eqnarray}   

\noindent with the error matrix: 

\begin{equation}
   \varepsilon = 
                 \left(
                    \begin{array}{c c} 
                        2.240\text{E-}4 & -1.653\text{E-}4 \\
                       -1.653\text{E-}4 &  1.351\text{E-}4 
                    \end{array} 
                 \right).
\end{equation}

\noindent The fit was performed in ROOT~\cite{ROOT}, and the extracted parameters and
their uncertainties were obtained from the Minuit minimization package~\cite{James:1975dr}.
This fit allowed the extraction of $f_{2}^{n}$ as described in Section~\ref{sec:res_cf}.  

      %===============================================================================
\section{Systematic Uncertainty Tables}  \label{sec:syst-err-tables}
%===============================================================================

This section contains tables of showing the contributing factors to the systematic 
uncertainties for the final unpolarized cross sections and double-spin asymmetries, 
$d_{2}^{n}$, $g_{1}^{^{3}\text{He}}$, $g_{2}^{^{3}\text{He}}$, $A_{1}^{^{3}\text{He}}$, 
$A_{1}^{n}$, $g_{1}^{^{3}\text{He}}/F_{1}^{^{3}\text{He}}$, $g_{1}^{n}/F_{1}^{n}$ 
and the flavor-separated ratios $(\Delta u + \Delta \bar{u})/(u + \bar{u})$
and $(\Delta d + \Delta \bar{d})/(d + \bar{d})$.  

%===============================================================================
\subsection{Final unpolarized cross section systematic uncertainties} \label{sec:xs-syst-err-tables}
%===============================================================================

The breakdown of the systematic uncertainty on the unpolarized cross sections 
is given in Tables~\ref{tab:xs-syst-err-4} and~\ref{tab:xs-syst-err-5}. 
The column ``Cuts'' indicates uncertainties due to event cuts, which includes the gas \v{C}erenkov,
$E/p$, and target cuts to remove the target windows; ``Background'' corresponds 
to uncertainties related to the positron and nitrogen background subtractions; ``Misc.'' 
refers to the uncertainties incurred from the beam charge calibration, and the density 
of nitrogen and $^{3}$He in the target cell; ``RC'' is the uncertainty due to the 
radiative corrections, as explained in Section~\ref{sec:xs_rc}.  The final column 
(``Total'') gives the in-quadrature sum of all uncertainties in the row. 

\begin{table}[hbt]
\centering
\caption{Systematic uncertainty breakdown for the unpolarized $^{3}$He cross section
         at E = 4.74\,GeV.  All uncertainties are in nb/GeV/sr.  
         See Appendix~\ref{sec:xs-syst-err-tables} for a discussion of the various
         contributions to the systematic uncertainty. 
         }
\label{tab:xs-syst-err-4}
\begin{ruledtabular}
\begin{tabular}{cccccc}
 $<x>$   & Cuts  & Background & Misc.  & RC     & Total \\
\hline
 0.214 &  1.700E-01 &  3.800E-01 &  2.440E-01 &  2.870E-01 &  5.610E-01 \\
 0.299 &  1.390E-01 &  1.170E-01 &  2.120E-01 &  2.400E-02 &  2.810E-01 \\
 0.456 &  6.400E-02 &  1.700E-02 &  1.000E-01 &  5.000E-03 &  1.210E-01 \\
 0.494 &  5.300E-02 &  1.200E-02 &  8.800E-02 &  1.000E-03 &  1.030E-01 \\
 0.533 &  4.700E-02 &  8.000E-03 &  7.000E-02 &  0.000E+00 &  8.400E-02 \\
 0.579 &  3.600E-02 &  6.000E-03 &  5.300E-02 &  2.000E-03 &  6.500E-02 \\
 0.629 &  2.800E-02 &  4.000E-03 &  4.000E-02 &  4.000E-03 &  5.000E-02 \\
 0.686 &  2.000E-02 &  3.000E-03 &  2.800E-02 &  5.000E-03 &  3.500E-02 \\
 0.745 &  1.500E-02 &  3.000E-03 &  2.100E-02 &  9.000E-03 &  2.800E-02 \\
\end{tabular}
\end{ruledtabular}
\end{table}

\begin{table}[hbt]
\centering
\caption{Systematic uncertainty breakdown for the unpolarized $^{3}$He cross section
         at E = 5.89\,GeV.  All uncertainties are in nb/GeV/sr.  
         See Appendix~\ref{sec:xs-syst-err-tables} for a discussion of the various
         contributions to the systematic uncertainty.
         }
\label{tab:xs-syst-err-5}
\begin{ruledtabular}
\begin{tabular}{cccccc}
 $<x>$    & Cuts  & Background & Misc.  & RC    & Total \\
\hline
 0.208 &  1.060E-01 &  3.710E-01 &  1.610E-01 &  2.600E-01 &  4.920E-01 \\
 0.247 &  1.120E-01 &  2.280E-01 &  1.710E-01 &  4.700E-02 &  3.100E-01 \\
 0.330 &  5.600E-02 &  6.100E-02 &  9.800E-02 &  1.700E-02 &  1.300E-01 \\
 0.434 &  4.400E-02 &  1.700E-02 &  6.300E-02 &  4.000E-03 &  7.900E-02 \\
 0.468 &  3.800E-02 &  1.100E-02 &  4.900E-02 &  4.000E-03 &  6.300E-02 \\
 0.503 &  2.900E-02 &  8.000E-03 &  4.200E-02 &  1.000E-03 &  5.200E-02 \\
 0.539 &  2.600E-02 &  6.000E-03 &  3.300E-02 &  0.000E+00 &  4.200E-02 \\
 0.580 &  1.800E-02 &  4.000E-03 &  2.700E-02 &  1.000E-03 &  3.300E-02 \\
 0.629 &  1.100E-02 &  3.000E-03 &  1.900E-02 &  2.000E-03 &  2.200E-02 \\
 0.679 &  9.000E-03 &  2.000E-03 &  1.300E-02 &  2.000E-03 &  1.600E-02 \\
 0.738 &  7.000E-03 &  2.000E-03 &  1.000E-02 &  4.000E-03 &  1.300E-02 \\
\end{tabular}
\end{ruledtabular}
\end{table}

%===============================================================================
\subsection{Final asymmetry systematic uncertainties} \label{sec:asym-syst-err-tables}
%===============================================================================

Tables~\ref{tab:AparSys_4},~\ref{tab:AperpSys_5},~\ref{tab:AparSys_5} and~\ref{tab:AperpSys_5} 
list the systematic uncertainties assigned to the double-spin physics asymmetries 
$A_{\parallel}$ and $A_{\perp}$ on $^{3}$He.  The systematic uncertainty
depends on the electron beam polarization $P_b$, the target polarization $P_b$,
the nitrogen dilution factor $D_{\text{N}_2}$, and contaminations in the BigBite
analysis due to $\pi^{-}$ ($f_1$), $\pi^{+}$ ($f_2$), and $e^{+}$ ($f_3$).  
Also there are contributions from the electron selection cuts (``Cuts'') and the 
radiative corrections (``RC'').  The final column (``Total'') is the in-quadrature 
sum of the uncertainties in each row.  Columns for all quantities except for 
the PID cuts have been omitted since they were very small for the lowest $x$-bins, 
and were negligible otherwise. 

\begin{table}[hbt]
\centering
\caption{Systematic uncertainties assigned to $A_{\parallel}^{^{3}\text{He}}$ at an incident beam energy of 4.74\,GeV.
         See Appendix~\ref{sec:asym-syst-err-tables} for a discussion of the 
         various contributions to the systematic uncertainty.}
\label{tab:AparSys_4}
\begin{ruledtabular} 
\begin{tabular}{ccc}
$<x>$   & Cuts    & Total   \\
\hline
0.277&  7.000E-03&      7.000E-03\\
0.325&  2.000E-03&      3.000E-03\\
0.374&  2.000E-03&      2.000E-03\\
0.424&  4.000E-03&      5.000E-03\\
0.473&  3.000E-03&      3.000E-03\\
0.523&  2.000E-03&      2.000E-03\\
0.574&  4.000E-03&      4.000E-03\\
0.623&  2.000E-03&      3.000E-03\\
0.673&  4.000E-03&      5.000E-03\\
0.723&  6.000E-03&      6.000E-03\\
0.773&  1.300E-02&      1.300E-02\\
0.823&  1.100E-02&      1.100E-02\\
0.874&  1.700E-02&      1.700E-02\\
\end{tabular}
\end{ruledtabular} 
\end{table}

\begin{table}[hbt]
\centering
\caption{Systematic uncertainties assigned to $A_{\perp}^{^{3}\text{He}}$ at an incident beam energy of 4.74\,GeV
        See Appendix~\ref{sec:asym-syst-err-tables} for a discussion of the 
        various contributions to the systematic uncertainty.}
\label{tab:AperpSys_4}
\begin{ruledtabular} 
\begin{tabular}{ccc}
$<x>$   & Cuts    & Total\\
\hline
0.277&  2.000E-03&      2.000E-03\\
0.325&  2.000E-03&      2.000E-03\\
0.374&  2.000E-03&      2.000E-03\\
0.424&  2.000E-03&      2.000E-03\\
0.473&  1.000E-03&      1.000E-03\\
0.523&  1.000E-03&      1.000E-03\\
0.574&  2.000E-03&      2.000E-03\\
0.623&  2.000E-03&      2.000E-03\\
0.673&  3.000E-03&      3.000E-03\\
0.723&  2.000E-03&      2.000E-03\\
0.773&  4.000E-03&      5.000E-03\\
0.823&  6.000E-03&      6.000E-03\\
0.874&  4.000E-03&      6.000E-03\\
\end{tabular}
\end{ruledtabular} 
\end{table}

\begin{table}[hbt]
\center
\caption{Systematic uncertainties assigned to $A_{\parallel}^{^{3}\text{He}}$ at an incident beam energy of 5.89\,GeV 
         See Appendix~\ref{sec:asym-syst-err-tables} for a discussion of the 
         various contributions to the systematic uncertainty.}
\label{tab:AparSys_5}
\begin{ruledtabular} 
\begin{tabular}{ccc}
$<x>$   & Cuts    & Total   \\
\hline
0.277&  9.000E-03&      1.000E-02\\
0.325&  3.000E-03&      3.000E-03\\
0.374&  2.000E-03&      2.000E-03\\
0.424&  2.000E-03&      3.000E-03\\
0.474&  6.000E-03&      6.000E-03\\
0.524&  2.000E-03&      2.000E-03\\
0.573&  3.000E-03&      3.000E-03\\
0.624&  5.000E-03&      5.000E-03\\
0.674&  5.000E-03&      5.000E-03\\
0.723&  5.000E-03&      5.000E-03\\
0.773&  8.000E-03&      8.000E-03\\
0.823&  1.400E-02&      1.400E-02\\
0.873&  1.700E-02&      1.700E-02\\
\end{tabular}
\end{ruledtabular}
\end{table}
 
\begin{table}[hbt]
\centering
\caption{Systematic uncertainties assigned to $A_{\perp}^{^{3}\text{He}}$ at an incident beam energy of 5.89\,GeV 
         See Appendix~\ref{sec:asym-syst-err-tables} for a discussion of the 
         various contributions to the systematic uncertainty.}
\label{tab:AperpSys_5}
\begin{ruledtabular} 
\begin{tabular}{ccc}
$<x>$   & Cuts      & Total   \\
\hline
0.277&  2.000E-03&      3.000E-03\\
0.325&  1.000E-03&      1.000E-03\\
0.374&  0.000E+00&      0.000E+00\\
0.424&  1.000E-03&      1.000E-03\\
0.474&  1.000E-03&      1.000E-03\\
0.524&  1.000E-03&      1.000E-03\\
0.573&  1.000E-03&      1.000E-03\\
0.624&  1.000E-03&      1.000E-03\\
0.674&  2.000E-03&      2.000E-03\\
0.723&  3.000E-03&      3.000E-03\\
0.773&  2.000E-03&      2.000E-03\\
0.823&  3.000E-03&      4.000E-03\\
0.873&  7.000E-03&      7.000E-03\\
\end{tabular}
\end{ruledtabular}
\end{table}

%===============================================================================
\subsection{Polarized spin structure function systematic uncertainties} \label{sec:sf-syst-err}
%===============================================================================

Tables~\ref{tab:g1Sys_4},~\ref{tab:g1Sys_5},~\ref{tab:g2Sys_4} and~\ref{tab:g2Sys_5} list 
the systematic uncertainties assigned to the polarized spin-structure functions.  A number 
of factors contribute to the systematic uncertainty, including kinematics (``Kin.'') 
and all of the systematic uncertainties found to contribute to the final 
asymmetries, namely $P_{b}$, $P_{t}$, $f_{1}$, $f_{2}$ $f_{3}$, PID cuts (``Cuts'') 
and radiative corrections (``RC'').  Other contributions include the systematic 
uncertainty on the unpolarized cross section ($\sigma_{0}$), in addition to the 
uncertainties associated with the interpolation or extrapolation of the data where 
necessary.  The value for each these uncertainties was determined by varying each of these 
contributions within reasonable limits and taking the corresponding change in the 
observable (either $g_{1}^{^3\text{He}}$ or $g_{2}^{^3\text{He}}$) as the uncertainty.
The last column (``Total'') is the in-quadrature sum of the uncertainties 
in each row.  All uncertainties are absolute.  Columns for all quantities except for 
the PID cuts have been omitted since they were very small for the lowest $x$-bins, 
and were negligible otherwise. 

\begin{table}[hbt]
\center
\caption{Systematic uncertainties assigned to $g_{1}^{^{3}\text{He}}$ at an incident beam energy of 4.74\,GeV
         See Appendix~\ref{sec:sf-syst-err} for a discussion of the 
         various contributions to the systematic uncertainty.}
\label{tab:g1Sys_4}
\begin{ruledtabular} 
\begin{tabular}{ccc}
$<x>$  & Cuts           & Total\\
\hline
0.277&  8.000E-03&      9.000E-03\\
0.325&  2.000E-03&      2.000E-03\\
0.374&  2.000E-03&      2.000E-03\\
0.424&  3.000E-03&      3.000E-03\\
0.473&  1.000E-03&      2.000E-03\\
0.523&  0.000E+00&      0.000E+00\\
0.574&  1.000E-03&      1.000E-03\\
0.623&  0.000E+00&      1.000E-03\\
0.673&  1.000E-03&      1.000E-03\\
0.723&  0.000E+00&      1.000E-03\\
0.773&  1.000E-03&      1.000E-03\\
0.823&  1.000E-03&      1.000E-03\\
0.874&  0.000E+00&      1.000E-03\\

\end{tabular}
\end{ruledtabular}
\end{table}

\begin{table}[hbt]
\centering
\caption{Systematic uncertainties assigned to $g_{2}^{^{3}\text{He}}$ at an incident beam energy of 4.74\,GeV
         See Appendix~\ref{sec:sf-syst-err} for a discussion of the 
         various contributions to the systematic uncertainty.}
\label{tab:g2Sys_4}
\begin{ruledtabular}
\begin{tabular}{ccc}
$<x>$   & Cuts    & Total\\
\hline
0.277&  9.000E-03&      9.000E-03\\
0.325&  6.000E-03&      6.000E-03\\
0.374&  4.000E-03&      5.000E-03\\
0.424&  2.000E-03&      2.000E-03\\
0.473&  0.000E+00&      1.000E-03\\
0.523&  1.000E-03&      1.000E-03\\
0.574&  1.000E-03&      1.000E-03\\
0.623&  1.000E-03&      1.000E-03\\
0.673&  1.000E-03&      1.000E-03\\
0.723&  0.000E+00&      1.000E-03\\
0.773&  0.000E+00&      1.000E-03\\
0.823&  0.000E+00&      0.000E+00\\
0.874&  0.000E+00&      1.000E-03\\
\end{tabular}
\end{ruledtabular}
\end{table}
 
\begin{table}[hbt]
\centering
\caption{Systematic uncertainties assigned to $g_{1}^{^{3}\text{He}}$ at an incident beam energy of 5.89\,GeV 
         See Appendix~\ref{sec:sf-syst-err} for a discussion of the 
         various contributions to the systematic uncertainty.}
\label{tab:g1Sys_5}
\begin{ruledtabular} 
\begin{tabular}{ccc}
$<x>$   & Cuts    & Total\\
\hline
0.277   &1.000E-02      &1.200E-02      \\
0.325   &2.000E-03      &3.000E-03      \\
0.374   &2.000E-03      &2.000E-03      \\
0.424   &1.000E-03      &2.000E-03      \\
0.474   &2.000E-03      &2.000E-03      \\
0.524   &1.000E-03      &1.000E-03      \\
0.573   &1.000E-03      &1.000E-03      \\
0.624   &1.000E-03      &1.000E-03      \\
0.674   &0.000E+00      &0.000E+00      \\
0.723   &0.000E+00      &0.000E+00      \\
0.773   &0.000E+00      &0.000E+00      \\
0.823   &0.000E+00      &0.000E+00      \\
0.873   &0.000E+00      &0.000E+00      \\
\end{tabular}
\end{ruledtabular}
\end{table}

\begin{table}[hbt]
\centering
\caption{Systematic uncertainties assigned to $g_{2}^{^{3}\text{He}}$ at an incident beam energy of 5.89\,GeV 
         See Appendix~\ref{sec:sf-syst-err} for a discussion of the 
         various contributions to the systematic uncertainty.}
\label{tab:g2Sys_5}
\begin{ruledtabular} 
\begin{tabular}{ccc}
$<x>$   & Cuts    & Total   \\
\hline
0.277&  1.400E-02&      1.900E-02\\
0.325&  5.000E-03&      6.000E-03\\
0.374&  1.000E-03&      2.000E-03\\
0.424&  2.000E-03&      2.000E-03\\
0.474&  2.000E-03&      2.000E-03\\
0.524&  1.000E-03&      1.000E-03\\
0.573&  1.000E-03&      1.000E-03\\
0.624&  0.000E+00&      0.000E+00\\
0.674&  0.000E+00&      0.000E+00\\
0.723&  0.000E+00&      0.000E+00\\
0.773&  0.000E+00&      0.000E+00\\
0.823&  0.000E+00&      1.000E-03\\
0.873&  0.000E+00&      0.000E+00\\
\end{tabular}
\end{ruledtabular}
\end{table}

%===============================================================================
\subsection{$d_{2}^{n}$ systematic uncertainties} \label{sec:d2-syst-err}
%===============================================================================

A breakdown of the $d_2^n$ systematic uncertainties is given in Table~\ref{tab:d2n_sys}, 
for each of the measured mean $Q^2$ points. This table includes the effects of all the 
uncertainties found in the preceding tables (i.e. $P_p$, $P_t$, etc.), referred to as 
experimental systematics (``Exp.''), in addition to radiative corrections, $d_2^p$, the proton 
and neutron polarizations ($\tilde{P}_p$ and $\tilde{P}_n$), and the unmeasured low $x$ contributions.
The value for each these uncertainties was determined by varying each of these 
contributions within reasonable limits and taking the corresponding change in $d_{2}^{n}$ 
as the uncertainty.  The two sources of uncertainty that dominate the $d_2^n$ 
systematic uncertainty are those from the experimental and low-$x$ contributions.  
However, the final $d_2^n$ measurement's statistical uncertainty is larger than its 
systematic uncertainty.

\begin{table*}[hbt] 
\centering
\caption{Systematic uncertainties assigned to different regions of $d_2^n$.  ``Res'' 
         indicates the contribution from the resonance region; in particular, from 
         data with $W < 2$\,GeV.  ``DIS'' represents the contribution due to data for $W > 2$\,GeV.  
         See Appendix~\ref{sec:d2-syst-err} for a discussion of the various contributions 
         to the systematic uncertainty. 
        }
\label{tab:d2n_sys}
\begin{ruledtabular} 
\begin{tabular}{ccccccccc}
Region  & $\left< Q^2 \right>$ $\left(\text{GeV}^{\text{2}}\right)$ & Exp. & RC & $d_2^p$ & $\tilde{P}_p$, $\tilde{P}_n$ (high error) 
& $\tilde{P}_p$, $\tilde{P}_n$ (low error) & Low $x$ & Total\\
\hline
DIS+Res & 3.21& 4.700E-04       &2.000E-05      &4.000E-05      &6.000E-05      &1.000E-04      &5.800E-04      &7.500E-04\\
DIS+Res & 4.32& 3.700E-04       &2.000E-05      &4.000E-05      &2.000E-05      &4.000E-05      &5.800E-04      &6.900E-04\\
\hline
DIS & 2.59&     3.600E-04       &1.000E-05      &5.000E-05      &3.000E-05      &5.000E-05      &---            &3.700E-04\\
DIS & 3.67&     2.900E-04       &2.000E-05      &4.000E-05      &2.000E-05      &2.000E-05      &---            &2.900E-04\\
\hline
Res &4.71&      2.200E-04       &0.000E+00      &4.000E-05      &3.000E-05      &5.000E-05      &---            &2.300E-04\\
Res &5.99&      1.100E-04       &0.000E+00      &2.000E-05      &1.000E-05      &3.000E-05      &---            &1.200E-04\\

\end{tabular}
\end{ruledtabular}
\end{table*}

%===============================================================================
\subsection{$a_{2}$ systematic uncertainties} \label{sec:a2-syst-err}
%===============================================================================

The systematic uncertainties for the measured $a_{2}^{^{3}\text{He}}$ are given in 
Table~\ref{tab:a2he3-syst-err}, where the column labeled $g_{1}^{^{3}\text{He}}$ 
corresponds to the uncertainty due to our $g_{1}^{^{3}\text{He}}$ data, and the column 
labeled $x$ is the uncertainty due to $x$ in the integration.  The value of these 
uncertainties was determined by varying each of these contributions within reasonable 
limits and taking the change in the $a_{2}^{n}$ as the uncertainty.  The in-quadrature sum of 
the two contributions is given as the column labeled ``Total.''  The systematic uncertainties 
for the $a_{2}^{n}$ extraction for the full $x$-range are presented in Table~\ref{tab:a2n-syst-err}.  
The columns labeled low-$x$ (high-$x$) correspond to the uncertainties due to the low-$x$ 
(high-$x$) regions.  The uncertainty  due to the effective proton (neutron) polarization 
is given by the column labeled $\tilde{P}_{p}$ ($\tilde{P}_{n}$).  The uncertainties due to 
$a_{2}^{p}$ and our measured $a_{2}^{^{3}\text{He}}$ are also given.  

\begin{table}[!ht]
\caption{The systematic uncertainties contributing to the $a_{2}^{^{3}\text{He}}$ result
         in the measured $x$-range.  See Appendix~\ref{sec:a2-syst-err} for 
         a discussion of the various contributions to the systematic uncertainty.
         }
\label{tab:a2he3-syst-err}
\center
\begin{ruledtabular}
\begin{tabular}{cccc}
$\left<Q^2\right>$ $\left(\text{GeV}^{\text{2}}\right)$ & $g_{1}^{^{3}\text{He}}$ & $x$ & Total \\
\hline
3.21 & 3.428E-05 & 8.803E-06 & 3.539E-05 \\ 
4.32 & 3.281E-05 & 6.681E-06 & 3.348E-05 \\ 
\end{tabular}
\end{ruledtabular}
\end{table}

\begin{table*}[!ht]
\caption{The systematic uncertainties contributing to the $a_{2}^{n}$ result over the
         full $x$-range. See Appendix~\ref{sec:a2-syst-err} for 
         a discussion of the various contributions to the systematic uncertainty listed below.
        }
\label{tab:a2n-syst-err}
\center
\begin{ruledtabular}
\begin{tabular}{cccccccc}
$\left<Q^2\right>$ $\left(\text{GeV}^{\text{2}}\right)$ & low-$x$ & high-$x$ 
& $\tilde{P}_{p}$ & $\tilde{P}_{n}$ & $a_{2}^{p}$ & $a_{2}^{^{3}\text{He}}$ & Total \\
\hline
3.21 & 1.374E-4 & 1.373E-04 & 2.439E-04 & 3.012E-04 & 3.068E-04 & 3.052E-04 & 6.125E-04 \\ 
4.32 & 8.878E-5 & 8.880E-05 & 2.518E-04 & 3.054E-04 & 3.118E-04 & 3.089E-04 & 6.042E-04 \\ 
\end{tabular}
\end{ruledtabular}
\end{table*}

%===============================================================================
\subsection{$A_{1}$ and $g_{1}/F_{1}$ systematic uncertainties} \label{sec:a1-g1f1-syst-err}
%===============================================================================

This section discusses the breakdown of the systematic uncertainties on $A_{1}^{^{3}\text{He}}$
and $A_{1}^{n}$.  The main factors that contribute to the uncertainties on the $^{3}$He data
are the physics asymmetries $A_{\parallel}$ and $A_{\perp}$, and the kinematic
factors $D$, $\eta$, $\xi$ and $d$ (see Section~\ref{sec:a1-intro}). 
Each physics asymmetry was varied within its uncertainty, and the change in $A_{1}^{^{3}\text{He}}$ was observed.
For the kinematics, the low-level variables of the electron momentum $p$ and scattering
angle $\theta$ were changed within their relative uncertainties of $1\%$ and $1.4\%$~\cite{DFlayThesis}
respectively, and the kinematic factors were re-evaluated, and the change in the $A_{1}$
asymmetry was observed.  The resulting contributions to the systematic uncertainty in $A_{1}^{^{3}\text{He}}$
are listed in Tables~\ref{tab:a1he3-syst-err-4} and~\ref{tab:a1he3-syst-err-5} for the 
4.74\,GeV and 5.89\,GeV runs, respectively. 

The systematic uncertainties for $A_{1}^{n}$ are listed in in Tables~\ref{tab:a1n-syst-err-4} 
and~\ref{tab:a1n-syst-err-5}. The inputs in the $A_{1}^{n}$ extraction that were varied consisted of 
$F_{2}^{n}$ and $F_{2}^{p}$, $F_{2}^{^{3}\text{He}}$, $A_{1}^{p}$ and our $A_{1}^{^{3}\text{He}}$
data.  For the neutron and proton $F_{2}$, various 
models~\cite{CTEQ6,Owens:2012bv,Bourrely:2001du,*Bourrely:2007if} were compared, and the 
largest difference in $A_{1}^{n}$ was taken as the uncertainty, listed in the $F_{2}^{n,p}$ 
column. The same procedure was used for $F_{2}^{^{3}\text{He}}$, where the models considered 
were F1F209~\cite{Bosted:2012qc} and NMC95~\cite{Arneodo:1995cq}.  This uncertainty is given 
in the $F_{2}^{^{3}\text{He}}$ column.  Our fit of the world $A_{1}^{p}$ data was varied within 
its uncertainty and the change in $A_{1}^{n}$ was taken as the uncertainty, listed in the 
$A_{1}^{p}$ column.  The values for the effective neutron (proton) polarization 
$\tilde{P}_{n}$ ($\tilde{P}_{p}$) were varied within their uncertainties, and the change in $A_{1}^{n}$ 
was taken as the uncertainty.  These contributions are listed as $\tilde{P}_{n}$ and 
$\tilde{P}_{p}$ for the neutron and proton, respectively.  We varied our $A_{1}^{^{3}\text{He}}$ data 
within their systematic uncertainties and observed the changes in the $A_{1}^{n}$ results, 
which were taken as the uncertainties, listed in the $A_{1}^{^{3}\text{He}}$ column.  
The in-quadrature sum of all contributions is given in the column labeled ``Total''.    

To evaluate the systematic uncertainties on $g_{1}/F_{1}$ data, shown in 
Tables~\ref{tab:g1f1he3-syst-err-4},~\ref{tab:g1f1he3-syst-err-5},~\ref{tab:g1f1n-syst-err-4} 
and~\ref{tab:g1f1n-syst-err-5}, the same procedure used for the $A_{1}$ data was applied.  
The same models for $F_{2}$ on the neutron, proton and $^{3}$He were used, in addition 
to the effective polarizations of the neutron and proton, $\tilde{P}_{n}$ and $\tilde{P}_{p}$.    

\clearpage

\begin{table*}[hbt]
\centering
\caption{Systematic uncertainties for $A_{1}^{^{3}\text{He}}$ data at E = 4.74\,GeV.
         See Appendix~\ref{sec:a1-g1f1-syst-err} for a discussion of the various
         contributions to the systematic uncertainty.}
\label{tab:a1he3-syst-err-4}
\begin{ruledtabular}
\begin{tabular}{ccccc}
 $<x>$ & $A_{\parallel}$ & $A_{\perp}$ &  Kin.   &  Total  \\  
\hline
 0.277 &  3.850E-03 &  1.900E-04 &  1.200E-04 &  3.860E-03 \\
 0.325 &  1.480E-03 &  1.400E-04 &  7.000E-05 &  1.480E-03 \\
 0.374 &  9.900E-04 &  2.100E-04 &  5.400E-04 &  1.150E-03 \\
 0.424 &  2.650E-03 &  1.700E-04 &  1.200E-04 &  2.660E-03 \\
 0.473 &  1.840E-03 &  1.100E-04 &  1.900E-04 &  1.850E-03 \\
 0.523 &  1.100E-03 &  1.600E-04 &  1.900E-04 &  1.130E-03 \\
 0.574 &  2.430E-03 &  3.300E-04 &  2.000E-04 &  2.460E-03 \\
 0.623 &  1.840E-03 &  3.600E-04 &  1.400E-04 &  1.880E-03 \\
 0.673 &  2.620E-03 &  5.000E-04 &  1.300E-04 &  2.670E-03 \\
 0.723 &  3.770E-03 &  4.400E-04 &  2.000E-04 &  3.800E-03 \\
 0.773 &  7.510E-03 &  8.900E-04 &  3.900E-04 &  7.570E-03 \\
 0.823 &  7.000E-03 &  1.180E-03 &  3.200E-04 &  7.100E-03 \\
 0.874 &  1.019E-02 &  1.250E-03 &  1.960E-03 &  1.045E-02 \\
\end{tabular}
\end{ruledtabular}
\end{table*}

\begin{table*}[hbt]
\centering
\caption{Systematic uncertainties for $A_{1}^{^{3}\text{He}}$ data at E = 5.89\,GeV.
         See Appendix~\ref{sec:a1-g1f1-syst-err} for a discussion of the various
         contributions to the systematic uncertainty.
        }
\label{tab:a1he3-syst-err-5}
\begin{ruledtabular}
\begin{tabular}{ccccc}
 $<x>$ & $A_{\parallel}$ & $A_{\perp}$ &  Kin.   &  Total  \\  
\hline
 0.277 &  5.520E-03 &  1.800E-04 &  5.900E-04 &  5.550E-03 \\
 0.325 &  1.850E-03 &  9.000E-05 &  2.400E-04 &  1.870E-03 \\
 0.374 &  1.230E-03 &  4.000E-05 &  4.000E-05 &  1.230E-03 \\
 0.424 &  1.510E-03 &  1.100E-04 &  2.200E-04 &  1.530E-03 \\
 0.474 &  3.270E-03 &  9.000E-05 &  4.000E-05 &  3.270E-03 \\
 0.524 &  1.200E-03 &  1.200E-04 &  1.000E-05 &  1.210E-03 \\
 0.573 &  1.800E-03 &  1.400E-04 &  1.300E-04 &  1.810E-03 \\
 0.624 &  3.120E-03 &  1.000E-04 &  1.600E-04 &  3.130E-03 \\
 0.674 &  2.770E-03 &  2.700E-04 &  7.000E-05 &  2.780E-03 \\
 0.723 &  2.740E-03 &  3.900E-04 &  1.300E-04 &  2.770E-03 \\
 0.773 &  4.430E-03 &  3.500E-04 &  1.100E-04 &  4.450E-03 \\
 0.823 &  7.870E-03 &  7.100E-04 &  6.600E-04 &  7.930E-03 \\
 0.873 &  9.270E-03 &  1.170E-03 &  1.400E-04 &  9.340E-03 \\
\end{tabular}
\end{ruledtabular}
\end{table*}

\begin{table*}[hbt]
\centering
\caption{Systematic uncertainties for $A_{1}^{n}$ at E = 4.74\,GeV.
         See Appendix~\ref{sec:a1-g1f1-syst-err} for a discussion of the various
         contributions to the systematic uncertainty.
        }
\label{tab:a1n-syst-err-4}
\begin{ruledtabular}
\begin{tabular}{cccccccc}
 $<x>$ & $F_{2}^{n,p}$  &  $F_{2}^{^{3}\text{He}}$ &  $\tilde{P}_{p}$  
&  $\tilde{P}_{n}$ & $A_{1}^{p}$ &  $A_{1}^{^{3}\text{He}}$ &  Total  \\  
\hline
 0.277 &  2.730E-03 &  7.000E-04 &  5.160E-03 &  1.500E-04 &  5.080E-03 &  2.680E-03 &  8.220E-03 \\
 0.325 &  2.690E-03 &  1.010E-03 &  6.160E-03 &  1.200E-04 &  6.050E-03 &  1.850E-03 &  9.280E-03 \\
 0.374 &  6.900E-03 &  6.500E-04 &  7.200E-03 &  1.690E-03 &  6.690E-03 &  6.370E-03 &  1.371E-02 \\
 0.424 &  5.980E-03 &  1.810E-03 &  8.310E-03 &  1.190E-03 &  7.710E-03 &  6.390E-03 &  1.449E-02 \\
 0.473 &  5.740E-03 &  1.710E-03 &  9.570E-03 &  8.500E-04 &  8.960E-03 &  5.280E-03 &  1.537E-02 \\
\end{tabular}
\end{ruledtabular}
\end{table*}

\begin{table*}[hbt]
\centering
\caption{Systematic uncertainties for $A_{1}^{n}$ at E = 5.89\,GeV.
         See Appendix~\ref{sec:a1-g1f1-syst-err} for a discussion of the various
         contributions to the systematic uncertainty.
        }
\label{tab:a1n-syst-err-5}
\begin{ruledtabular}
\begin{tabular}{cccccccc}
 $<x>$ & $F_{2}^{n,p}$  &  $F_{2}^{^{3}\text{He}}$ &  $\tilde{P}_{p}$  
&  $\tilde{P}_{n}$ & $A_{1}^{p}$ &  $A_{1}^{^{3}\text{He}}$ &  Total  \\  
\hline
 0.277 &  4.500E-03 &  4.000E-04 &  5.140E-03 &  2.070E-03 &  5.130E-03 &  3.370E-02 &  3.483E-02 \\
 0.325 &  2.440E-03 &  1.310E-03 &  6.140E-03 &  5.900E-04 &  5.810E-03 &  3.060E-03 &  9.430E-03 \\
 0.374 &  2.650E-03 &  6.400E-04 &  7.190E-03 &  5.200E-04 &  6.700E-03 &  1.470E-03 &  1.032E-02 \\
 0.424 &  2.900E-03 &  2.120E-03 &  8.300E-03 &  7.400E-04 &  7.940E-03 &  3.740E-03 &  1.263E-02 \\
 0.474 &  3.940E-03 &  1.350E-03 &  9.560E-03 &  8.400E-04 &  9.240E-03 &  9.750E-03 &  1.702E-02 \\
 0.524 &  9.950E-03 &  0.000E+00 &  1.096E-02 &  1.720E-03 &  1.060E-02 &  5.050E-03 &  1.897E-02 \\
 0.573 &  1.300E-02 &  1.500E-04 &  1.247E-02 &  2.140E-03 &  1.207E-02 &  8.000E-03 &  2.321E-02 \\
\end{tabular}
\end{ruledtabular}
\end{table*}

\begin{table*}[hbt]
\centering
\caption{Systematic uncertainties for $g_{1}^{^{3}\text{He}}/F_{1}^{^{3}\text{He}}$ data at E = 4.74\,GeV.
         See Appendix~\ref{sec:a1-g1f1-syst-err} for a discussion of the various
         contributions to the systematic uncertainty.
        }
\label{tab:g1f1he3-syst-err-4}
\begin{ruledtabular}
\begin{tabular}{ccccc}
 $<x>$ & $A_{\parallel}$ & $A_{\perp}$ &  Kin.   &  Total  \\  
\hline
 0.277 &  3.680E-03 &  5.300E-04 &  4.000E-05 &  3.720E-03 \\
 0.325 &  1.340E-03 &  3.500E-04 &  4.000E-05 &  1.380E-03 \\
 0.374 &  9.200E-04 &  4.900E-04 &  4.000E-05 &  1.040E-03 \\
 0.424 &  2.560E-03 &  3.300E-04 &  1.300E-04 &  2.580E-03 \\
 0.473 &  1.710E-03 &  2.000E-04 &  1.100E-04 &  1.720E-03 \\
 0.523 &  1.030E-03 &  2.600E-04 &  2.000E-05 &  1.060E-03 \\
 0.574 &  2.200E-03 &  4.700E-04 &  3.000E-05 &  2.260E-03 \\
 0.623 &  1.610E-03 &  5.000E-04 &  1.700E-04 &  1.690E-03 \\
 0.673 &  2.430E-03 &  6.700E-04 &  1.400E-04 &  2.530E-03 \\
 0.723 &  3.350E-03 &  5.100E-04 &  2.000E-04 &  3.390E-03 \\
 0.773 &  6.410E-03 &  9.700E-04 &  1.600E-04 &  6.490E-03 \\
 0.823 &  5.760E-03 &  1.260E-03 &  2.800E-04 &  5.900E-03 \\
 0.874 &  8.590E-03 &  1.230E-03 &  3.200E-04 &  8.680E-03 \\
\end{tabular}
\end{ruledtabular}
\end{table*}

\begin{table*}[hbt]
\centering
\caption{Systematic uncertainties for $g_{1}^{^{3}\text{He}}/F_{1}^{^{3}\text{He}}$ data at E = 5.89\,GeV.
         See Appendix~\ref{sec:a1-g1f1-syst-err} for a discussion of the various
         contributions to the systematic uncertainty.
        }
\label{tab:g1f1he3-syst-err-5}
\begin{ruledtabular}
\begin{tabular}{ccccc}
 $<x>$ & $A_{\parallel}$ & $A_{\perp}$ &  Kin.   &  Total  \\  
\hline
 0.277 &  5.320E-03 &  2.570E-03 &  5.000E-05 &  5.910E-03 \\
 0.325 &  1.720E-03 &  1.070E-03 &  4.000E-05 &  2.030E-03 \\
 0.374 &  1.170E-03 &  1.900E-04 &  1.000E-05 &  1.190E-03 \\
 0.424 &  1.350E-03 &  9.500E-04 &  4.000E-05 &  1.650E-03 \\
 0.474 &  3.160E-03 &  1.100E-04 &  5.000E-05 &  3.160E-03 \\
 0.524 &  1.200E-03 &  1.100E-04 &  0.000E+00 &  1.200E-03 \\
 0.573 &  1.640E-03 &  6.800E-04 &  1.000E-05 &  1.780E-03 \\
 0.624 &  2.720E-03 &  9.100E-04 &  2.000E-05 &  2.870E-03 \\
 0.674 &  2.420E-03 &  4.400E-04 &  1.000E-05 &  2.460E-03 \\
 0.723 &  2.370E-03 &  1.070E-03 &  2.000E-05 &  2.600E-03 \\
 0.773 &  3.900E-03 &  1.120E-03 &  4.000E-05 &  4.060E-03 \\
 0.823 &  7.090E-03 &  1.068E-02 &  1.800E-04 &  1.282E-02 \\
 0.873 &  8.480E-03 &  2.030E-03 &  1.000E-04 &  8.720E-03 \\
\end{tabular}
\end{ruledtabular}
\end{table*}

\begin{table*}[hbt]
\centering
\caption{Systematic uncertainties for $g_{1}^{n}/F_{1}^{n}$ at E = 4.74\,GeV.
         See Appendix~\ref{sec:a1-g1f1-syst-err} for a discussion of the various
         contributions to the systematic uncertainty.
        }
\label{tab:g1f1n-syst-err-4}
\begin{ruledtabular}
\begin{tabular}{cccccccc}
 $<x>$ & $F_{2}^{n,p}$  & $F_{2}^{^{3}\text{He}}$ & $\tilde{P}_{p}$  
& $\tilde{P}_{n}$ & $g_{1}^{p}/F_{1}^{p}$ &  $g_{1}^{^{3}\text{He}}/F_{1}^{^{3}\text{He}}$ &  Total   \\  
\hline
 0.277 &  2.530E-03 &  7.100E-04 &  5.320E-03 &  1.200E-04 &  2.950E-03 &  7.630E-03 &  1.011E-02 \\
 0.325 &  2.480E-03 &  9.800E-04 &  6.270E-03 &  1.300E-04 &  3.650E-03 &  3.060E-03 &  8.310E-03 \\
 0.374 &  5.640E-03 &  2.900E-04 &  7.230E-03 &  1.080E-03 &  4.680E-03 &  2.470E-03 &  1.064E-02 \\
 0.424 &  5.770E-03 &  1.730E-03 &  8.190E-03 &  1.100E-03 &  6.020E-03 &  6.360E-03 &  1.346E-02 \\
 0.473 &  6.020E-03 &  1.700E-03 &  9.240E-03 &  9.700E-04 &  7.600E-03 &  5.050E-03 &  1.445E-02 \\
\end{tabular}
\end{ruledtabular}
\end{table*}

\begin{table*}[hbt]
\centering
\caption{Systematic uncertainties for $g_{1}^{n}/F_{1}^{n}$ at E = 5.89\,GeV.
         See Appendix~\ref{sec:a1-g1f1-syst-err} for a discussion of the various
         contributions to the systematic uncertainty.
        }
\label{tab:g1f1n-syst-err-5}
\begin{ruledtabular}
\begin{tabular}{cccccccc}
 $<x>$ & $F_{2}^{n,p}$  & $F_{2}^{^{3}\text{He}}$ & $\tilde{P}_{p}$  
& $\tilde{P}_{n}$ & $g_{1}^{p}/F_{1}^{p}$ &  $g_{1}^{^{3}\text{He}}/F_{1}^{^{3}\text{He}}$ &  Total   \\  
\hline
 0.277 &  4.500E-03 &  4.500E-04 &  5.100E-03 &  2.450E-03 &  2.970E-03 &  1.193E-02 &  1.427E-02 \\
 0.325 &  2.180E-03 &  1.160E-03 &  6.030E-03 &  3.300E-04 &  3.760E-03 &  4.510E-03 &  8.780E-03 \\
 0.374 &  2.420E-03 &  6.200E-04 &  6.950E-03 &  5.300E-04 &  4.790E-03 &  2.920E-03 &  9.290E-03 \\
 0.424 &  3.110E-03 &  2.110E-03 &  7.890E-03 &  8.400E-04 &  6.020E-03 &  4.410E-03 &  1.152E-02 \\
 0.474 &  3.770E-03 &  1.260E-03 &  8.900E-03 &  7.600E-04 &  7.280E-03 &  9.170E-03 &  1.525E-02 \\
 0.524 &  9.150E-03 &  0.000E+00 &  9.980E-03 &  1.680E-03 &  9.620E-03 &  3.940E-03 &  1.716E-02 \\
 0.573 &  1.240E-02 &  1.500E-04 &  1.111E-02 &  2.280E-03 &  1.183E-02 &  7.140E-03 &  2.176E-02 \\
\end{tabular}
\end{ruledtabular}
\end{table*}

%===============================================================================
\subsection{Flavor decomposition systematic uncertainties} \label{sec:flavor-syst-err}
%===============================================================================

Tables~\ref{tab:u-syst-err-4} and~\ref{tab:d-syst-err-4} give a breakdown of the
systematic uncertainties at $E = 4.74$\,GeV for the up and down quark ratios, respectively.
Tables~\ref{tab:u-syst-err-5} and~\ref{tab:d-syst-err-5} list the uncertainties for $E = 5.89$\,GeV.
The columns of the table represent the contribution due to our $g_{1}^{n}/F_{1}^{n}$ data,
our fit to world $g_{1}^{p}/F_{1}^{p}$ data, the $(d + \bar{d})/(u + \bar{u})$ parameterization,
and the strange uncertainty, respectively.  The value of these uncertainties, except 
for the strange uncertainty (see Section~\ref{sec:res_flavor}), was determined by varying 
each of these contributions within reasonable limits and taking the change in the quark ratio 
as the uncertainty.  This was done for both the up and down quark ratios.  The in-quadrature 
sum of all contributions is displayed in the last column, labeled ``Total.''

\begin{table*}[!ht]
\centering
\caption{Systematic uncertainties for $(\Delta u + \Delta \bar{u})/(u + \bar{u})$ at $E = 4.74$\,GeV.
         See Appendix~\ref{sec:flavor-syst-err} for a discussion of the various
         contributions to the systematic uncertainty.
        }
\label{tab:u-syst-err-4}
\begin{ruledtabular}
\begin{tabular}{cccccc}
 $<x>$ & $g_{1}^{n}/F_{1}^{n}$ & $g_{1}^{p}/F_{1}^{p}$ & $(d + \bar{d})/(u + \bar{u})$ & $s$     & Total \\
\hline
0.277 &  9.900E-04 &  2.918E-02 &  3.260E-03 &  8.700E-03 &  3.064E-02 \\
0.325 &  7.800E-04 &  3.556E-02 &  3.590E-03 &  6.450E-03 &  3.633E-02 \\
0.374 &  9.300E-04 &  4.243E-02 &  3.610E-03 &  4.580E-03 &  4.284E-02 \\
0.424 &  1.090E-03 &  4.967E-02 &  4.620E-03 &  3.280E-03 &  5.000E-02 \\
0.473 &  1.100E-03 &  6.300E-02 &  5.070E-03 &  2.350E-03 &  6.326E-02 \\
\end{tabular}
\end{ruledtabular}
\end{table*}

\begin{table*}[!ht]
\centering
\caption{Systematic uncertainties for $(\Delta d + \Delta \bar{d})/(d + \bar{d})$ at $E = 4.74$\,GeV.
         See Appendix~\ref{sec:flavor-syst-err} for a discussion of the various
         contributions to the systematic uncertainty.
        }
\label{tab:d-syst-err-4}
\begin{ruledtabular}
\begin{tabular}{cccccc}
 $<x>$ & $g_{1}^{n}/F_{1}^{n}$ & $g_{1}^{p}/F_{1}^{p}$ & $(d + \bar{d})/(u + \bar{u})$ & $s$     & Total \\
\hline
0.277 &  8.240E-03 &  1.518E-02 &  1.455E-02 &  1.673E-02 &  2.811E-02 \\
0.325 &  7.140E-03 &  2.027E-02 &  1.886E-02 &  1.363E-02 &  3.167E-02 \\
0.374 &  9.290E-03 &  2.647E-02 &  2.280E-02 &  1.071E-02 &  3.770E-02 \\
0.424 &  1.193E-02 &  3.399E-02 &  3.467E-02 &  8.140E-03 &  5.066E-02 \\
0.473 &  1.343E-02 &  4.814E-02 &  4.754E-02 &  5.940E-03 &  6.923E-02 \\
\end{tabular}
\end{ruledtabular}
\end{table*}

\begin{table*}[!ht]
\centering
\caption{Systematic uncertainties for $(\Delta u + \Delta \bar{u})/(u + \bar{u})$ at $E = 5.89$\,GeV.
         See Appendix~\ref{sec:flavor-syst-err} for a discussion of the various
         contributions to the systematic uncertainty.
        }
\label{tab:u-syst-err-5}
\begin{ruledtabular}
\begin{tabular}{cccccc}
 $<x>$ & $g_{1}^{n}/F_{1}^{n}$ & $g_{1}^{p}/F_{1}^{p}$ & $(d + \bar{d})/(u + \bar{u})$ & $s$     & Total \\
\hline
0.277 &  1.380E-03 &  3.088E-02 &  1.980E-03 &  8.300E-03 &  3.207E-02 \\
0.325 &  8.100E-04 &  3.675E-02 &  3.750E-03 &  6.190E-03 &  3.746E-02 \\
0.374 &  7.800E-04 &  4.430E-02 &  3.800E-03 &  4.390E-03 &  4.469E-02 \\
0.424 &  9.400E-04 &  5.109E-02 &  4.460E-03 &  3.110E-03 &  5.139E-02 \\
0.474 &  1.160E-03 &  6.267E-02 &  4.810E-03 &  2.220E-03 &  6.290E-02 \\
0.524 &  1.220E-03 &  6.976E-02 &  4.130E-03 &  1.600E-03 &  6.991E-02 \\
0.573 &  1.430E-03 &  8.454E-02 &  3.860E-03 &  1.240E-03 &  8.465E-02 \\
\end{tabular}
\end{ruledtabular}
\end{table*}

\begin{table*}[!ht]
\centering
\caption{Systematic uncertainties for $(\Delta d + \Delta \bar{d})/(d + \bar{d})$ at $E = 5.89$\,GeV.
         See Appendix~\ref{sec:flavor-syst-err} for a discussion of the various
         contributions to the systematic uncertainty.
        }
\label{tab:d-syst-err-5}
\begin{ruledtabular}
\begin{tabular}{cccccc}
 $<x>$ & $g_{1}^{n}/F_{1}^{n}$ & $g_{1}^{p}/F_{1}^{p}$ & $(d + \bar{d})/(u + \bar{u})$ & $s$     & Total \\
\hline
0.277 &  1.158E-02 &  1.622E-02 &  8.640E-03 &  1.631E-02 &  2.716E-02 \\
0.325 &  7.470E-03 &  2.116E-02 &  1.973E-02 &  1.339E-02 &  3.274E-02 \\
0.374 &  7.850E-03 &  2.792E-02 &  2.503E-02 &  1.036E-02 &  3.969E-02 \\
0.424 &  1.039E-02 &  3.539E-02 &  3.438E-02 &  7.810E-03 &  5.102E-02 \\
0.474 &  1.436E-02 &  4.863E-02 &  4.929E-02 &  5.660E-03 &  7.094E-02 \\
0.524 &  1.731E-02 &  6.181E-02 &  5.298E-02 &  3.580E-03 &  8.331E-02 \\
0.573 &  2.341E-02 &  8.633E-02 &  6.641E-02 &  2.130E-03 &  1.114E-01 \\
\end{tabular}
\end{ruledtabular}
\end{table*}

   \end{appendix}    

   \clearpage

   % \bibliography{references}  

\end{document}